\documentclass[12pt]{article}
\usepackage{amssymb,amsmath}
\makeatletter

\usepackage{arydshln}

\newcount\comment@nesting

\begingroup
\xdef\comment@begincomment{\string\\begin\string\{comment\string\}}
\xdef\comment@endcomment{\string\\end\string\{comment\string\}}
\endgroup

\begingroup
\catcode`\^^M=\active
\def\@temp{\endgroup\def\comment@processline##1^^M}%
\@temp{%
    \def\comment@curline{#1}%
    \let\@next=\comment@processline
    \ifx\comment@curline\comment@endcomment
        \ifnum\comment@nesting=0
            \def\@next{\end{comment}}%
        \else
            \global\advance\comment@nesting by -1
        \fi
    \else
        \ifx\comment@curline\comment@begincomment
            \global\advance\comment@nesting by 1
        \fi
    \fi
    \@next
}

\makeatother

\usepackage[dvipdfmx]{graphicx}
\usepackage{bm}
\usepackage{color}
\usepackage{here}
\usepackage{enumerate}
\usepackage{here}
\usepackage[vcentermath]{youngtab}
\usepackage{subfigure}
\usepackage{mathrsfs}
\usepackage{dsfont}
\usepackage{tikz}
\usepackage{tikz-cd}
\usepackage{hyperref}

\DeclareMathOperator*{\Tr}{{\rm Tr}}

\renewcommand{\Re}{\operatorname{Re}}
\renewcommand{\Im}{\operatorname{Im}}
\newcommand{\Pf}{\operatorname{Pf}}

\numberwithin{equation}{section}

\allowdisplaybreaks[1]

\definecolor{mygreen}{rgb}{0,0.714,0.286}

\begin{document}

\thispagestyle{empty}

\begin{flushright}
%
YITP-25-184

\end{flushright}
\vskip1.5cm
\begin{center}
{\Large \bf 
Supersymmetric zeta functions and determinants \\
}

\vskip1.5cm
Yu Nakayama\footnote{yu.nakayama@yukawa.kyoto-u.ac.jp} 
and Tadashi Okazaki\footnote{tokazaki@seu.edu.cn}

\bigskip
{\it Center for Gravitational Physics and Quantum Information, Yukawa Institute for Theoretical Physics, Kyoto University,\\
	Kitashirakawa Oiwakecho, Sakyo-ku, Kyoto 606-8502, Japan

School of Physics and Shing-Tung Yau Center, Southeast University,\\
Yifu Architecture Building, No.2 Sipailou, Xuanwu district, \\
Nanjing, Jiangsu, 210096, China

}

\end{center}

\vskip1cm
\begin{abstract}
We define supersymmetric zeta functions and supersymmetric determinants, which can reveal spectral properties complementary to those captured by the supersymmetric indices. They play a crucial role in analyzing the Cardy-like behaviors of the supersymmetric indices and the supersymmetric Casimir energies associated with the supersymmetric partition functions. We investigate a variety of examples of the supersymmetric zeta functions and determinants for two-, four-, and six-dimensional supersymmetric field theories.
\end{abstract}
\newpage
\setcounter{tocdepth}{3}
\tableofcontents

\section{Introduction and conclusions}
In any theory of quantum systems, the most fundamental question is to understand the spectrum of operators (e.g., the Hamiltonian), specifically their eigenvalues and distributions. From them, we can construct various spectral functions. In physics, the most familiar spectral functions are the partition function and, when the theory is equipped with nilpotent fermionic symmetries, the index. The former is 
of fundamental interest 
because it encodes the thermodynamic properties of the system, and the latter, if available, 
is a highly useful and, more importantly, computable quantity 
that can be regarded as an invariant of the system under small perturbations and can be used to analyze dualities. The main advantage of studying the partition function or index is that it can be reformulated in terms of the path integral; this path integral expression, when compared with the spectral definition, sometimes yields remarkable formulae 
for certain geometric and topological invariants, such as the Atiyah-Singer index theorem \cite{Atiyah:1968mp,Alvarez-Gaume:1983zxc} and the Jones polynomial \cite{Witten:1988hf}.

In mathematics, other spectral functions are also of interest, such as
spectral zeta functions and spectral determinants. 
One advantage these alternative tools offer over partition functions is their potential for larger analyticity domains. Analytic continuation of the partition function in the "temperature" is highly non-trivial.
For instance, if the theory has an infinite tower of excited states, the partition function takes the form of the an infinite sum, which converges on the half-plane $\mathrm{Re} \beta>0$ and  may be singular at the high temperature limit $\beta \to 0$. 
On the other hand, the zeta function can have a meromorphic extension to the whole complex plane with only zeros and simple poles when the spectrum is discrete. 
The determinant takes the form of an infinite product with only isolated zeros and poles corresponding to the locations of the eigenvalues. 
Such structures can make complex analysis much easier. 
These different types 
of spectral functions can be connected to one another via a certain integral transform. Such an integral transform can map the original function space into another one, thereby providing a powerful method for finding analytical solutions to problems that might be difficult or impossible to solve by other methods.
The resulting form is often particularly suitable 
for the saddle point approximation in the presence of large parameters.
In particular, since the zeta functions and the partition functions are related via the Mellin transform, the structure of the zeros and poles of the zeta functions 
can encode the asymptotic behavior of the partition functions. 

The asymptotic behavior of the partition functions is typically associated with the geometry underlying the spectrum. 
Thus, the zeta functions encode such geometric information in a naturally regularized manner. 
In practice, when computing the partition functions, the quantum field theories and their path integrals are full of divergences. 
Nevertheless, 
the zeta functions provide a powerful tool for obtaining finite numbers while revealing the underlying geometry. 
This regularization, for instance, will be useful for understanding emergent geometries in large $N$ theories via holography, such as the AdS/CFT correspondence \cite{Maldacena:1997re}. 
The essential philosophy here is analogous to the classic question ``Can one hear the shape of a drum?'', posed by Kac \cite{MR201237}. 
In fact, Martelli, Sparks and Yau \cite{Martelli:2005tp,Martelli:2006yb,Martelli:2006vh} demonstrated that 
the volume and other geometric properties of the Sasaki-Einstein manifolds
can be read off from certain spectral data of the holographically dual 4d $\mathcal{N}=1$ superconformal field theories. 

The main motivation of this work is to investigate new spectral functions for physical theories.\footnote{A certain kind of spectral functions for a non-supersymmetric two-dimensional quantum field theory was  introduced and investigated by Fateev, Lukyanov and Zamolodchikov \cite{Fateev:2009jf}. Also see \cite{Litvinov:2024riz,Artemev:2025cev,Litvinov:2025geb,Meshcheriakov:2025lsx} for more generalizations. } 
We begin with the supersymmetric indices and introduce new types of spectral functions for the supersymmetric theories, 
\textit{supersymmetric zeta functions} and \textit{supersymmetric determinants}. Supersymmetric indices \cite{Witten:1982df,Kinney:2005ej,Romelsberger:2005eg} are well-established spectral functions that encode the BPS spectra of the supersymmetric theories. 
Defined as a trace over the BPS states, they 
are 
used to count the BPS local operators in the theory by reading off their expansion coefficients. 

In our initial exploration of these spectral functions for supersymmetric theories, we focus on the supersymmetric field theories in $d=2$, $4$ and $6$ dimensions. 
For such even-dimensional field theories, there exist local anomalies encoded by the anomaly polynomial $(d+2)$-form $\mathcal{A}_{d+2}$ that is a polynomial in the Chern and Pontryagin classes of gauge and gravity fields. 
It is known that the Cardy-like limits (high temperature limits) of the supersymmetric indices and the supersymmetric Casimir energy (zero temperature limits) of the supersymmetric partition functions on $S^1\times S^{d-1}$ 
have a universal relationship with the anomaly coefficients for the superconformal field theories. 
We find that these properties (e.g., the Cardy-like limits) can be understood as consequences of universal aspects of the supersymmetric zeta functions we introduce.
We propose that residues and special values of the supersymmetric zeta functions 
can be universally expressed in terms of the anomaly coefficients or central charges for the superconformal field theories. 
Conversely, the anomaly coefficients and the central charges for the superconformal field theories 
can be calculated from the supersymmetric zeta functions. 

\subsection{Structure}
In section \ref{sec_Szeta} we define the supersymmetric zeta functions. 
They are linked to the supersymmetric indices via the Mellin transform. 
We present formulae for the Cardy-like limit of the supersymmetric indices and the asymptotic degeneracy in terms of the supersymmetric zeta functions. 
Also, we propose that the special supersymmetric zeta values can encode the supersymmetric Casimir energies. 
In section \ref{sec_Sdet} we define the supersymmetric determinants as the third variant of the spectral functions. 
The vacuum exponents, which appear as their special values in the asymptotic formulae, can provide criteria for identifying a supersymmetry-breaking vacuum or an unstable vacuum. 
In section \ref{sec_2d}, section \ref{sec_4d} and section \ref{sec_6d}, we provide various examples 
of the supersymmetric zeta functions and supersymmetric determinants for 2d, 4d and 6d supersymmetric field theories. 
We present some conventions of the $q$-shifted factorials in Appendix \ref{app_convention}. 
Several useful formulae for the residues and the special values of the zeta functions are summarized 
in Appendices \ref{app_res} and \ref{app_values}. 

\subsection{Future works}

\begin{itemize}

\item It would be interesting to explore the universal structure of the residues and special values 
of the supersymmetric zeta functions for other settings of supersymmetric field theories, 
including the theories in odd dimensions, the configurations with defect operators and the deformations such as the topological twisting and the Higgsing. 
We hope to report on our results in future work. 

\item For the supersymmetric field theories with gravity duals the large $N$ limits of the supersymmetric indices capture the Kaluza-Klein modes of the supergravity.  
The associated supersymmetric zeta functions can be used to analyze the asymptotic behaviors, which typically grow much faster than those at finite $N$. 
We leave the detailed analysis of large $N$ results to future work. 

\item The supersymmetric zeta functions and supersymmetric determinants also contain several other theory-dependent data, 
such as zeros and their behavior at large values, which are also expected to encode certain physical information. At present, these aspects are less understood, and we hope to report on them in future work.

\item The supersymmetric indices may satisfy interesting relations under certain transformations, 
e.g. the elliptic and modular transformations \cite{Kawai:1993jk,Spiridonov:2012ww,Razamat:2012uv}, the $T$-reflection \cite{Berkovich:1995nx,Basar:2014mha}. 
From the perspective of the supersymmetric determinants, such properties are translated into certain functional equations. In turn, the supersymmetric zeta functions will obey certain algebraic relations. These relations will provide us with interesting avenues for further study. 

\item In mathematics, the arithmetic nature of the special zeta values is often discussed. 
The values of the Riemann zeta function at any positive even integer are transcendental due to the well-known fact that $\pi$ is transcendental \cite{MR1510165}. 
Although a proof is still lacking, it is conjectured that the zeta values at odd integers are algebraically independent over $\mathbb{Q}$ and have higher transcendentality (see e.g. \cite{MR3363457,MR1787183,MR1861452} for the irrationality). In physics, it is conjectured \cite{Intriligator:2003jj,Rastelli:2023sfk} that the central charges such as $a$ and $c$ take algebraic numbers or rational numbers, depending on the number of supersymmetries. 
It would be interesting to further understand the arithmetic nature of our supersymmetric zeta values that encode the central charges and connections to the ``zeta world" of number theory.

\item While the main focus of our paper is the asymptotic behaviors of the supersymmetric indices on the first sheet, it would be interesting to study the so-called second sheet (or higher sheet) indices by studying their multi-valued structures. 
Such indices turn out to play an important role in accounting for the black hole microstates, 
see e.g. \cite{Choi:2018hmj,Honda:2019cio,ArabiArdehali:2019tdm,Kim:2019yrz,Cabo-Bizet:2019osg,Amariti:2019mgp,ArabiArdehali:2019orz,GonzalezLezcano:2020yeb,Amariti:2020jyx,Cassani:2021fyv}. 
We leave the analysis of the associated supersymmetric zeta functions and determinants as an interesting direction for future work.

\end{itemize}
 
\section{Supersymmetric zeta functions}
\label{sec_Szeta}

\subsection{Definition}
We begin with the supersymmetric index of a supersymmetric theory. 
It can be defined by
\begin{align}
\label{index_DEF}
\mathbb{I}(q)={\Tr}_{\mathcal{H}} (-1)^F q^{\Delta}, 
\end{align}
where the trace is taken over the cohomology $\mathcal{H}$ of the chosen supercharges in the theory. 
$F$ is the fermion number operator, and $\Delta$ is a certain appropriate combination of the Cartan generators of global symmetries in the theory 
that commute with the chosen supercharges. 
When one expands the index 
\begin{align}
\label{index_expansion}
\mathbb{I}(q)&=\sum_{n}d(n)q^{n}, 
\end{align}
the expansion coefficients $d(n)$ can be identified with the \textit{degeneracy} of the BPS operators with charge (or ``energy") $n$ 
as they count the BPS bosonic and fermionic operators or states with positive and negative signs, respectively. 
When we consider a supersymmetric quantum field theory with a UV Lagrangian, the supersymmetric index can be computed as a certain supersymmetric partition function on the product space of a circle and a sphere $S^1 \times S^{d-1}$ (or more generally, on $S^{d-1}$ bundle over $S^1$ when angular momentum chemical potential is introduced) up to the supersymmetric Casimir energy factor. 

Here we assume that the degeneracy $d(n)$ exhibits the polynomial growth $d(n)$ $=$ $\mathcal{O}(n^\alpha)$ 
(see section \ref{sec_expgrowth} for the exponential growth). 
Let us write $q=e^{-\beta}$ and consider the asymptotic behavior of the number of states for the large charge $n$. 
By using the continuum approximation for the sum, 
one can view the expression (\ref{index_expansion}) as the Laplace transform of the degeneracy $d(n)$. 
Thus, we can obtain the asymptotic behavior of the degeneracy via the inverse Laplace transform of the supersymmetric index.
The inverse Laplace transform can be easily performed by taking the Laurent series expansion of the supersymmetric index of the form
\begin{align}
\label{index_beta_exp}
\mathbb{I}(\beta)&=\sum_{k}C_k \beta^k. 
\end{align}
The asymptotic behavior of the degeneracy with polynomial growth 
$d(n)$ $=$ $\mathcal{O}(n^\alpha)$ arises from the negative powers  of $\beta$. 
Performing the inverse Laplace transform, one finds 
\begin{align}
\label{asymp_d}
d(n)&\sim \sum_{k=1}^{\alpha+1}\frac{C_{-k}}{(k-1)!} n^{k-1}. 
\end{align}
Also, the asymptotic number of the BPS operators with charge less than or equal to $x$ is estimated as
\begin{align}
\label{asymp_Omega}
\Omega(x)&=\sum_{n\le x} d(n)
\nonumber\\
&\sim 
\int_{0}^x dn\ d(n)=
\sum_{k=1}^{\alpha+1}\frac{C_{-k}}{k!} x^k. 
\end{align}
Sometimes, $\Omega(x)$ may be better-behaved than $d(n)$ as a function of $x$ when $d(n)$ exhibits an oscillatory behavior. 

We now define our central quantities. 
We introduce an arithmetic generating function by defining the \textit{supersymmetric zeta function} by
\begin{align}
\label{Szeta_DEF}
\mathbb{Z}(s,z)&={\Tr}_{\mathcal{H}} (-1)^F \frac{1}{(\Delta+z)^s}
\nonumber\\
&=\sum_{n}\frac{d(n)}{(n+z)^{s}}. 
\end{align}

It can be regarded as the generalized Dirichlet series associated with the BPS degeneracy $d(n)$ of the supersymmetric theory. 
For the series to be well-defined as a function of the complex variable $s$, we need to consider its convergence properties.  
In general, if the degeneracy $d(n)$ exhibits polynomial growth $d(n)$ $=$ $\mathcal{O}(n^\alpha)$, the Dirichlet series (\ref{Szeta_DEF}) converges absolutely to an analytic function on the half plane $\mathrm{Re}(s)>\alpha+1$ \cite{MR256993}. 
In addition, the supersymmetric zeta functions can be analytically continued to meromorphic functions on the whole complex $s$-plane, 
which are holomorphic everywhere except for simple poles. 

\subsection{Mellin transform}
The supersymmetric zeta function can be obtained from the supersymmetric index 
by applying the following integral transform:  
\begin{align}
\label{Mellin_transf1}
\mathbb{Z}(s,z)&=\frac{1}{\Gamma(s)}\int_{1}^{\infty}dq\ \mathbb{I}(q^{-1}) q^{-z-1} (\log q)^{s-1}
\nonumber\\
&=\frac{1}{\Gamma(s)}\int_{1}^{\infty}dq\ \sum_{n}d(n) q^{-(n+z)-1} (\log q)^{s-1}
\nonumber\\
&=\sum_n d(n)
\frac{1}{\Gamma(s)} (n+z)^{-s}
\int_{0}^{\infty}dt\ t^{s-1} e^{-t}
\nonumber\\
&=\sum_{n}\frac{d(n)}{(n+z)^s}. 
\end{align}
Here, in the third line, we have interchanged the order of summation and integration (assuming this is valid) and made the substitution $q=e^{\frac{t}{n+z}}$. 
In the fourth line, we have used the Euler integral of the second kind
\begin{align}
\Gamma(w)&=\int_{0}^{\infty}dt\ t^{w-1}e^{-t}. 
\end{align}

Alternatively, we have
\begin{align}
\label{Mellin_transf2}
\mathbb{Z}(s,z)&=
\frac{1}{\Gamma(s)}\int_{0}^{\infty}d\beta\ 
\mathbb{I}(\beta)e^{-\beta z}\beta^{s-1}. 
\end{align}
This is the Mellin transform of the product $\mathbb{I}(\beta)e^{-\beta z}$. 
(Here $z$ can be viewed as an additional contribution to the charge of the ``vacuum state''.)
By splitting the integral (\ref{Mellin_transf2}) at $\beta=1$, 
we can use the Laurent expansion to perform the integration for small $\beta$, 
whereas the integration over the large $\beta$ region gives an entire function of $s$. 
It leads to the relation 
\begin{align}
C_{-k}&=(k-1)! \mathrm{Res}_{s=k}\ \mathbb{Z}(s,0) \qquad \textrm{for $k>0$}, 
\end{align}
where $\mathrm{Res}_{s=k}\ \mathbb{Z}(s,0)$ is the residue at a simple pole $s=k$ of 
the supersymmetric zeta function with $z=0$. 
By plugging this into (\ref{asymp_d}) and (\ref{asymp_Omega}), 
we get
\begin{align}
\label{asymp_d_2}
d(n)&\sim \sum_{k=1}^{\alpha+1} \mathrm{Res}_{s=k}\ \mathbb{Z}(s,0)\ n^{k-1}
\end{align}
and 
\begin{align}
\label{asymp_Omega_2}
\Omega(x)&\sim \sum_{k=1}^{\alpha+1} \frac{\mathrm{Res}_{s=k}\ \mathbb{Z}(s,0) }{k} x^k. 
\end{align}
The formula (\ref{asymp_Omega_2}) is a result of 
the Wiener-Ikehara Tauberian theorem \cite{MR1546472,ikehara1931extension} for the arithmetic generating function.  

If the supersymmetric zeta function (\ref{Szeta_DEF}) can be analytically continued to non-positive integers, 
its special values give rise to the regularized ``weighted moments'' of the charge $\Delta$
\begin{align}
\label{sZeta_moments}
\mathbb{Z}(-k,z)&={\Tr}_{\mathcal{H}}(-1)^F(\Delta+z)^k
\nonumber\\
&=\sum_{n} (n+z)^k d(n). 
\end{align}
For superconformal field theories, the charge $\Delta$ can be identified with the (R-symmetry twisted) Hamiltonian obtained from the radial quantization 
so that the special supersymmetric zeta values can be viewed as the regularized energy moments. 
It follows that these values appear as the non-negative power coefficients in the Laurent series (\ref{index_beta_exp}) for the supersymmetric index in terms of $\beta$
\begin{align}
C_k&=\frac{(-1)^k\mathbb{Z}(-k,0)}{k!} \qquad \textrm{for $k\ge 0$}. 
\end{align}

As a simple example, let us consider the half-BPS index for 4d $\mathcal{N}=4$ $U(N)$ super Yang-Mills (SYM) theory
\begin{align}
\label{ind_4dN4H_uN}
\mathcal{I}_{\textrm{$\frac12$BPS}}^{\textrm{4d $\mathcal{N}=4$ $U(N)$}}(q)
&=\prod_{n=1}^N \frac{1}{(1-q^n)}
=\sum_{n_1,\cdots, n_N\ge0} q^{n_1+2n_2+\cdots+Nn_N}. 
\end{align}
The supersymmetric spectral zeta function is given by
\begin{align}
\label{Szeta_4dN4H_uN}
\mathbb{Z}_{\textrm{$\frac12$BPS}}^{\textrm{4d $\mathcal{N}=4$ $U(N)$}}(s,z)
&=\zeta_{N}(s,z;1,2,\cdots,N), 
\end{align}
where
\begin{align}
\label{Bmultiplezeta}
\zeta_N(s,z;\omega_1,\cdots,\omega_N)
&=\sum_{n_1,\cdots, n_N\ge0}(n_1\omega_1+\cdots+n_N\omega_N+z)^{-s}
\end{align}
is the Barnes multiple zeta function \cite{barnes1904theory}. 
It has simple poles at $s=1,\cdots, N$ with residues
\begin{align}
\label{res_multipleBzeta}
\mathrm{Res}_{s=N-i}\ 
\zeta_N(s,z;\omega_1, \cdots, \omega_N)
&=\frac{F_i(e_1(\omega),\cdots, e_k(\omega);z)}{\Gamma(N-i)}
\prod_{j=1}^{N}\frac{1}{\omega_j}, 
\end{align}
with $i=0,1,\cdots, N-1$. 
Here  
\begin{align}
F_i(e_1(\omega),\cdots,e_k(\omega);z)&=\sum_{k=0}^{i}T_k(e_1(\omega),\cdots, e_k(\omega))
\frac{(-z)^{i-k}}{(i-k)!}
\end{align}
is a function of the elementary symmetric functions $e_1(\omega)$, $\cdots$, $e_k(\omega)$ and $z$. 
The functions $T_k$ $(e_1(\omega),\cdots, e_k(\omega))$ are the $k$-th Todd polynomials \cite{MR82174} 
which can be defined by
\begin{align}
\prod_{i=1}^N \frac{tx_i}{1-e^{-tx_i}}
&=\sum_{k=0}^{\infty}
T_k(e_1(x),\cdots, e_k(x))t^k. 
\end{align}
The residue formula (\ref{res_multipleBzeta}) is obtained from the integral representation 
\begin{align}
\zeta_N(s,z;\omega_1,\cdots,\omega_N)
&=\frac{1}{\Gamma(s)}
\int_0^{\infty}dt\ t^{s-1} \frac{e^{-zt}}{(1-e^{-\omega_1 t})\cdots (1-e^{-\omega_N t})}
\end{align}
of the Barnes multiple zeta function. 
For example, 
\begin{align}
T_0&=1, \\
T_1(e_1)&=\frac12 e_1(x), \\
T_2(e_1,e_2)&=\frac{1}{12}(e_1(x)^2+e_2(x)), \\
T_3(e_1,e_2,e_3)&=\frac{1}{24}e_1(x)e_2(x), \\
T_4(e_1,e_2,e_3,e_4)&=\frac{1}{720}(-e_1(x)^4+4e_1(x)^2e_2(x)+3e_2(x)^2+e_1(x)e_3(x)-e_4(x)). 
\end{align}
The Barnes multiple zeta function (\ref{Bmultiplezeta}) converges absolutely for $\mathrm{Re}(s)>N$. 
It follows that the supersymmetric zeta function (\ref{Szeta_4dN4H_uN}) has simple poles at $s=N-i$, $i$ $=$ $0,\cdots, N-1$   
with residues 
\begin{align}
\label{Res_4dN4H_uN}
&
\mathrm{Res}_{s=N-i}\ \mathbb{Z}_{\textrm{$\frac12$BPS}}^{\textrm{4d $\mathcal{N}=4$ $U(N)$}}(s,z)
\nonumber\\
&=\frac{1}{\Gamma(N-i)N!}
\sum_{k=0}^{i}T_k(|s(N+1,N)|,\cdots, |s(N+1,N+1-k)|)
\frac{(-z)^{i-k}}{(i-k)!}, 
\end{align}
where $s(n,m)$ is the Stirling number of the first kind. 
The residues (\ref{Res_4dN4H_uN}) encode the terms with $\beta^{-N+i}$ 
in the Laurent series expansion of the supersymmetric index (\ref{ind_4dN4H_uN}). 
By the inverse Laplace transform, we find the asymptotic degeneracy
\begin{align}
d_{\textrm{$\frac12$BPS}}^{\textrm{4d $\mathcal{N}=4$ $U(N)$}}(n)&\sim 
\sum_{k=0}^{N-1}
\frac{T_{N-k-1}(|s(N+1,N)|,\cdots,|s(N+1,N+1-k)|)}{N!k!}n^k
\nonumber\\
&=\frac{1}{N! (N-1)!}n^{N-1}+
\frac{N+1}{4(N-1)!(N-2)!}n^{N-2}
\nonumber\\
&+\frac{(N+1)(9N^2+5N-2)}{288 (N-1)!(N-3)!}n^{N-3}
+\frac{N(N+1)^2(3N+2)}{1152(N-2)!(N-4)!}n^{N-4}+\cdots
\end{align}
The number of the BPS states with charge less than or equal to $x$ is
\begin{align}
\Omega_{\textrm{$\frac12$BPS}}^{\textrm{4d $\mathcal{N}=4$ $U(N)$}}(x)&\sim 
\sum_{k=0}^{N-1}
\frac{T_{N-k-1}(|s(N+1,N)|,\cdots,|s(N+1,N+1-k)|)}{N!(k+1)!}x^{k+1}
\nonumber\\
&=\frac{1}{(N!)^2}x^{N}
+\frac{N+1}{4 ((N-1)!)^2}x^{N-1}
\nonumber\\
&+\frac{(N+1)(9N^2+5N-2)}{288 (N-1)!(N-2)!}x^{N-2}
+\frac{N(N+1)^2(3N+2)}{1152(N-2)!(N-3)!}x^{N-3}+\cdots
\end{align}
The special values of the Barnes zeta function at negative integers are given by \cite{MR3231418}
\begin{align}
\zeta_{N}(-k,z;\omega_1,\cdots, \omega_N)
&=\frac{(-1)^N k!}{(k+N)!}B_{k+N}(z;\omega_1,\cdots,\omega_N), 
\end{align}
where the multiple Bernoulli polynomial is defined by 
\begin{align}
\frac{x^n e^{xz}}{(e^{\omega_1 x}-1)\cdots(e^{\omega_n x}-1)}
&=\sum_{k\ge 0}B_k(z;\omega_1,\cdots,\omega_n)\frac{x^k}{k!}, 
\end{align}
the energy moments are given by
\begin{align}
\mathbb{Z}_{\textrm{$\frac12$BPS}}^{\textrm{4d $\mathcal{N}=4$ $U(N)$}}(-k,0)
&=\frac{(-1)^N k!}{(k+N)!}B_{k+N}(0;1,\cdots,N). 
\end{align}

\subsection{Multiple supersymmetric zeta functions}
\label{sec_multiplezeta}
Our supersymmetric zeta functions can be extended to multiple supersymmetric zeta functions by introducing additional fugacities 
coupled to the Cartan generators of global symmetries in a straightforward fashion. 

For the supersymmetric field theories in $d=4$, $5$ and $6$ dimensions, there exist multiple Cartan generators of the rotational symmetry group. 
While these Cartan generators are coupled to independent fugacities $\{q_I\}$ for the supersymmetric indices, 
we first replace them with the exponential variables $\{\omega_I\}$ with the same base $q$ for the supersymmetric zeta functions, 
i.e. $q_I=q^{\omega_I}$. 
Then we obtain from (\ref{Mellin_transf1}) the multiple supersymmetric zeta functions depending on $\{\omega_I\}$ 
as the Mellin transform of the supersymmetric indices. 
We will give a variety of examples of the multiple supersymmetric zeta functions for 4d and 6d supersymmetric field theories 
in section \ref{sec_4d} and section \ref{sec_6d}. 

Furthermore, when the theories have other global symmetries, the supersymmetric indices can be refined by introducing equivariant parameters, also known as flavor fugacities $\{x_{\alpha}\}$ coupled to their Cartan generators. 
The resulting indices, which are referred to as the flavored supersymmetric indices, are formal multivariable Taylor series in the fugacities $\{q_i\}$ 
whose coefficients are Laurent series in the flavor fugacities $\{x_{\alpha}\}$. 
In this case, we obtain the flavored supersymmetric zeta functions depending on the flavor fugacities $\{x_{\alpha}\}$ 
by performing the Mellin transform (\ref{Mellin_transf1}) of the flavored supersymmetric indices. 
In this work, we specialize to the unflavored case for the supersymmetric zeta functions. 
We leave a full discussion on the flavored supersymmetric zeta functions to future work. 

\subsection{Degeneracy with exponential growth}
\label{sec_expgrowth}
If the supersymmetric indices have expansion coefficients of polynomial growth, the supersymmetric zeta functions can be directly defined by taking the Mellin transform (\ref{Mellin_transf1}) of the supersymmetric indices. 
However, the supersymmetric indices $\mathcal{I}(q)$ in supersymmetric field theories are typically defined as certain traces over the Fock space for a multi-particle system 
so that the degeneracy $d(n)$ may not exhibit polynomial growth. 
In such cases, we define the supersymmetric zeta functions $\mathfrak{Z}(s,z)$ by the Mellin transform  
of the ``single-particle indices $i(q)$'', which we obtain by taking the plethystic logarithm of the supersymmetric indices $\mathcal{I}(q)$. 
As we will discuss below, the resulting supersymmetric zeta functions turn out to play a crucial role in studying the asymptotic Cardy-like behaviors 
and the supersymmetric Casimir energies. 

\subsubsection{Cardy-like limit}
\label{sec_Cardylim}
Consider the single-particle index $i(q)$ that is defined by the plethystic logarithm \cite{MR1601666} of the supersymmetric index $\mathcal{I}(q)$
\begin{align}
i(q)&=\mathrm{PL}[\mathcal{I}(q)]
=\sum_{d\ge1}\frac{\mu(d)}{d} \log \mathcal{I}(q^d), 
\end{align}
where $\mu(d)$ is the M\"{o}bius function. 
Conversely, one can recover the supersymmetric index $\mathcal{I}(q)$ by taking the plethystic exponential of the single-particle index $i(q)$
\begin{align}
\mathcal{I}(q)&=\mathrm{PE}[i(q)]
=\exp\left[\sum_{k=1}^{\infty} \frac{i(q^k)}{k}\right]. 
\end{align}
When the single-particle index $i(q)$ can be expanded as
\begin{align}
i(q)&=\sum_{\nu} \delta(\nu) q^{\nu}, 
\end{align}
where $\delta(\nu)$ may not be integers, we define the supersymmetric zeta function as
\begin{align}
\label{Szeta_DEF_sind}
\mathfrak{Z}(s,z)
&=\sum_{\nu}\frac{\delta(\nu)}{(\nu+z)^s}. 
\end{align}
As discussed in section \ref{sec_multiplezeta}, this definition can be straightforwardly generalized to multiple supersymmetric zeta functions by introducing additional variables. 

We assume that the supersymmetric zeta function (\ref{Szeta_DEF_sind}) has simple poles at $s=\alpha_i$ and that it absolutely converges for $\Re(s)$ $>$ $\alpha_*$, 
where $\alpha_*$ corresponds to the rightmost pole. 
Also, we assume that it can be analytically continued to the region $\Re(s)$ $\ge$ $-C_0$ $(0<C_0<1)$. 
Furthermore, we assume that as $|\Im(s)|$ $\rightarrow$ $\infty$, 
\begin{align}
\label{Zeta_assumpO}
\mathfrak{Z}(s,z)&=\mathcal{O}(|\Im(s)|^{C_1})
\end{align}
uniformly on $\Re(s)$ $\ge$ $-C_0$ for some positive number $C_1$. 
The number $C_1$ can be viewed as the order of growth of the supersymmetric zeta function, 
which generalizes the least possible exponent (order of growth) \cite{MR1994094,MR882550} of the Riemann zeta function. 

We note that the logarithm of the supersymmetric index $\mathcal{I}(q)$ is given by
\begin{align}
\label{log_ind1}
\log \mathcal{I}(\beta)
&=\sum_{k=1}^{\infty} \frac{i(e^{-k \beta})}{k}
\nonumber\\
&=\sum_{k=1}^{\infty}\sum_{\nu}\frac{\delta(\nu)}{k}e^{-\nu k \beta}, 
\end{align}
where we have set $q=e^{-\beta}$. 
Using the Mellin inversion formula
\begin{align}
e^{-\beta}&=\frac{1}{2\pi i}\int_{\sigma_0-i\infty}^{\sigma_0+i\infty} 
ds\ \beta^{-s}\Gamma(s)
\end{align}
for $\Re \beta>0$ and $\sigma_0>0$, 
the exponential function in (\ref{log_ind1}) can be rewritten. 
Consequently, we get 
\begin{align}
\label{log_ind2}
\log \mathcal{I}(\beta)
&=\frac{1}{2\pi i}
\int_{\alpha_*+1-i\infty}^{\alpha_*+1+i\infty}ds\ 
\beta^{-s} \Gamma(s) \zeta(s+1) \mathfrak{Z}(s,0). 
\end{align}
Here we have interchanged the order of summation and integration, which is justified by the assumed absolute convergence of the series.
By shifting the line of integration from $\Re(s)=\alpha_*+1$ to $\Re(s)=-C_0$, 
we can evaluate the integral (\ref{log_ind2}) by picking up the residues
\begin{align}
\beta^{-\alpha_i}\Gamma(\alpha_i)\zeta(\alpha_i+1)\mathrm{Res}_{s=\alpha_i}\mathfrak{Z}(s,0)
\end{align}
at simple poles $s=\alpha_i$ 
and the residue
\begin{align}
\mathfrak{Z}(0,0)'-\mathfrak{Z}(0,0)\log \beta
\end{align}
at a double pole $s=0$, 
where $\mathfrak{Z}(0,0)'$ $=$ $\frac{\partial}{\partial s}\mathfrak{Z}(s,z)|_{s=0,z=0}$. 
Then we find that 
\begin{align}
\log \mathcal{I}(\beta)
&=\sum_{i}\frac{\mathrm{Res}_{s=\alpha_i}\mathfrak{Z}(s,0)\Gamma(\alpha_i)\zeta(\alpha_i+1)}{\beta^{\alpha_i}}
-\mathfrak{Z}(0,0)\log\beta+\mathfrak{Z}(0,0)'
\nonumber\\
&+\frac{1}{2\pi i}
\int_{-C_0-i\infty}^{-C_0+i\infty} ds\ 
\beta^{-s}\Gamma(s)\zeta(s+1)\mathfrak{Z}(s,0). 
\end{align}
The final integral can be bounded as follows: 
\begin{align}
&\left|\frac{1}{2\pi i}
\int_{-C_0-i\infty}^{-C_0+i\infty} ds\ 
\beta^{-s}\Gamma(s)\zeta(s+1)\mathfrak{Z}(s,0)\right|
\nonumber\\
&=\mathcal{O}\left(
|\beta|^{C_0} \int_{-\infty}^{\infty}dt 
\exp\left(-\frac{\pi}{2} |t| \right)
|t|^{C_1+C_2+C_3}
\right)
\nonumber\\
&=\mathcal{O}\left(|\beta|^{C_0}\right), 
\end{align}
where we have used the assumption (\ref{Zeta_assumpO}), 
the least possible exponent (order of growth) of the Riemann zeta function \cite{MR1994094,MR882550}
\begin{align}
\zeta(s+1)
&=\mathcal{O}(|\mathrm{Im}(s)|^{C_2})
\end{align}
and the asymptotic estimate of the gamma function \cite{MR698779}
\begin{align}
\Gamma(s)&\sim \sqrt{2\pi}|\mathrm{Im}(s)|^{\mathrm{Re}(s)-\frac12}\exp \left(-\frac{\pi}{2}|\mathrm{Im}(s)|\right),
\end{align} 
which implies $\Gamma(s) =\mathcal{O}\left(\exp\left(-\frac{\pi}{2}|\mathrm{Im}(s)|\right)|\mathrm{Im}(s)|^{C_3} \right)$ 
as $|\mathrm{Im}(s)|$ $\rightarrow$ $\infty$. 

Hence, we obtain the formula for the asymptotic behavior as $\beta\rightarrow 0$
\begin{align}
\label{Cardy_lim0}
&
\log \mathcal{I}(\beta)
\nonumber\\
&=\sum_{i} \frac{\mathrm{Res}_{s=\alpha_i}\mathfrak{Z}(s,0)\Gamma(\alpha_i)\zeta(\alpha_i+1)}{\beta^{\alpha_i}}
-\mathfrak{Z}(0,0)\log\beta+\mathfrak{Z}(0,0)'
+\mathcal{O}\left(|\beta|^{C_0}\right). 
\end{align}
The asymptotic formula (\ref{Cardy_lim0}) applies to numerous supersymmetric indices with exponential growth. 
We refer to it as the Cardy-like limit of the supersymmetric index 
(see \cite{Cardy:1986ie} for the original Cardy formula for the partition function of a 2d CFT). 
The formula (\ref{Cardy_lim0}) asserts that the terms $\beta^{-\alpha_i}$ with $\alpha_i>0$ in the Cardy-like limit of the supersymmetric indices 
correspond to simple poles at $\alpha_i$ of the supersymmetric zeta functions $\mathfrak{Z}(s,z)$, and that their coefficients are encoded by the residues at these poles with $z=0$. 
For superconformal field theories in 2d, 4d, and 6d, it has been pointed out that these coefficients can be universally determined by the anomaly coefficients \cite{Gadde:2013lxa,DiPietro:2014bca} 
(see e.g. \cite{Buican:2015ina,ArabiArdehali:2015ybk,Cecotti:2015lab,Choi:2018hmj,Honda:2019cio,ArabiArdehali:2019tdm,Kim:2019yrz,Cabo-Bizet:2019osg,Cassani:2021fyv,Nahmgoong:2019hko} for further variants). 
In such cases, there exist universal relations between the residues of the supersymmetric zeta functions and the anomaly coefficients, 
as we will see in section \ref{sec_2d}, section \ref{sec_4d} and section \ref{sec_6d}. 

There may exist a logarithmic correction in (\ref{Cardy_lim0}) if the supersymmetric zeta value $\mathfrak{Z}(0,0)$ is non-trivial. 
We shall call $\mathfrak{Z}(0,0)$ the \textit{Zeta-index}. 
Note that it differs from the ordinary Witten index as the former is associated with the plethystic logarithm of a supersymmetric index that exhibits exponential growth.  
It has been argued \cite{DiPietro:2014bca,ArabiArdehali:2015ybk} that the logarithmic term in the Cardy-like limit of the 4d $\mathcal{N}=1$ supersymmetric index arises 
due to the presence of the unlifted zero modes parameterizing the quantum Coulomb branch of the 3d $\mathcal{N}=2$ gauge theory 
obtained upon compactifying the 4d $\mathcal{N}=1$ gauge theory on a circle. 
Accordingly, the Zeta-index for the 4d $\mathcal{N}=1$ gauge theory is expected to encode 
the complex dimension of the quantum Coulomb branch of the resulting 3d theory. 

Moreover, the constant term in (\ref{Cardy_lim0}) is given by the $s$-derivative of the supersymmetric zeta function evaluated at the origin ($s=0$, $z=0$). 
This is naturally characterized by a special value of the supersymmetric determinant discussed in section \ref{sec_Sdet}. 

\subsubsection{Effective Field Theory interpretation}
At this point, it is instructive to discuss the effective field theory interpretation of the formula \eqref{Cardy_lim0}. 
The supersymmetric index $\mathcal{I}$ is, in fact, a specific partition function on $S^1 \times S^{d-1}$ (twisted by R-symmetry, etc.). Therefore, the effective field theory analysis in the $\beta \to 0$ limit can be applied to $\mathcal{I}(\beta)$ just as it would be to a standard partition function.
The idea is that the supersymmetric index, or more specifically the superconformal index, can be regarded as a type of partition function on $S^1\times S^{d-1}$. On the other hand, we can predict the behavior of the partition function in the small $\beta$ limit by using the effective field theory techniques. The comparison gives a physical interpretation of the asymptotic behavior of the free energy as a function of $\beta^{-1}$.

Suppose we want to study the partition function of a certain conformal field theory on $S^1 \times S^{d-1}$. It can be realized as a path integral over $S^1 \times S^{d-1}$ with appropriate boundary conditions on the Euclidean time circle of radius $\beta$. 
Let us assume that the boundary condition is such that there is no zero mode, and the low-energy effective field theory on $S^{d-1}$ becomes gapped. In this situation, the partition function, more precisely the free energy (i.e., the logarithm of the partition function), must be given by a local functional of the background metric $\hat{g}_{\mu\nu} = \beta^{-2} g_{\mu\nu}$ on $S^{d-1}$ (up to Weyl anomaly). This is the consequence of the Weyl invariance because the original metric $ds^2 = \beta^2 d\tau^2 + g_{\mu\nu} dx^\mu dx^\nu$ is Weyl equivalent to $d\hat{s}^2 = d\tau^2 + \beta^{-2} g_{\mu\nu} dx^\mu dx^\nu$.  In more complicated situations, we may further introduce the background Kaluza-Klein graviphoton to describe the spin and the other gauge fields to describe chemical potentials of global symmetries.

If we accept the effective field theory analysis, we should have
\begin{align}
Z(\beta)_{\mathrm{CFT}_d} = Z(\beta)_{\mathrm{gapped}_{d-1}} \sim e^{-S_{d-1}[\hat{g}]}
\end{align}
with the effective action given by the derivative expansion:
\begin{align}
S[g] = \int d^{d-1}x \sqrt{\hat{g}} (-f + c_1 \hat{R}+ c_2 \hat{R}^2 + \cdots).  
\end{align}
Substituting the metric $\hat{g}$ for $S^{d-1}$ explicitly, we find 
\begin{align}
S[g] = \mathrm{Vol} S^{d-1} \left(-f \beta^{1-d} + (d-2)(d-1) c_1 \beta^{3-d} + (d-2)^2(d-1)^2 c_2 \beta^{5-d} \cdots \right) \ .
\end{align}
It should be compared with \eqref{Cardy_lim0}. In particular,  the residues and special values of $\mathfrak{Z}(s,0)$ are directly related to the effective field theory Wilson coefficients $f$, $c_1$, $c_2$, and so on.

For a more precise comparison, we need to consider several additional features. First of all, our target is the supersymmetric index and not the partition function. Furthermore, our definition of ``time evolution" may involve extra contributions from the gauge field background. For instance, the ``Hamiltonian" defining the superconformal index contains rotation and the R-symmetry chemical potential, and the extra terms in the effective action, such as the Chern-Simons terms for the background gauge field, will also contribute. In addition, from the viewpoint of the effective field theory, the supersymmetry introduces further constraints: the supersymmetry forbids the appearance of several terms in the effective action, such as the ``cosmological constant" $f$, or the supersymmetry relates various terms, e.g., the Einstein-Hilbert term with the R-symmetry Chern-Simons term in $d=4$.

A systematic analysis of the effective field theory for the superconformal index can be found, e.g., in the work \cite{DiPietro:2014bca} of di Pietro and Komargodski. 
They, for example, showed that, after taking into account the R-symmetry twisting, 
the Wilson coefficient $c_1$ is proportional to a particular combination of the central charges $c-a$ in $d=4$. 
They also studied the prediction of several coefficients in $d=6$ from the anomaly polynomial, and further confirmation was provided in the work \cite{Chang:2019uag} of Chang et al. by studying the 5d effective action. 
We will use these facts in later sections.

If the $S^1$ compactification leaves the ``gapless" degrees of freedom or zero modes on $S^{d-1}$, the effective field theory expansion can fail. Its manifestation is the presence of the $\log \beta$ term, which we called the Zeta-index. We have already mentioned the claim that the coefficient in 4d encodes the complex dimension of the quantum Coulomb branch of the compactified 3d theory. Such a contribution cannot be written as a local functional of the metric and gauge fields. 

Finally, when $d$ is even, the Weyl anomaly exists, so performing the Weyl transformation from $ds^2 = \beta^{2} d\tau^2 + g_{\mu\nu} dx^\mu dx^\nu$ to $d\tilde{s}^2 = d\tau^2 + \beta^{-2} g_{\mu\nu} dx^\mu dx^\nu$ introduces extra terms proportional to $\beta$, which are related to the Casimir energy that we will discuss in section \ref{sec_Casimir}. In the work \cite{Benjamin:2023qsc} of Benjamin et al., it was argued that if we could choose a scheme where the Casimir energy in the $d\tilde{s}^2$ metric is zero, the  Casimir energy in the $ds^2$ metric should be entirely determined from the Weyl anomaly. We will come back to this point in the context of the superconformal index in section \ref{sec_Casimir}.

\subsubsection{Asymptotic degeneracy}
\label{sec_asymp_Dege}
We next consider the asymptotic behavior of the degeneracy $d(n)$. 
Here we assume that it grows exponentially as $n\rightarrow \infty$.\footnote{An important caveat is that if the degeneracy $d(n)$ does not exhibit exponential growth (say, if it alternates between positive and negative values), the following asymptotic analysis will not work in general.} 
According to the Cauchy integral theorem, 
we have
\begin{align}
\label{Cauchy_int1}
d(n)&=\frac{1}{2\pi i}\int_{\beta_0}^{\beta_0+2\pi i} d\beta\ \mathcal{I}(\beta) e^{n\beta}. 
\end{align}
With $\beta$ $=$ $y+2\pi ix$, it can be written as
\begin{align}
\label{Cauchy_int2}
d(n)&=\int_{-\frac12}^{\frac12}
dx\ \mathcal{I}(y+2\pi ix) e^{ny+2\pi i n x}. 
\end{align}

Our strategy is to apply the saddle point approximation to the integral (\ref{Cauchy_int2}). 
As the supersymmetric index $\mathcal{I}(\beta)$ has a singularity at $\beta=0$, 
we wish to evaluate the dominant saddle by making use of the Cardy-like limit (\ref{Cardy_lim0}) as $\beta\rightarrow 0$. 
The absolute value of the integrand can be approximated as
\begin{align}
&
|\mathcal{I}(y+2\pi ix)e^{ny+2\pi inx}|
\nonumber\\
&\sim 
\left|
\exp \left[ 
\mathrm{Res}_{s=\alpha_*}\mathfrak{Z}(s,0)\Gamma(\alpha_*) \zeta(\alpha_*+1)(y+2\pi ix)^{-\alpha_*}
+ny\right] 
\right|, 
\end{align}
where $\alpha_*$ is the location of the rightmost pole as discussed in section \ref{sec_Cardylim}. 
Provided that the residue $\mathrm{Res}_{s=\alpha_*}\mathfrak{Z}(s,0)$ at the rightmost pole for $z=0$ is positive, 
the maximum absolute value of the integrand in (\ref{Cauchy_int2}) appears when $x=0$. 
Then the integrand with $x=0$ can be well approximated by
\begin{align}
\exp\left[
\mathrm{Res}_{s=\alpha_*}\mathfrak{Z}(s,0)
\Gamma(\alpha_*)\zeta(\alpha_*+1)y^{-\alpha_*}+ny
\right], 
\end{align}
Applying the saddle point approximation, 
we should choose $y$ by solving 
\begin{align}
\frac{\partial}{\partial y}
\left\{
\exp\left[
\mathrm{Res}_{s=\alpha_*}\mathfrak{Z}(s,0)
\Gamma(\alpha_*)\zeta(\alpha_*+1)y^{-\alpha_*}+ny
\right]
\right\}&=0. 
\end{align}
Then we find
\begin{align}
\label{y_value}
y&=n^{-\frac{1}{\alpha_*+1}}\left[
\mathrm{Res}_{s=\alpha_*}\mathfrak{Z}(s,0)\Gamma(\alpha_*+1)\zeta(\alpha_*+1)
\right]^{\frac{1}{\alpha_*+1}}. 
\end{align}

Substituting (\ref{y_value}) into (\ref{Cauchy_int2}), 
we obtain
\begin{align}
\label{Cauchy_int3}
d(n)&\sim e^{m}\int_{-\frac12}^{\frac12}
dx\ \mathcal{I}(y+2\pi ix) e^{2\pi i n x}, 
\end{align}
where we have defined 
\begin{align}
m&=n^{\frac{\alpha_*}{\alpha_*+1}}
\left[
\mathrm{Res}_{s=\alpha_*}\mathfrak{Z}(s,0)
\Gamma(\alpha_*+1)\zeta(\alpha_*+1)
\right]^{\frac{1}{\alpha_*+1}}. 
\end{align}
To proceed, we also note that the integral (\ref{Cauchy_int3}) can be further estimated 
by specifying a path where the angle $\arg(\beta)$ is fixed. 
Suppose that $|x|$ $\le$ $\frac{y}{2\pi}$. 
Then we have the condition $|\arg(\beta)|\le \frac{\pi}{4}$. 
In this case, as $y$ is sufficiently small so that we can split the integral (\ref{Cauchy_int3}) into two parts 
i) $0\le |x|\le y^{\delta}$ and ii) $y^{\delta}\le |x|\le \frac12$ for some $\delta$ with $1+\frac{\alpha^*}{3}$ $<$ $\delta$ $<$ $1+\frac{\alpha^*}{2}$ 
for which the former region is dominant in the large $n$ limit. 
This process is similar to the Hardy-Ramanujan Farey dissection that divides the unit circle into specific segments or arcs in the Hardy-Ramanujan-Littlewood circle method \cite{MR1575586}. 
The dominant integral takes the form
\begin{align}
\label{Cauchy_int4}
d(n)&\sim 
\exp\left[ m\left(1+\frac{1}{\alpha_*}\right)+\mathfrak{Z}(0,0)' \right]
\nonumber\\
&\times 
\int_{-(\frac{m}{n})^{\delta}}^{(\frac{m}{n})^{\delta}}
dx\ 
\exp\Biggl\{
\frac{m}{\alpha_*}\left[
-1+\left(1+\frac{2\pi in}{m}x\right)^{-\alpha_*}
\right]
+2\pi i nx-\mathfrak{Z}(0,0) \log\left(\frac{m}{n}+2\pi ix\right)
\Biggr\}, 
\end{align}
where we have used the Cardy-like limit (\ref{Cardy_lim0}). 
After the variable change $2\pi x$ $=$ $\frac{m}{n}\omega$, 
the expression can be rewritten as
\begin{align}
\label{Cauchy_int5}
d(n)&\sim 
\exp\left[
m\left(1+\frac{1}{\alpha_*}\right)+\mathfrak{Z}(0,0)'
-(\mathfrak{Z}(0,0)-1)\log\left(\frac{m}{n}\right)
-\log(2\pi)
\right]
\nonumber\\
&\times 
\int_{-2\pi (\frac{m}{n})^{\delta-1}}^{2\pi (\frac{m}{n})^{\delta-1}} 
d\omega
\exp\left[
m\left(
\frac{1}{\alpha_*(1+i\omega)^{\alpha}}
-\frac{1}{\alpha_*}+i\omega
\right)
-\mathfrak{Z}(0,0)\log(1+i\omega)
\right]. 
\end{align}
Again taking the limit $m\rightarrow \infty$ (or equivalently $n \rightarrow \infty$), the integral in the second line is approximated as
\begin{align}
\label{2nd_int}
&
\int_{-2\pi (\frac{m}{n})^{\delta-1}}^{2\pi (\frac{m}{n})^{\delta-1}} 
d\omega
\exp\left[
m\left(
\frac{1}{\alpha_*(1+i\omega)^{\alpha_*}}
-\frac{1}{\alpha_*}+i\omega
\right)
-\mathfrak{Z}(0,0)\log(1+i\omega)
\right]
\nonumber\\
&=\int_{-Cm^{\frac{1-\delta}{\alpha_*}}}^{Cm^{\frac{1-\delta}{\alpha_*}}}
d\omega
\exp\Biggl[
-\frac{m(\alpha_*+1)}{2}\omega^2
\nonumber\\
&+
m\left(
\frac{1}{\alpha_*(1+i\omega)^{\alpha_*}}
-\frac{1}{\alpha_*}+i\omega+\frac{\alpha_*+1}{2}\omega^2
\right)
-\mathfrak{Z}(0,0)\log(1+i\omega)
\Biggr]
\nonumber\\
&\sim\int_{-Cm^{\frac{1-\delta}{\alpha_*}}}^{Cm^{\frac{1-\delta}{\alpha_*}}}
d\omega
\exp\Biggl[
-\frac{m(\alpha_*+1)}{2}\omega^2
+\mathcal{O}(m|\omega|^3)-\mathfrak{Z}(0,0) \mathcal{O}(|\omega|)
\Biggr], 
\end{align}
where 
\begin{align}
C&=2\pi \left[
\mathrm{Res}_{s=\alpha_*}\mathfrak{Z}(s,0)
\Gamma(\alpha_*+1)\zeta(\alpha_*+1)
\right]^{\frac{\delta-1}{\alpha_*}}. 
\end{align}
As $m\rightarrow \infty$, 
we have $\mathcal{O}(|\omega|)$ $=$ $\mathcal{O}(m^{\frac{1-\delta}{\alpha_*}})$ and therefore the dominant integral is further evaluated as
\begin{align}
\label{2nd_int_final}
&\int_{-Cm^{\frac{1-\delta}{\alpha_*}}}^{Cm^{\frac{1-\delta}{\alpha_*}}}
d\omega
\exp\Biggl[
-\frac{m(\alpha_*+1)}{2}\omega^2
\Biggr]
\nonumber\\
&=
\sqrt{\frac{2}{m(\alpha_*+1)}}
\int_{-C'm^{\frac12+\frac{1-\delta}{\alpha^*}}}^{C'm^{\frac12+\frac{1-\delta}{\alpha^*}}}
d\omega' \exp(-{\omega'}^2)
\nonumber\\
&\sim 
\left(\frac{2\pi}{m(\alpha_*+1)}\right)^{\frac12}. 
\end{align}
Combining (\ref{Cauchy_int5}) and (\ref{2nd_int_final}), we finally obtain
\begin{align}
\label{Asymptotic_Dege}
d(n)&\sim 
\exp\left[
\left(1+\frac{1}{\alpha_*}\right)m
+\mathfrak{Z}(0,0)'
\right]
\left(
\frac{n}{m}
\right)^{\mathfrak{Z}(0,0)-1}
(2\pi m(\alpha_*+1))^{-\frac12}. 
\end{align}
This asymptotic formula (\ref{Asymptotic_Dege}) provides the leading asymptotic behavior of the degeneracy for a supersymmetric zeta function $\mathfrak{Z}(s,0)$ with a simple pole at $s=\alpha_*>0$, which can be viewed as an extension of the Meinardus theorem \cite{MR62781}. 
For example, if the supersymmetric zeta function has a simple pole at $s=1$, we have
\begin{align}
d(n)&\sim \frac12 \left(\frac{R_1}{6n^3}\right)^{\frac14}
\left(\frac{6n}{\pi^2 R_1}\right)^{\frac{\mathfrak{Z}(0,0)}{2}}
\exp
\left[
2\pi\sqrt{\frac{R_1}{6}}n^{\frac12}
+\mathfrak{Z}(0,0)'
\right]
\end{align}
with $R_1$ $=$ $\mathrm{Res}_{s=1}\mathfrak{Z}(s,0)$. 

When the supersymmetric zeta functions have multiple simple poles, the subleading terms contribute correction terms to the asymptotic formula (\ref{Asymptotic_Dege}), although the leading term corresponding to the rightmost pole is still well approximated. 
A general method to evaluate the subleading terms was mathematically formulated in \cite{MR2958955}. 
For equidistant simple poles at $s=$ $i$, 
with $i=0,1,\cdots,r$ and $r=\alpha_*$, the asymptotic formula for $d(n)$ will take the form 
\begin{align}
\label{Asymptotic_Dege2}
d(n)&\sim 
\exp \Biggl[
n\delta_n+\sum_{j=1}^r \left(\mathrm{Res}_{s=j}\Gamma(j)\zeta(j+1)\delta_n^{-j}\right)
+\mathfrak{Z}(0,0)'
\Biggr]
\left(\frac{n}{m}\right)^{\mathfrak{Z}(0,0)-1}
(2\pi m (\alpha_*+1))^{-\frac12}, 
\end{align}
where 
\begin{align}
\delta_n&=\left[
\mathrm{Res}_{s=\alpha_*}\mathfrak{Z}(s,0)
\Gamma(\alpha_*+1)\zeta(\alpha_*+1)
\right]^{\frac{1}{\alpha_*+1}}
\sum_{j=0}^{r}\psi_j z^{j+1}\Biggl|_{z\rightarrow n^{-\frac{1}{r+1}}}
\end{align}
and $\psi_j$ can be determined by solving the recursive equation
\begin{align}
h_r\left(\sum_{l=0}^{s}\psi_l z^l\right)^{r+1}
&=\sum_{k=0}^{r}h_k h_r^{\frac{r-k}{r+1}}z^{r-k}
\left(\sum_{l=0}^s \psi_l z^l\right)^{r-k}
\end{align}
with 
\begin{align}
h_i&=\mathrm{Res}_{s=i} \mathfrak{Z}(s,0) \Gamma(i+1)\zeta(i+1). 
\end{align}
For example, when the supersymmetric zeta function has simple poles at $s=2$ and $1$, 
we find
\begin{align}
\label{Asymptotic_Dege2A}
d(n)&\sim 
\frac{1}{\sqrt{6\pi}}
\left(
\frac{2\zeta(3)R_2}{n^4}
\right)^{\frac16}
\left(
\frac{n}{2\zeta(3)R_2}
\right)^{\frac{\mathfrak{Z}(0,0)}{3}}
\nonumber\\
&\times \exp
\left[
\frac32 (2\zeta(3)R_2)^{\frac13}n^{\frac23}
+\frac{R_1\pi^2}{6}\left(\frac{1}{2\zeta(3)R_2}\right)^{\frac13}n^{\frac13}
-\frac{\pi^4R_1^2}{432\zeta(3)R_2}
+\mathfrak{Z}(0,0)'
\right], 
\end{align}
where we have defined $R_i$ $=$ $\mathrm{Res}_{s=i}\mathfrak{Z}(s,0)$. 
Such growth appears in the large $N$ indices of the 3d supersymmetric field theories describing a stack of $N$ M2-branes \cite{Okazaki:2022sxo,Hayashi:2023txz}. 
When the supersymmetric zeta function has simple poles at $s=3,2$ and $1$, we obtain
\begin{align}
\label{Asymptotic_Dege2B}
d(n)&\sim 
\frac{1}{2\sqrt{2}}
\left(\frac{R_3}{15n^{5}}\right)^{\frac18}
\left(\frac{15n}{\pi^4 R_3}\right)^{\frac{\mathfrak{Z}(0,0)}{4}}
\nonumber\\
&\times 
\exp\Biggl[
\frac{4\pi}{3}\left(\frac{R_3}{15}\right)^{\frac14}n^{\frac34}
+\frac{\zeta(3)R_2}{\pi^2}\left(\frac{15}{R_3}\right)^{\frac12}n^{\frac12}
+\frac{\pi^6 R_1R_3-45\zeta(3)^2R_2^2}{2\pi^5}
\left(\frac{5}{27R_3^5}\right)^{\frac14}n^{\frac14}
\nonumber\\
&-\frac{5(\pi^6\zeta(3)R_1R_2R_3-60\zeta(3)^3R_2^3)}{4\pi^8 R_3^2}
+\mathfrak{Z}(0,0)'
\Biggr]. 
\end{align}
Furthermore, when there are simple poles at $s=4,3,2$ and $1$, we find
\begin{align}
\label{Asymptotic_Dege2C}
d(n)&\sim 
\frac{1}{\sqrt{10\pi}}
\left(\frac{24\zeta(5)R_4}{n^6}\right)^{\frac{1}{10}}
\left(\frac{n}{24\zeta(5)R_4}\right)^{\frac{\mathfrak{Z}(0,0)}{5}}
\nonumber\\
&\times 
\exp\Biggl[
\frac54\left(24\zeta(5)R_4\right)^{\frac15}n^{\frac45}
+\frac{R_3\pi^4}{45}\left(\frac{1}{24\zeta(5)R_4}\right)^{\frac35}n^{\frac35}
\nonumber\\
&+\frac{54000\zeta(3)\zeta(5)R_2R_4-\pi^8R_3^2}{108000}
\left(
\frac{1}{18\zeta(5)^7R_4^7}
\right)^{\frac15}n^{\frac25}
\nonumber\\
&+\frac{\pi^{12}R_3^3-54000\pi^4\zeta(3)\zeta(5)R_2R_3R_4
+8100000\pi^2\zeta(5)^2R_1R_4^2}
{48600000}
\left(\frac{1}{24\zeta(5)^{11}R_4^{11}}\right)^{\frac15}n^{\frac15}
\nonumber\\
&-\frac{1}{41990400000\zeta(5)^3R_4^3}
(\pi^{16}R_3^4-64800\pi^8\zeta(3)\zeta(5)R_2R_3^2R_4
\nonumber\\
&+3888000\pi^6\zeta(5)^2R_1R_3R_4^2+699840000\zeta(3)^2\zeta(5)^2R_2^2R_4^2)+\mathfrak{Z}(0,0)'
\Biggr]. 
\end{align}
When the supersymmetric zeta function has simple poles at $s=5,4,3,2$ and $1$, we find
\begin{align}
\label{Asymptotic_Dege2D}
d(n)&\sim 
\frac{1}{2\sqrt{3}}
\left(\frac{8R_5}{63n^7}\right)^{\frac{1}{12}}
\left(\frac{63 n}{8\pi^6 R_5}\right)^{\frac{\mathfrak{Z}(0,0)}{6}}
\nonumber\\
&\times \exp\Biggl[
\frac{6\pi}{5}\left(\frac{8R_5}{63}\right)^{\frac16} n^{\frac56}
+\frac{6\zeta(5)}{\pi^4}\left(\frac{63}{8R_5}\right)^{\frac23}n^{\frac23}
\nonumber\\
&+\frac{\pi^{10}R_3R_5-17010\zeta(5)^2R_4^2}{30\pi^9}\left(\frac{7}{2R_5^3}\right)^{\frac12}n^{\frac12}
\nonumber\\
&+\frac{10\pi^{12}\zeta(3)R_2R_5^2-21\pi^{10}\zeta(5)R_3R_4R_5+317520\zeta(5)^3R_4^3}
{20\pi^4}
\left(\frac{63}{R_5^7}\right)^{\frac13}n^{\frac13}
\nonumber\\
&-\frac{1}{800\pi^{19}}
\Bigl( \pi^{20}R_5^2(7R_3^2-400R_1R_5)+151200\pi^{12}\zeta(3)\zeta(5)R_2R_4R_5^2
\nonumber\\
&-555660\pi^{10}\zeta(5)^2R_3R_4R_5+6826283100\zeta(5)^4 R_4^4 \Bigr)
\left(\frac{7}{648R_5^{19}}\right)^{\frac16}n^{\frac16}
\nonumber\\
&-\frac{7}{800\pi^{24}R_5^4}
\Bigl(
20\pi^{22}\zeta(3)R_2R_3R_5^3+3\pi^{20}\zeta(5)R_4R_5^2(-21R_3^2+200R_1R_5)
\nonumber\\
&-680400\pi^{12}\zeta(3)\zeta(5)^2R_2R_4^2R_5^2
+2857680\pi^{10}\zeta(5)^3R_3R_4^3R_5
\nonumber\\
&-29165482080\zeta(5)^5 R_4^5
\Bigr)+\mathfrak{Z}(0,0)'
\Biggr]. 
\end{align}

Although we focus on the asymptotic BPS degeneracy for the supersymmetric theories in this work, our asymptotic formulae will be applicable to larger classes of counting functions with exponential growth by introducing appropriate zeta functions. 
For example, the asymptotic density of states in $d$-dimensional CFT can be extracted 
by analyzing its partition function on $S^1\times S^{d-1}$, which is a weighted sum over states on $S^{d-1}$. 
In such cases some of the residues of the associated zeta functions may vanish 
so that our formulae reduce to the formulae obtained in \cite{Benjamin:2023qsc} 
by using the thermal effective action. We note that related ideas involving zeta functions have also appeared 
in the context of asymptotic analyses of two-dimensional CFTs \cite{Benjamin:2022pnx,Lei:2024oij,Benjamin:2025kvm,Perlmutter:2025ngj}.\footnote{Also see \cite{Henning:2017fpj,Melia:2020pzd} for a survey of the asymptotic degeneracy of the Hilbert series by means of the effective field theory. } 

It is worth noting that the above asymptotic formulae for the degeneracy $d(n)$ may not be valid if some of the residues are non-positive values. 
The extent of its validity and the error terms are not yet known exactly. 
Also we remark that the fact that the discrete spectrum may be characterized by different finite spacing 
needs further consideration for the determination of the exact constant term in the asymptotic degeneracy $d(n)$. 
Nevertheless, our asymptotic formulae for the degeneracy $d(n)$ may be used to provide a good approximation for 
the asymptotic number $\Omega(x)$ of the BPS operators with charge less than or equal to $x$, even though the formulae themselves are invalid. 

\subsubsection{Supersymmetric Casimir energies}
\label{sec_Casimir}
The supersymmetric index $\mathcal{I}$ for the $d$-dimensional supersymmetric quantum field theory 
is intimately related to the supersymmetric partition function $\mathcal{Z}_{S^1\times S^{d-1}}$ on $S^1\times S^{d-1}$. 
It is known that they can be distinguished by the supersymmetric Casimir energy factor 
\begin{align}
\mathcal{Z}_{S^1\times S^{d-1}}&=e^{-\beta E_{\textrm{SUSY}}}\mathcal{I}, 
\end{align}
where the supersymmetric Casimir energy $E_{\textrm{SUSY}}$ can be obtained from the large circle limit (zero temperature limit) 
\begin{align}
E_{\textrm{SUSY}}&=-\lim_{\beta\rightarrow\infty} \frac{\partial}{\partial\beta} \log \mathcal{Z}_{S_{\beta}^1\times M_{d-1}}^{\textrm{SUSY}}. 
\end{align}
In general, the Casimir energies may depend on the regularization scheme; however, it has been argued that they are scheme-independent and enjoy the universal structure \cite{Kim:2009wb,Kim:2012ava,Assel:2014paa,Assel:2014tba,Cassani:2014zwa,Lorenzen:2014pna,Assel:2015nca,Bobev:2015kza,Martelli:2015kuk} 
(also see e.g. \cite{Brunner:2016nyk,BenettiGenolini:2016qwm,BenettiGenolini:2016tsn,Papadimitriou:2017kzw,ArabiArdehali:2018mil,Panopoulos:2023cdg} for further study). 
In fact, the statement was demonstrated in \cite{Assel:2015nca} for 4d $\mathcal{N}=1$ superconformal field theories, 
for which the supersymmetric partition functions are scheme-independent. 
Subsequently, it was proposed by Bobev, Bullimore and Kim \cite{Bobev:2015kza} that the supersymmetric Casimir energies for even-dimensional superconformal field theories 
are evaluated as the equivariant integrals of the anomaly polynomials. 

The main focus of our paper is the supersymmetric zeta functions, and it is not immediately obvious if we can say anything about the supersymmetric Casimir energies. 
However, by comparing the supersymmetric Casimir energies studied in the literature with our computation of the supersymmetric zeta values in various examples, we find that a non-trivial relation exists:
\begin{align}
\label{zeta-1_Cas}
E_{\textrm{SUSY}}
&=
\frac12 \mathfrak{Z}(-1,0), 
\end{align}
where the supersymmetric zeta function $\mathfrak{Z}(s,z)$ is associated with the plethystic logarithm of the supersymmetric indices. 
This is suggestive because the supersymmetric zeta values at negative integers $s$ can be interpreted as certain regularized energy moments for superconformal field theories, but it is highly non-trivial because it implies that the supersymmetric zeta values somehow encode information about the supersymmetric partition functions rather than just the index.

We will directly compute the supersymmetric Casimir energies according to the formula (\ref{zeta-1_Cas}) for many examples in later sections. 
For free theories with UV Lagrangian descriptions, we can easily check that the relation (\ref{zeta-1_Cas}) reproduces the formulae obtained via the supersymmetric localization technique, which were originally pointed out by Kim \cite{Kim:2009wb},  
where the supersymmetric Casimir energy shift in the effective action can be formally written as 
\begin{align}
\label{Cas_localization1}
E_{\textrm{SUSY}}&=\frac12 \Tr (-1)^F \Delta. 
\end{align}
Here the trace is taken over the single particle Hilbert space in such a way that 
$\Tr (-1)^F q^{\Delta}$ is the single-particle index $i(q)$. 
$\Delta$ is an appropriate combination of the dilatation operator and the Cartan generators of the R-symmetry 
that commute with the chosen supercharges for the definition of the supersymmetric index. 
The summation in (\ref{Cas_localization1}) should require regularization. 
It can be evaluated from the single-particle index $i(q)$ as \cite{Kim:2009wb,Kim:2012ava} 
\begin{align}
\label{Cas_localization2}
E_{\textrm{SUSY}}&=\frac12 \lim_{q\rightarrow 1^{-}}
q \frac{\partial}{\partial q} i(q), 
\end{align}
which turns out to be compatible with (\ref{zeta-1_Cas}). 

In Lagrangian field theories where the localization computation is possible, the supersymmetric Casimir energy evaluated in \cite{Kim:2012ava} agrees with our computation whenever the comparison is possible. Note that beyond free theories, the way the Casimir energy was computed there is different from ours, but the final result agrees.
Remarkably, our formula can also be applied to the non-Lagrangian theories. 
In fact, as we will see in sections \ref{sec_AD}, \ref{sec_ADmatter} and \ref{sec_ADmatterS}, by combining the supersymmetric Casimir energies evaluated from the formula (\ref{zeta-1_Cas}) for the non-Lagrangian theories with the universal formula proposed in \cite{Bobev:2015kza} for the supersymmetric Casimir energies, we obtain the correct central charges.



We now suggest that the universal properties of the Cardy-like behaviors and the supersymmetric Casimir energies result from those of the supersymmetric zeta functions. An important byproduct of the supersymmetric zeta functions is a novel approach to extract anomaly coefficients or central charges for superconformal field theories from the BPS spectral data.

\section{Supersymmetric determinants}
\label{sec_Sdet}

\subsection{Definition}
The properties of the spectrum can also be encoded in the spectral determinants. 
Here we define the \textit{supersymmetric determinant} by
\begin{align}
\label{sDet_DEF}
\mathbb{D}(z)&=\exp\left[\frac{\partial}{\partial s} \mathbb{Z}(s,z) \right] \Biggl|_{s=0}. 
\end{align}
It can be written as
\begin{align}
\label{sDet_product}
\mathbb{D}(z)&=\prod_n (n+z)^{-d(n)}. 
\end{align}
The supersymmetric spectral determinant is a function of the ``charge variable'' $z$, 
which is related to certain energy eigenvalues in the radial quantization for the superconformal field theories. 
The poles (resp. zeros) at $z=-n$ of order $|d(n)|$ correspond to the $|d(n)|$ bosonic (resp. $|d(n)|$ fermionic) operators of charge $n$ in the theory. 

When the degeneracy $d(n)$ does not grow polynomially, we use the supersymmetric zeta function $\mathfrak{Z}(s,z)$ 
given by (\ref{Szeta_DEF_sind}) in the definition (\ref{sDet_DEF}) and the $d(n)$ in the product expression (\ref{sDet_product}) is replaced by $\delta(n)$. 
We denote the resulting supersymmetric determinant by 
\begin{align}
\mathfrak{D}(z)&=\exp\left[\frac{\partial}{\partial s} \mathfrak{Z}(s,z) \right] \Biggl|_{s=0}
\nonumber\\
&=\prod_n (n+z)^{-\delta(n)}. 
\end{align}

\subsection{Laplace transformation}
The spectral determinant (\ref{sDet_DEF}) is related to the supersymmetric index via a Laplace transformation. 
From (\ref{Mellin_transf2}) and (\ref{sDet_DEF}), the logarithm of the spectral determinant can be written as
\begin{align}
\label{Laplace_transf}
\log \mathbb{D}(z)
&=\frac{\partial}{\partial s}\mathbb{Z}(s,z)\Biggl|_{s=0} \nonumber\\
&=\frac{\partial}{\partial s} 
\left[ \frac{1}{\Gamma(s)}\int_{0}^{\infty}d\beta\ \mathbb{I}(\beta) e^{-\beta z}\beta^{s-1}\right]\Biggl|_{s=0}
\nonumber\\
&=
\left[
-\frac{1}{\Gamma(s)}\int_0^{\infty}d\beta\ \mathbb{I}(\beta) e^{-\beta z}\beta^{s-1} \psi(s)
+\frac{1}{\Gamma(s)}\int_0^{\infty}d\beta\ \mathbb{I}(\beta) e^{-\beta z}\beta^{s-1} \log \beta
\right]
\Biggl|_{s=0}
\nonumber\\
&=\int_{0}^{\infty} d\beta\ {\frac{\mathbb{I}(\beta)}{\beta} e^{-\beta z} }^*, 
\end{align}
where $\psi(s)$ is the digamma function. 
In the last line, we have used 
\begin{align}
\lim_{s\rightarrow0}\frac{\psi(s)}{\Gamma(s)}&=-1
\end{align}
and $1/\Gamma(0)=0$. 
The notation $*$ in (\ref{Laplace_transf}) stands for  
an appropriate regularization that subtracts the divergent terms of the integrand as $\beta\rightarrow 0$. 
Alternatively, the Laplace transform of the supersymmetric index is equal to 
the log-derivative of the supersymmetric determinant 
\begin{align}
-\frac{\partial}{\partial z}\log \mathbb{D}(z)
&=\int_{0}^{\infty} d\beta\ {\mathbb{I}(\beta) e^{-\beta z} }^*. 
\end{align}

\subsection{Vacuum exponents}
When the variable $z$ is set to zero, one obtains the special value of the supersymmetric determinants $\mathfrak{D}(z)$ at $z=0$ 
\begin{align}
\label{vac_exp}
\mathfrak{D}(0)&=\exp[\mathfrak{Z}(0,0)']=\prod_n n^{-\delta(n)}. 
\end{align}
One of the important aspects of this value lies in its role in the asymptotic formulae of the supersymmetric indices. 
It appears in the Cardy-like limit (\ref{Cardy_lim0}) and the asymptotic degeneracy (\ref{Asymptotic_Dege}) as a constant term. 
The analytic behavior of the supersymmetric determinant near $z=0$ is controlled by the distribution of eigenvalue near $n=0$, i.e. the vacuum. 
We refer to it as the \textit{vacuum exponent}. 
It is known that the supersymmetric index vanishes for theories with spontaneously broken supersymmetry, 
while it diverges in the unflavored limit for theories admitting non-normalizable vacua. 
Both features admit a reformulation in terms of the behavior of the supersymmetric determinant. 
The crucial point is the asymptotic expansion (\ref{Asymptotic_Dege}) of the index coefficients $d(n)$. 
The vacuum exponent $\mathfrak{D}(0)=\exp[\mathfrak{Z}(0,0)']$ appears as an overall multiplicative factor in the full asymptotic expansion. 
If the vacuum exponent vanishes at $z=0$, the asymptotic coefficients vanish identically and therefore the index itself should be zero. 
Conversely, if the supersymmetric determinant diverges, the coefficients grow without bound, reproducing the divergence of the index.\footnote{Strictly speaking, the vanishing or divergence of the asymptotic coefficients should be interpreted as a probe of the underlying saddle structure 
rather than as a global analytic structure of the index itself. 
Nevertheless, we propose to regard this asymptotic behavior as a practical diagnostic criterion 
while keeping in mind that this criterion is conditional and may not capture more subtle analytic phenomena. }
We propose that 
\begin{align}
\label{vacexp_criteria}
\begin{cases}
\textrm{SUSY breaking vacuum}&\rightarrow  \qquad \mathfrak{D}(0)=0\cr
\textrm{no stable vacuum}&\rightarrow \qquad \mathfrak{D}(0)=\infty 
\end{cases}
\end{align}
In other words, if supersymmetry is broken, then the vacuum exponent will vanish, although the converse is not necessarily true. 
Similarly, if there is no normalizable vacuum due to the non-compact degrees of freedom, then the vacuum exponent will diverge. 

\subsection{Functional equations}
We make a further remark that 
certain transformations and relations satisfied by the supersymmetric indices 
may reduce to functional equations for the supersymmetric determinants. 

Suppose that the supersymmetric index satisfies the relation
\begin{align}
\label{index_AUT}
\mathbb{I}(q^{-1})&=Cq^{D} \mathbb{I}(q), 
\end{align}
with $C$ $=$ $\pm1$ and $D\in \mathbb{R}$. 
Such a transformation is called the $T$-reflection \cite{Basar:2014mha} (also see \cite{Berkovich:1995nx}) 
as it corresponds to a reflection of the temperature. 
It is then tempting to explore a reflection relation for the supersymmetric determinant 
that relates $\mathbb{D}(z)$ to $\mathbb{D}(D-z)$. 
If we define a function 
\begin{align}
\mathbb{C}(z)&:=\mathbb{D}(D-z)^C \mathbb{D}(z)^{-1}, 
\end{align}
this function will be crucial for understanding the behavior and analytic properties of the supersymmetric determinant, such as its poles and zeros,
by relating its values on either side of the line $\mathrm{Re}(z)=D/2$.

As an example, let us consider the half-BPS index (\ref{ind_4dN4H_uN}) of 4d $\mathcal{N}=4$ $U(N)$ SYM theory. 
It obeys
\begin{align}
\label{ind_4dN4H_uN_AUT}
\mathcal{I}_{\textrm{$\frac12$BPS}}^{\textrm{4d $\mathcal{N}=4$ $U(N)$}}(q^{-1})
&=(-1)^{N}q^{\frac{N(N+1)}{2}}\mathcal{I}_{\textrm{$\frac12$BPS}}^{\textrm{4d $\mathcal{N}=4$ $U(N)$}}(q). 
\end{align}
The supersymmetric determinant is given by 
\begin{align}
\mathbb{D}_{\textrm{$\frac12$BPS}}^{\textrm{4d $\mathcal{N}=4$ $U(N)$}}(z)
&=\Gamma_{N}(z;1,2,\cdots,N)
\nonumber\\
&=\prod_{n_1,\cdots,n_N\ge0}\frac{1}{(n_1+2n_2+\cdots+Nn_N+z)}, 
\end{align}
where
\begin{align} 
\Gamma_r(z;\omega_1,\cdots, \omega_r)
&:=\exp\left[
\frac{\partial}{\partial s}\zeta_r(s,z;\omega_1,\cdots, \omega_r)\Biggl|_{s=0}
\right]
\end{align}
is the Barnes multiple gamma function \cite{barnes1904theory}. 
We have
\begin{align}
&
\mathbb{C}_{\textrm{$\frac12$BPS}}^{\textrm{4d $\mathcal{N}=4$ $U(N)$}}(z)
\nonumber\\
&=\Gamma_{N}\left(z;1,2,\cdots,N\right)^{-1}
\Gamma_{N}\left(\frac{N(N+1)}{2}-z;1,2,\cdots,N\right)^{(-1)^N}
\nonumber\\
&=S_N(z;1,2,\cdots, N), 
\end{align}
where
\begin{align}
S_r(z;\omega_1,\omega_2,\cdots, \omega_r)
&=\Gamma_r(z;\omega_1,\omega_2,\cdots, \omega_r)^{-1}
\Gamma_r(\omega_1+\cdots+\omega_r-z;\omega_1,\omega_2,\cdots,\omega_r)^{(-1)^r}
\end{align}
is the multiple sine function \cite{MR1105522,MR2010282,Jimbo:1996ss}. 
Thus, we have the reflection formula 
\begin{align}
\mathbb{C}_{\textrm{$\frac12$BPS}}^{\textrm{4d $\mathcal{N}=4$ $U(N)$}}(z)
&=\mathbb{D}_{\textrm{$\frac12$BPS}}^{\textrm{4d $\mathcal{N}=4$ $U(N)$}}\left(D-z\right)^{C}
\mathbb{D}_{\textrm{$\frac12$BPS}}^{\textrm{4d $\mathcal{N}=4$ $U(N)$}}(z)^{-1}, 
\end{align}
where $D=N(N+1)/2$ and $C=(-1)^N$. 
When $N=1$, this reduces to
\begin{align}
S_1(z;\omega_1)&=
\Gamma_1\left(1-z;\omega_1\right)^{-1}
\Gamma_1\left(z;\omega_1\right)^{-1}
\nonumber\\
&=\frac{2\pi}{\Gamma\left(\frac{z}{\omega_1}\right)\Gamma\left(1-\frac{z}{\omega_1}\right)}
\nonumber\\
&=2\sin \left(\frac{\pi z}{\omega_1}\right). 
\end{align}
This is nothing but the Euler reflection formula
\begin{align}
\Gamma(z)\Gamma(1-z)=\frac{\pi}{\sin(\pi z)}. 
\end{align}

To conclude this section, we have identified the following related families of supersymmetric spectral functions for these theories: supersymmetric indices, supersymmetric zeta functions, and supersymmetric determinants.

\

\begin{tikzpicture}
\node [fill=yellow!20] (ind) at (0,0) {\textrm{Supersymmetric index $\mathbb{I}(q)$}};
\node [fill=red!20] (zeta) at (7,2) {\textrm{Supersymmetric zeta function $\mathbb{Z}(s,z)$}};
\node [fill=blue!20] (det) at (7,-2) {\textrm{Supersymmetric determinant $\mathbb{D}(z)$}};
\draw[->] (ind) -- node[auto=left]{\textrm{Mellin transf.}}(zeta);
\draw[->] (ind) --node[auto=right]{\textrm{Laplace transf.}}(det);
\draw[->] (zeta) -- node[auto=left]{\textrm{$\exp[\partial_s|_{s=0}]$}}(det);
\end{tikzpicture}

\

For theories whose degeneracies $d(n)$ may not grow polynomially, we take their plethystic logarithm $\mathbb{I}(q)$ (whose coefficients $\delta(n)$ may themselves grow polynomially) to define the supersymmetric zeta function $\mathfrak{Z}(s,z)$ and the supersymmetric determinant $\mathfrak{D}(z)$.
In the following sections, we illustrate these functions by considering a variety of 2d, 4d, and 6d supersymmetric field theories.

\section{2d supersymmetric field theories}
\label{sec_2d}

\subsection{Zeta functions for 2d $\mathcal{N}=(0,2)$ elliptic genera}

The elliptic genera for 2d $\mathcal{N}=(0,2)$ supersymmetric field theories in the Ramond-Ramond (RR) sector and Neveu-Schwarz-Neveu-Schwarz (NSNS) sector are defined by 
\begin{align}
\mathcal{I}_{\textrm{R}}^{\textrm{2d $(0,2)$}}(x;q)&={\Tr}_{\textrm{R}} (-1)^F q^{H_L} \overline{q}^{H_R}\prod_{\alpha}x_{\alpha}^{f_{\alpha}}, \\
\mathcal{I}_{\textrm{NS}}^{\textrm{2d $(0,2)$}}(x;q)&={\Tr}_{\textrm{NS}} (-1)^F q^{H_L} \overline{q}^{H_R-\frac{R}{2}}\prod_{\alpha}x_{\alpha}^{f_{\alpha}}. 
\end{align}
Here $H_L=(H+iP)/2$ and $H_R=(H-iP)/2$ are the left- and right-moving Hamiltonians in the Euclidean signature 
in terms of the Hamiltonian $H$ and the momentum $P$, respectively. 
$R$ stands for the $U(1)_R$ R-charge and $f_{\alpha}$ are the Cartan generators of other global symmetries.  
While only right-moving ground states with $H_R=0$ contribute to the elliptic genus in the RR sector, 
right-moving chiral primary states with $H_R=R/2$ contribute to the elliptic genus in the NSNS sector. A prescription for computing the matrix integrals of the elliptic genera of 2d $\mathcal{N}=(0,2)$ supersymmetric gauge theories was given in \cite{Benini:2013nda,Benini:2013xpa}. 

One has the $4$-form anomaly polynomial of the form (see e.g. \cite{Benini:2013cda})
\begin{align}
\label{2d_anomaly}
\mathcal{A}_4&=\frac12 k_{RR}c_1(R)^2-\frac{1}{24}k p_1(T), 
\end{align}
where $c_1(R)$ is the first Chern class of the $U(1)_R$ background gauge field 
and $p_1(T)$ is the first Pontryagin class of the tangent bundle $T$ of the two-dimensional spacetime. 
The 't Hooft anomaly coefficient $k_{RR}$ of the $U(1)_R$ R-symmetry can be computed as
\begin{align}
k_{RR}&=\Tr \gamma_3 U(1)_R^2,
\end{align}
where $\gamma_3$ is the two-dimensional chirality matrix that takes value $+1$ on the right-handed fermions and $-1$ on the left-handed fermions. 
For the superconformal field theories, 
the right-moving central charge $c_R$ is related to the 't Hooft anomaly for the $U(1)_R$ R-symmetry
\begin{align}
c_R&=3k_{RR}. 
\end{align}
The gravitational anomaly coefficient $k$ is given by
\begin{align}
k&=\Tr \gamma_3. 
\end{align}
For the superconformal field theories, the left-moving central charge $c_L$ can be determined by the gravitational anomaly, which yields the relation:
\begin{align}
\label{cR-cL_2d}
c_R-c_L&=k. 
\end{align}

We define the supersymmetric zeta functions $\mathfrak{Z}_{\textrm{NS}}^{\textrm{2d $(0,2)$}}(s,z)$ for the 2d $\mathcal{N}=(0,2)$ supersymmetric field theories 
by the Mellin transform (\ref{Mellin_transf1}) of the plethystic logarithms of the elliptic genera in the NSNS sector. 
We also define from (\ref{sDet_DEF}) the supersymmetric determinants $\mathfrak{D}_{\textrm{NS}}^{\textrm{2d $(0,2)$}}(z)$ 
associated with the supersymmetric zeta functions $\mathfrak{Z}_{\textrm{NS}}^{\textrm{2d $(0,2)$}}(s,z)$. 
The supersymmetric zeta functions $\mathfrak{Z}_{\textrm{NS}}^{\textrm{2d $(0,2)$}}(s,z)$ can have a simple pole at $s=1$. 

According to the asymptotic formulae (\ref{Cardy_lim0}) and (\ref{Asymptotic_Dege}), we obtain
\begin{align}
\label{2d02_asymp1}
\log \mathcal{I}_{\textrm{NS}}^{\textrm{2d $(0,2)$}}(\beta)
&\sim \frac{\pi^2 \mathrm{Res}_{s=1}\mathfrak{Z}_{\textrm{NS}}^{\textrm{2d $(0,2)$}}(s,0)}{6\beta}
-\mathfrak{Z}_{\textrm{NS}}^{\textrm{2d $(0,2)$}}(0,0) \log\beta
+\log \mathfrak{D}_{\textrm{NS}}^{\textrm{2d $(0,2)$}}(0), \\
\label{2d02_asymp2}
d_{\textrm{NS}}^{\textrm{2d $(0,2)$}}(n)
&\sim \left(\frac{\mathrm{Res}_{s=1}\mathfrak{Z}_{\textrm{NS}}^{\textrm{2d $(0,2)$}}(s,0)}{96n^3}\right)^{\frac14}
\left(\frac{6n}{\pi^2 \mathrm{Res}_{s=1}\mathfrak{Z}_{\textrm{NS}}^{\textrm{2d $(0,2)$}}(s,0)}\right)^{\frac{\mathfrak{Z}_{\textrm{NS}}^{\textrm{2d $(0,2)$}}(0,0)}{2}}
\nonumber\\
&\quad \times \mathfrak{D}_{\textrm{NS}}^{\textrm{2d $(0,2)$}}(0)\exp\left[
2\pi \sqrt{\frac{\mathrm{Res}_{s=1}\mathfrak{Z}_{\textrm{NS}}^{\textrm{2d $(0,2)$}}(s,0)}{6}} n^{\frac12}
\right], 
\end{align}
where $d_{\textrm{NS}}^{\textrm{2d $(0,2)$}}(n)$ are the expansion coefficients of the elliptic genus in the NSNS sector
\begin{align}
\mathcal{I}_{\textrm{NS}}^{\textrm{2d $(0,2)$}}(x;q)&=\sum_n d_{\textrm{NS}}^{\textrm{2d $(0,2)$}}(n) q^n. 
\end{align}

For the superconformal field theories, we observe that the residues at a simple pole $s=1$ of the supersymmetric zeta functions $\mathfrak{Z}_{\textrm{NS}}^{\textrm{2d $(0,2)$}}(s,z)$ 
are given by the gravitational anomaly coefficients  
\begin{align}
\label{Cardy_2dEG}
\mathrm{Res}_{s=1}\mathfrak{Z}_{\textrm{NS}}^{\textrm{2d $(0,2)$}}(s,0)
&=2k=2(c_R-c_L). 
\end{align}
Then the formulae (\ref{2d02_asymp1}) and (\ref{2d02_asymp2}) reduce to\footnote{A similar asymptotic formula is found in Appendix A of \cite{Gadde:2013lxa}.} 
\begin{align}
\label{2d_asymp1}
\log \mathcal{I}_{\textrm{NS}}^{\textrm{2d $(0,2)$}}(\beta)
&\sim \frac{\pi^2(c_R-c_L)}{3\beta}
-\mathfrak{Z}_{\textrm{NS}}^{\textrm{2d $(0,2)$}}(0,0)\log\beta
+\log \mathfrak{D}_{\textrm{NS}}^{\textrm{2d $(0,2)$}}(0), \\
\label{2d_asymp2}
d_{\textrm{NS}}^{\textrm{2d $(0,2)$}}(n)
&\sim \left(\frac{c_R-c_L}{48n^3}\right)^{\frac14}
\left(\frac{3n}{\pi^2 (c_R-c_L)}\right)^{\frac{\mathfrak{Z}_{\textrm{NS}}^{\textrm{2d $(0,2)$}}(0,0)}{2}}
\nonumber\\
&\quad \times \mathfrak{D}_{\textrm{NS}}^{\textrm{2d $(0,2)$}}(0)\exp\left[
2\pi \sqrt{\frac{c_R-c_L}{3}} n^{\frac12}
\right]. 
\end{align}

From the various examples we have examined, we find that the supersymmetric zeta values for $s=-1$ and $z=0$ are given by
\begin{align}
\label{Cas_2dEG}
\mathfrak{Z}_{\textrm{NS}}^{\textrm{2d $(0,2)$}}(-1,0)&=-\frac{c_L}{12}. 
\end{align}
This universal structure, determined by the left-moving central charge $c_L$, is compatible with the supersymmetric Casimir energies of the 2d $\mathcal{N}=(0,2)$ theories. 

For the superconformal field theories, the central charges are obtainable from (\ref{Cardy_2dEG}) and (\ref{Cas_2dEG}) in terms of the supersymmetric zeta functions
\begin{align}
\label{cL_zeta}
c_L&=-12\mathfrak{Z}_{\textrm{NS}}^{\textrm{2d $(0,2)$}}(-1,0), \\
\label{cR_zeta}
c_R&=\frac12 \mathrm{Res}_{s=1}\mathfrak{Z}_{\textrm{NS}}^{\textrm{2d $(0,2)$}}(s,0)-12\mathfrak{Z}_{\textrm{NS}}^{\textrm{2d $(0,2)$}}(-1,0). 
\end{align}
The $4$-form anomaly coefficients in (\ref{2d_anomaly}) are given by
\begin{align}
\label{02k_zeta}
k&=\frac12 \mathrm{Res}_{s=1}\mathfrak{Z}_{\textrm{NS}}^{\textrm{2d $(0,2)$}}(s,0), \\
\label{02kRR_zeta}
k_{RR}&=\frac16 \mathrm{Res}_{s=1}\mathfrak{Z}_{\textrm{NS}}^{\textrm{2d $(0,2)$}}(s,0)-4\mathfrak{Z}_{\textrm{NS}}^{\textrm{2d $(0,2)$}}(-1,0). 
\end{align}

\subsubsection{2d $\mathcal{N}=(0,2)$ chiral multiplet}

The elliptic genus of the 2d $\mathcal{N}=(0,2)$ chiral multiplet of R-charge $r$ reads
\begin{align}
\label{ind_2d(0,2)cm}
\mathcal{I}_{\textrm{NS}}^{\textrm{2d $(0,2)$ chiral}_r}(x;q)
&=\frac{1}{(q^{\frac{r}{2}}x;q)_{\infty}(q^{1-\frac{r}{2}}x^{-1};q)_{\infty}}. 
\end{align}
The single-particle index is
\begin{align}
\label{sind_2d(0,2)cm}
i_{\textrm{NS}}^{\textrm{2d $(0,2)$ chiral}_r}(q)
&=\frac{q^{\frac{r}{2}}x}{1-q}+\frac{q^{1-\frac{r}{2}}x^{-1}}{1-q}. 
\end{align}

The supersymmetric zeta function is given by
\begin{align}
\label{zeta_2d(0,2)cm}
\mathfrak{Z}_{\textrm{NS}}^{\textrm{2d $(0,2)$ chiral}_r}(s,z)
&=\zeta\left(s,\frac{r}{2}+z\right)+\zeta\left(s,1-\frac{r}{2}+z\right), 
\end{align}
where 
\begin{align}
\zeta(s,a)&=\sum_{n=0}^{\infty}\frac{1}{(n+a)^s} 
\end{align}
is the Hurwitz zeta function. 
The supersymmetric zeta function (\ref{zeta_2d(0,2)cm}) has a simple pole at $s=1$ with residue 
\begin{align}
\label{Res1_2d(0,2)cm}
\mathrm{Res}_{s=1}\mathfrak{Z}_{\textrm{NS}}^{\textrm{2d $(0,2)$ chiral}_r}(s,z)&=2. 
\end{align}
The Zeta-index vanishes 
\begin{align}
\label{0_2d(0,2)cm}
\mathfrak{Z}_{\textrm{NS}}^{\textrm{2d $(0,2)$ chiral}_r}(0,0)&=0. 
\end{align}
The supersymmetric zeta value with $s=-1$ and $z=0$ is given by 
\begin{align}
\label{-1_2d(0,2)cm}
\mathfrak{Z}_{\textrm{NS}}^{\textrm{2d $(0,2)$ chiral}_r}(-1,0)
&=-\frac{3(r-1)^2-1}{12}. 
\end{align}
Plugging (\ref{Res1_2d(0,2)cm}) and (\ref{-1_2d(0,2)cm}) into the formulae 
(\ref{02k_zeta}) and (\ref{02kRR_zeta}), we obtain the gravitational and 't Hooft anomalies 
\begin{align}
k&=1, \\
k_{RR}&=(r-1)^2. 
\end{align}
When the R-charge is chosen as the superconformal anomaly, they agree with the central charge of the superconformal field theories due to the 't Hooft anomaly matching. 
From (\ref{cL_zeta}) and (\ref{cR_zeta}), 
we find 
\begin{align}
c_L&=3(r-1)^2-1, \\
c_R&=3(r-1)^2. 
\end{align}
They agree with the anomalies contributed from the 2d $\mathcal{N}=(0,2)$ chiral multiplet of R-charge $r$ \cite{Gadde:2013lxa}. 

\begin{figure}
\begin{center}
\includegraphics[width=10cm]{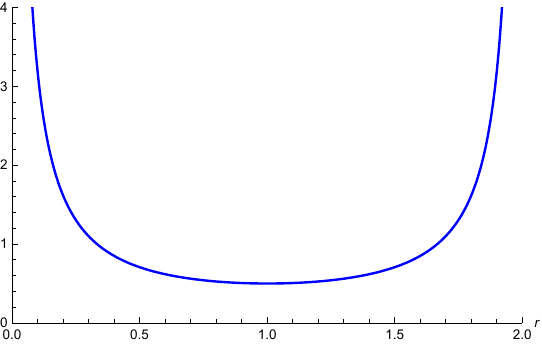}
\caption{The vacuum exponent $\mathfrak{D}_{\textrm{NS}}^{\textrm{2d $(0,2)$ chiral}_r}(0)$.}
\label{fig_vacexp_2d02cm}
\end{center}
\end{figure}

The supersymmetric determinant is 
\begin{align}
\mathfrak{D}_{\textrm{NS}}^{\textrm{2d $(0,2)$ chiral}_r}(z)
&=\frac{\Gamma\left(\frac{r}{2}+z\right)\Gamma\left(1-\frac{r}{2}+z\right)}{2\pi}. 
\end{align}
The vacuum exponent is 
\begin{align}
\mathfrak{D}_{\textrm{NS}}^{\textrm{2d $(0,2)$ chiral}_r}(0)
&=\frac{\Gamma\left(\frac{r}{2}\right)\Gamma\left(1-\frac{r}{2}\right)}{2\pi}
\nonumber\\
&=\frac{1}{2\sin(\frac{\pi r}{2})}. 
\end{align}
It is shown in Figure \ref{fig_vacexp_2d02cm}. 
The vacuum exponent has a simple pole at $r=0$, 
for which we have a theory of non-compact free bosons with the non-normalizable vacuum. 
The degeneracy $d^{\textrm{2d $(0,2)$ chiral}_r}(n)$ obeys the exponential growth for $0<r<2$. 
We obtain from (\ref{2d_asymp2}) the asymptotic growth 
\begin{align}
\label{growth_2dcm1}
d^{\textrm{2d $(0,2)$ chiral}_{r=1}}(n)
&\sim \frac{1}{3^{\frac14}\cdot 4 n^{\frac34}\sin\left(\frac{\pi r}{2}\right)}
\exp\left[
\frac{2\pi}{3^{\frac12}}n^{\frac12}
\right]. 
\end{align}
For example, for $r=1$ we show the exact numbers $d_{\textrm{exact}}^{\textrm{2d $(0,2)$ chiral}_{r=1}}(n)$ 
and the values $d_{\textrm{asymp}}^{\textrm{2d $(0,2)$ chiral}_{r=1}}(n)$ 
evaluated from (\ref{growth_2dcm1}) in the following: 
\begin{align}
\begin{array}{c|c|c}
n&d_{\textrm{exact}}^{\textrm{2d $(0,2)$ chiral}_{r=1}}(n)&d_{\textrm{asymp}}^{\textrm{2d $(0,2)$ chiral}_{r=1}}(n)\\ \hline 
10&1598&3240 \\
100&1.70092\times 10^{13}&3.41288\times 10^{13}\\
1000&3.5254\times 10^{46}&7.05732\times 10^{46}\\
\end{array}. 
\end{align}
As discussed in section \ref{sec_asymp_Dege}, 
the asymptotic formula (\ref{growth_2dcm1}) generally involves ambiguity in the constant term. 
The numerical computation suggests that 
the ratio of the estimated value to the actual value approaches $2$ as $n$ becomes large. 

\subsubsection{2d $\mathcal{N}=(0,2)$ Fermi multiplet}

The elliptic genus of the 2d $\mathcal{N}=(0,2)$ Fermi multiplet of R-charge $r$ is given by
\begin{align}
\mathcal{I}_{\textrm{NS}}^{\textrm{2d $(0,2)$ Fermi}_r}(x;q)
&=(q^{\frac{1+r}{2}}x;q)_{\infty}(q^{\frac{1-r}{2}}x^{-1};q)_{\infty}. 
\end{align}
The single-particle index is
\begin{align}
i_{\textrm{NS}}^{\textrm{2d $(0,2)$ Fermi}_r}(x;q)
&=-\frac{q^{\frac{1+r}{2}}x}{1-q}-\frac{q^{\frac{1-r}{2}}x^{-1}}{1-q}. 
\end{align}

The supersymmetric zeta function is 
\begin{align}
\mathfrak{Z}_{\textrm{NS}}^{\textrm{2d $(0,2)$ Fermi}_r}(s,z)
&=-\zeta\left(s,\frac{1+r}{2}+z\right)-\zeta\left(s,\frac{1-r}{2}+z\right). 
\end{align}
It has a simple pole at $s=1$ with residue 
\begin{align}
\label{Res1_2d(0,2)fm}
\mathrm{Res}_{s=1}\mathfrak{Z}_{\textrm{NS}}^{\textrm{2d $(0,2)$ Fermi}_r}(s,z)&=-2. 
\end{align}
The Zeta-index vanishes 
\begin{align}
\label{0_2d(0,2)fm}
\mathfrak{Z}_{\textrm{NS}}^{\textrm{2d $(0,2)$ Fermi}_r}(0,0)&=0. 
\end{align}
The supersymmetric zeta value with $s=-1$ and $z=0$ is 
\begin{align}
\label{-1_2d(0,2)fm}
\mathfrak{Z}_{\textrm{NS}}^{\textrm{2d $(0,2)$ Fermi}_r}(-1,0). 
&=\frac{3r^2-1}{12}. 
\end{align}
Substituting (\ref{Res1_2d(0,2)fm}) and (\ref{-1_2d(0,2)fm}) into the formulae (\ref{02k_zeta}) and (\ref{02kRR_zeta}), 
we get the gravitational and 't Hooft anomalies
\begin{align}
k&=-1, \\
k_{RR}&=-r^2. 
\end{align}
When the theory is superconformal we find from (\ref{cL_zeta}) and (\ref{cR_zeta}) the central charges
\begin{align}
c_L&=1-3r^2,\\
c_R&=-3r^2. 
\end{align}
They are the expected contributions from the $\mathcal{N}=(0,2)$ Fermi multiplet of R-charge $r$ \cite{Gadde:2013lxa}. 

\begin{figure}
\begin{center}
\includegraphics[width=10cm]{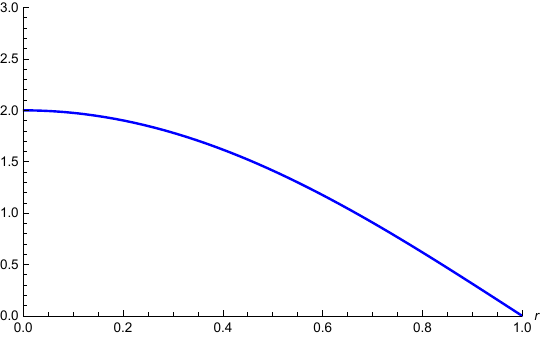}
\caption{The vacuum exponent $\mathfrak{D}_{\textrm{NS}}^{\textrm{2d $(0,2)$ Fermi}_r}(0)$.}
\label{fig_vacexp_2d02fm}
\end{center}
\end{figure}

The supersymmetric determinant is
\begin{align}
\mathfrak{D}_{\textrm{NS}}^{\textrm{2d $(0,2)$ Fermi}_r}(z)
&=\frac{2\pi}{\Gamma\left(\frac{1+r}{2}+z\right)\Gamma\left(\frac{1-r}{2}+z\right)}. 
\end{align}
The vacuum exponent is 
\begin{align}
\mathfrak{D}_{\textrm{NS}}^{\textrm{2d $(0,2)$ Fermi}_r}(0)
&=\frac{2\pi}{\Gamma\left(\frac{1+r}{2}\right)\Gamma\left(\frac{1-r}{2}\right)}
\nonumber\\
&=2\cos(\frac{\pi r}{2}). 
\end{align}
We show the vacuum exponent in Figure \ref{fig_vacexp_2d02fm}. 
%
%
%
%
%

\subsubsection{2d $\mathcal{N}=(0,2)$ vector multiplet}

The elliptic genus of the 2d $\mathcal{N}=(0,2)$ $U(1)$ vector multiplet is 
\begin{align}
\mathcal{I}_{\textrm{NS}}^{\textrm{2d $(0,2)$ vec}}(q)&=(q)_{\infty}^2. 
\end{align}
The single-particle index is
\begin{align}
i_{\textrm{NS}}^{\textrm{2d $(0,2)$ vec}}(q)
&=-\frac{2q}{1-q}. 
\end{align}

We have the supersymmetric zeta function
\begin{align}
\label{zeta_2d(0,2)vm}
\mathfrak{Z}_{\textrm{NS}}^{\textrm{2d $(0,2)$ vec}}(s,z)
&=-2\zeta(s,1+z). 
\end{align}
It has a simple pole at $s=1$ with residue 
\begin{align}
\label{Res1_2d(0,2)vm}
\mathrm{Res}_{s=1}\mathfrak{Z}_{\textrm{NS}}^{\textrm{2d $(0,2)$ vec}}(s,z)&=-2. 
\end{align}
The Zeta-index is 
\begin{align}
\label{0_2d(0,2)vm}
\mathfrak{Z}_{\textrm{NS}}^{\textrm{2d $(0,2)$ vec}}(0,0)&=1. 
\end{align}
The supersymmetric zeta value with $s=-1$ and $z=0$ is 
\begin{align}
\label{-1_2d(0,2)vm}
\mathfrak{Z}_{\textrm{NS}}^{\textrm{2d $(0,2)$ vec}}(-1,0)&=\frac16. 
\end{align}
According to the formulae (\ref{02k_zeta}) and (\ref{02kRR_zeta}) 
as well as the residue (\ref{Res1_2d(0,2)vm}) and the zeta value (\ref{-1_2d(0,2)vm}), we get the gravitational and 't Hooft anomalies
\begin{align}
k&=-1, \\
k_{RR}&=-1. 
\end{align}
Correspondingly, we have
\begin{align}
c_L&=-2, \\
c_R&=-3. 
\end{align}
which are identified with the expected contributions from the 2d $\mathcal{N}=(0,2)$ $U(1)$ vector multiplet \cite{Gadde:2013lxa}. 

The supersymmetric determinant is
\begin{align}
\label{det_2d(0,2)vm}
\mathfrak{D}_{\textrm{NS}}^{\textrm{2d $(0,2)$ vec}}(z)&=\frac{2\pi}{[ \Gamma(1+z) ]^2}. 
\end{align}
The vacuum exponent is 
\begin{align}
\label{D_2d(0,2)vm}
\mathfrak{D}_{\textrm{NS}}^{\textrm{2d $(0,2)$ vec}}(0)&=2\pi. 
\end{align}

\subsubsection{2d $\mathcal{N}=(0,2)$ SQED with chirals $(\Phi,\bar{\Phi}_a)$ and Fermis $(\Psi_{\alpha},\Gamma_a)$}
Consider the 2d $\mathcal{N}=(0,2)$ SQED studied in \cite{Gadde:2013sca,Gadde:2013lxa}, 
that is a $U(1)$ gauge theory with a chiral multiplet $\Phi$ of charge $+1$, $N_f-1$ chiral multiplets $\bar{\Phi}_a$ of charge $-1$,  
$N_f$ Fermi multiplets $\Psi_{\alpha}$ of charge $-1$ and $N_f-1$ neutral Fermi multiplets $\Gamma_a$. 
The theory is conjectured to be dual to a theory of $N_f$ free Fermi multiplets $\Lambda_{\alpha}'$ \cite{Gadde:2013lxa}. 
The field content is summarized as 
\begin{align}
\begin{array}{c|cccc}
&U(1)&SU(N_f)&SU(N_f-1)&U(1)_R\\ \hline 
\Phi&+1&\mathbf{1}&\mathbf{1}&0 \\
\bar{\Phi}_a&-1&\mathbf{1}&\mathbf{N_f-1}&\frac{N_f-2}{N_f-1} \\
\Psi_{\alpha}&-1&\mathbf{N_f}&\mathbf{1}&0 \\ 
\Gamma_a&0&\mathbf{1}&\mathbf{N_f-1}&\frac{1}{N_f-1} \\ \hline
 \Lambda_{\alpha}'&0&\mathbf{N_f}&\mathbf{1}&0 \\
\end{array}
\end{align}

The elliptic genus of the $U(1)$ gauge theory in the NSNS sector is evaluated as
\begin{align}
\label{ind_2d(0,2)SQED}
&
\mathcal{I}_{\textrm{NS}}^{\textrm{2d $(0,2)$ SQED $+(\Phi,\bar{\Phi}_a;\Psi_{\alpha},\Gamma_a)$}}(x_{\alpha};q)
\nonumber\\
&=(q)_{\infty}^2\oint_{\textrm{JK}} \frac{ds}{2\pi is}
\frac{1}{(s;q)_{\infty}(qs^{-1};q)_{\infty}}
\prod_{\alpha=1}^{N_f}
(q^{\frac12}s^{\pm}x_{\alpha}^{\mp};q)_{\infty}
\nonumber\\
&\times \prod_{a=1}^{N_f-1}
\frac{(q^{\frac12+\frac{1}{2(N_f-1)}}y_{a}^{-1};q)_{\infty}(q^{\frac12-\frac{1}{2(N_f-1)}}y_{a};q)_{\infty}}
{(q^{\frac12-\frac{1}{2(N_f-1)}}s^{-1}y_{a};q)_{\infty}(q^{\frac12+\frac{1}{2(N_f-1)}}sy_{a}^{-1};q)_{\infty}}. 
\end{align}
Here the contour integral $\oint_{\textrm{JK}}$ is evaluated by applying the Jeffery-Kirwan (JK) residue prescription \cite{MR1318878}. 
It agrees with the elliptic genus of $N_f$ free Fermi multiplets
\begin{align}
\label{ind_2d(0,2)SQEDdual}
\mathcal{I}_{\textrm{NS}}^{\textrm{2d $(0,2)$ $\Lambda_{\alpha}'$}}(x_{\alpha};q)
&=\prod_{\alpha=1}^{N_f}(q^{\frac12}x_{\alpha}^{\pm};q)_{\infty}. 
\end{align}
The single-particle index is
\begin{align}
\label{sind_2d(0,2)SQED}
i_{\textrm{NS}}^{\textrm{2d $(0,2)$ SQED $+(\Phi,\bar{\Phi}_a;\Psi_{\alpha},\Gamma_a)$}}(x_{\alpha};q)
&=-\sum_{\alpha=1}^{N_f} \frac{q^{\frac12}(x_{\alpha}+x_{\alpha}^{-1})}{1-q}. 
\end{align}

The supersymmetric zeta function is 
\begin{align}
\label{zeta_2d(0,2)SQED}
\mathfrak{Z}_{\textrm{NS}}^{\textrm{2d $(0,2)$ SQED $+(\Phi,\bar{\Phi}_a;\Psi_{\alpha},\Gamma_a)$}}(s,z)
&=-2N_f \zeta\left(s,\frac12+z\right). 
\end{align}
It has a simple pole at $s=1$ with residue
\begin{align}
\label{Res1_2d(0,2)SQED}
\mathrm{Res}_{s=1}\mathfrak{Z}_{\textrm{NS}}^{\textrm{2d $(0,2)$ SQED $+(\Phi,\bar{\Phi}_a;\Psi_{\alpha},\Gamma_a)$}}(s,z)
&=-2N_f. 
\end{align}
As the theory is dual to the free Fermi multiplets, the Zeta-index vanishes
\begin{align}
\label{0_2d(0,2)SQED}
\mathfrak{Z}_{\textrm{NS}}^{\textrm{2d $(0,2)$ SQED $+(\Phi,\bar{\Phi}_a;\Psi_{\alpha},\Gamma_a)$}}(0,0)&=0. 
\end{align}
The supersymmetric zeta value with $s=-1$ and $z=0$ is
\begin{align}
\label{-1_2d(0,2)SQED}
\mathfrak{Z}_{\textrm{NS}}^{\textrm{2d $(0,2)$ SQED $+(\Phi,\bar{\Phi}_a;\Psi_{\alpha},\Gamma_a)$}}(-1,0)
&=-\frac{N_f}{12}. 
\end{align}
From the residue (\ref{Res1_2d(0,2)SQED}) and the zeta value (\ref{-1_2d(0,2)SQED}) we find the anomalies
\begin{align}
k&=-N_f, \\
k_{RR}&=0, 
\end{align}
and
\begin{align}
c_L&=N_f, \\
c_R&=0, 
\end{align}
which agree with the central charges of the 2d $\mathcal{N}=(0,2)$ SQED in \cite{Gadde:2013lxa}. 

The supersymmetric determinant is
\begin{align}
\mathfrak{D}^{\textrm{2d $(0,2)$ SQED $+(\Phi,\bar{\Phi}_a;\Psi_{\alpha},\Gamma_a)$}}(z)
&=\frac{(2\pi)^{N_f}}{\Gamma\left(\frac12+z\right)^{2N_f}}. 
\end{align}
The vacuum exponent is 
\begin{align}
\mathfrak{D}^{\textrm{2d $(0,2)$ SQED $+(\Phi,\bar{\Phi}_a;\Psi_{\alpha},\Gamma_a)$}}(0)
&=2^{N_f}. 
\end{align}

\subsubsection{2d $\mathcal{N}=(0,2)$ $U(N_c)$ with chirals $(\Phi_s,\bar{\Phi}_a)$ 
and Fermis $(\Psi_{\alpha},\Omega,\Gamma_{s,a})$}
The 2d $\mathcal{N}=(0,2)$ $U(N_c)$ gauge theory with $N_c$ fundamental chiral multiplets $\Phi_s$, 
$N_f+N_c$ antifundamental chiral multiplets $\bar{\Phi}_a$, 
$N_f$ antifundamental Fermi multiplets $\Psi_{\alpha}$, $2$ determinant Fermi multiplets $\Omega_{1,2}$ and $N_f+N_c$ neutral Fermi multiplets $\Gamma_{s,a}$ 
is conjectured to be dual to the theory of $N_cN_f+2$ free fermions $\Gamma'_{\alpha,s}$ and $\Omega'_{1,2}$ \cite{Gadde:2013lxa}. 
The field content and the charges are 
\begin{align}
\begin{array}{c|cccccc}
&U(N_c)&SU(N_f)&SU(N_c)'&SU(N_f+N_c)&SU(2)&U(1)_R\\ \hline
\Phi_s&\mathbf{N_c}&\mathbf{1}&\overline{\mathbf{N_c}}&\mathbf{1}&\mathbf{1}&0 \\
\bar{\Phi}_a&\overline{\mathbf{N_c}}&\mathbf{1}&\mathbf{1}&\mathbf{N_f+N_c}&\mathbf{1}&\frac{N_f}{N_f+N_c} \\
\Psi_{\alpha}&\overline{\mathbf{N_c}}&\mathbf{N_f}&\mathbf{1}&\mathbf{1}&\mathbf{1}&0 \\
\Omega_{1,2}&\mathrm{det}&\mathbf{1}&\mathbf{1}&\mathbf{1}&\mathbf{2}&0 \\
\Gamma_{s,a}&\mathbf{1}&\mathbf{1}&\mathbf{N_c}&\overline{\mathbf{N_f+N_c}}&\mathbf{1}&\frac{N_c}{N_f+N_c} \\ \hline
\Gamma'_{\alpha,s}&\mathbf{1}&\mathbf{N_f}&\overline{\mathbf{N_c}}&\mathbf{1}&\mathbf{1}&0 \\ 
\Omega'_{1,2}&\mathbf{1}&\mathbf{1}&\mathbf{1}&\mathbf{1}&\mathbf{2}&0 \\ 
\end{array}
\end{align}

The elliptic genus of the $U(N_c)$ gauge theory is evaluated as
\begin{align}
\label{ind_2d(0,2)uNGGP}
&
\mathcal{I}_{\textrm{NS}}^{\textrm{2d $(0,2)$ $U(N_c)$ $+(\Phi_s,\bar{\Phi}_a;\Psi_{\alpha},\Omega,\Gamma_{s,a})$}}(x_{\alpha},u_s,v;q)
\nonumber\\
&=(q)_{\infty}^{2N_c}\oint_{\textrm{JK}} \prod_{i=1}^{N_c}\frac{ds_i}{2\pi is_i}
\prod_{i\neq j}(s_i^{\pm}s_j^{\mp};q)_{\infty}(qs_i^{\pm}s_j^{\mp};q)_{\infty}
\prod_{i=1}^{N_c}
\prod_{\alpha=1}^{N_f} (q^{\frac12}s_{i}^{\mp}x_{\alpha}^{\mp};q)_{\infty}
\nonumber\\
&\times 
\prod_{i=1}^{N_c}
\prod_{s=1}^{N_c}
\frac{1}
{(s_i u_s^{-1};q)_{\infty}(qs_i^{-1} u_s;q)_{\infty}}
\prod_{a=1}^{N_f+N_c}
\frac{\prod_{s=1}^{N_c}(q^{\frac12\pm \frac{N_c}{2(N_f+N_c)}}u_s^{\pm}y_a^{\mp};q)_{\infty}}
{\prod_{i=1}^{N_c}(q^{\frac{N_f}{2(N_c+N_f)}}s_i^{-1}y_a;q)_{\infty}(q^{1-\frac{N_f}{2(N_c+N_f)}}s_iy_a^{-1};q)_{\infty}}
\nonumber\\
&\times 
(q^{\frac12} (\prod_{i=1}^{N_c} s_i v)^{\pm};q)_{\infty}
(q^{\frac12} (\prod_{i=1}^{N_c} s_i v^{-1})^{\pm};q)_{\infty}. 
\end{align}
It is shown \cite{Gadde:2013lxa} 
that the expression (\ref{ind_2d(0,2)uNGGP}) is equal to the elliptic genus 
\begin{align}
\label{ind_2d(0,2)uNGGP2}
\mathcal{I}_{\textrm{NS}}^{\textrm{2d $(0,2)$ $\Gamma'_{\alpha,s}+\Omega'$}}(x_{\alpha},u_s,v;q)
&=
(q^{\frac12}v^{\pm};q)_{\infty}^2
\prod_{\alpha=1}^{N_f}\prod_{s=1}^{N_c}
(q^{\frac12}x_{\alpha}^{\pm}u_{s}^{\mp};q)_{\infty}
\end{align}
of the theory of $N_cN_f+2$ free fermions. 
The single-particle index reads
\begin{align}
\label{sind_2d(0,2)uNGGP}
&
i_{\textrm{NS}}^{\textrm{2d $(0,2)$ $U(N_c)$ $+(\Phi_s,\bar{\Phi}_a;\Psi_{\alpha},\Omega,\Gamma_{s,a})$}}(x_{\alpha},u_s,w;q)
\nonumber\\
&=-\frac{2q^{\frac12}(v+v^{-1})}{1-q}
-\sum_{\alpha=1}^{N_f}\sum_{s=1}^{N_c}
\frac{q^{\frac12}(x_{\alpha}u_s^{-1}+x_{\alpha}^{-1}u_s)}{1-q}. 
\end{align}

The supersymmetric zeta function is given by
\begin{align}
\label{zeta_2d(0,2)uNGGP}
\mathfrak{Z}_{\textrm{NS}}^{\textrm{2d $(0,2)$ $U(N_c)$ $+(\Phi_s,\bar{\Phi}_a;\Psi_{\alpha},\Omega,\Gamma_{s,a})$}}(s,z)
&=-2(N_cN_f+2)\zeta\left(s,\frac12+z\right). 
\end{align}
The residue at a simple pole $s=1$ is 
\begin{align}
\mathrm{Res}_{s=1}\mathfrak{Z}_{\textrm{NS}}^{\textrm{2d $(0,2)$ $U(N_c)$ $+(\Phi_s,\bar{\Phi}_a;\Psi_{\alpha},\Omega,\Gamma_{s,a})$}}(s,z)
&=-4-2N_cN_f. 
\end{align}
For $s=0$ and $z=0$ we have 
\begin{align}
\mathfrak{Z}_{\textrm{NS}}^{\textrm{2d $(0,2)$ $U(N_c)$ $+(\Phi_s,\bar{\Phi}_a;\Psi_{\alpha},\Omega,\Gamma_{s,a})$}}(0,0)&=0. 
\end{align}
When $s=-1$ and $z=0$ we get 
\begin{align}
\mathfrak{Z}_{\textrm{NS}}^{\textrm{2d $(0,2)$ $U(N_c)$ $+(\Phi_s,\bar{\Phi}_a;\Psi_{\alpha},\Omega,\Gamma_{s,a})$}}(-1,0)
&=-\frac16-\frac{N_cN_f}{12}.  
\end{align}
According to the formulae (\ref{02k_zeta}) and (\ref{02kRR_zeta}) we get the anomalies 
\begin{align}
k&=-2-N_cN_f, \\
k_{RR}&=0. 
\end{align}
Also from (\ref{cL_zeta}) and (\ref{cR_zeta}), we find 
\begin{align}
c_L&=2+N_cN_f, \\
c_R&=0.
\end{align}
They agree with the central charges obtained in \cite{Gadde:2013lxa}. 

The supersymmetric determinant is
\begin{align}
\mathfrak{D}^{\textrm{2d $(0,2)$ $U(N_c)$ $+(\Phi_s,\bar{\Phi}_a;\Psi_{\alpha},\Omega,\Gamma_{s,a})$}}(z)
&=\frac{(2\pi)^{N_cN_f+2}}{\Gamma\left(\frac12+z\right)^{2(N_cN_f+2)}}. 
\end{align}
The vacuum exponent is 
\begin{align}
\mathfrak{D}^{\textrm{2d $(0,2)$ $U(N_c)$ $+(\Phi_s,\bar{\Phi}_a;\Psi_{\alpha},\Omega,\Gamma_{s,a})$}}(0)
&=2^{N_cN_f+2}. 
\end{align}

\subsubsection{2d $\mathcal{N}=(0,2)$ $SU(2)$ with $4$ chirals $\Phi_{\alpha}$}
Consider the 2d $\mathcal{N}=(0,2)$ $SU(2)$ gauge theory with $4$ fundamental chiral multiplets $\Phi_{\alpha}$. 
The theory is conjectured to be dual to the Landau-Ginzburg (LG) theory 
with the Fermi multiplet $\Psi'$ and $6$ chiral multiplets $\Phi'$ interacting with the J-type Pfaffian superpotential \cite{Putrov:2015jpa,Gadde:2015wta,Dedushenko:2017osi}
\begin{align}
\mathcal{W}&=\Psi' \mathrm{Pf}(\Phi'). 
\end{align}
The R-charge of the J-term superpotential is required to be equal to $+1$ 
since the fermionic measure $d\theta^+$ has the R-charge $-1$. 
The field content is summarized as
\begin{align}
\begin{array}{c|cccc} 
&SU(2)&SU(4)&U(1)_x&U(1)_R\\ \hline
\Phi_{\alpha}&\bm{2}&\bm{4}&+1&0\\ \hline 
\Phi'&\bm{1}&\bm{6}&+2&0\\ 
\Psi'&\bm{1}&\bm{1}&-4&+1\\
\end{array}
\end{align}

The elliptic genus of the $SU(2)$ gauge theory reads
\begin{align}
\label{ind_2d02su2+4cm}
&
\mathcal{I}_{\textrm{NS}}^{\textrm{2d $(0,2)$ $SU(2)$ $+\Phi_{\alpha}$}}(x_{\alpha},x;q)
\nonumber\\
&=\frac12 (q)_{\infty}^{2}\oint \frac{ds}{2\pi is}
(s^{\pm 2};q)_{\infty}(qs^{\pm 2};q)_{\infty}
\nonumber\\
&\times 
\prod_{\alpha=1}^{4}
\frac{1}
{(sx_{\alpha}x;q)_{\infty}(qs^{-1}x_{\alpha}^{-1}x^{-1};q)_{\infty}
(s^{-1}x_{\alpha}x;q)_{\infty}(qsx_{\alpha}^{-1}x^{-1};q)_{\infty}}. 
\end{align}
It is equal to the elliptic genus of the dual LG theory
\begin{align}
\mathcal{I}_{\textrm{NS}}^{\textrm{2d $(0,2)$ $\Phi'+\Psi'$}}(x_{\alpha},x;q)
&=
(x^4;q)_{\infty}(qx^{-4};q)_{\infty}
\prod_{\alpha<\beta}^{4}
\frac{1}{(x_{\alpha}x_{\beta}x^2;q)_{\infty}(qx_{\alpha}^{-1}x_{\beta}^{-1}x^{-2};q)_{\infty}}. 
\end{align}
The single-particle index is
\begin{align}
\label{sind_2d02su2+4cm}
&
i_{\textrm{NS}}^{\textrm{2d $(0,2)$ $SU(2)$ $+\Phi_{\alpha}$}}(x_{\alpha},x;q)
\nonumber\\
&=-\frac{x^4}{1-q}-\frac{x^{-4}q}{1-q}
+\sum_{\alpha<\beta}^{4}
\left[
\frac{x_{\alpha}x_{\beta}x^2}{1-q}+\frac{x_{\alpha}^{-1}x_{\beta}^{-1}x^{-2}q}{1-q}
\right]. 
\end{align}
When we turn off the $SU(4)$ flavor fugacities $x_{\alpha}$ and set $x=q^{R_x}$, we get
\begin{align}
i_{\textrm{NS}}^{\textrm{2d $(0,2)$ $SU(2)$ $+\Phi_{\alpha}$}}(x_{\alpha}=1,x=q^{R_x};q)
&=-\frac{q^{4R_x}}{1-q}-\frac{q^{1-4R_x}}{1-q}+\frac{6q^{2R_x}}{1-q}+\frac{6q^{1-2R_x}}{1-q}. 
\end{align}

The supersymmetric zeta function is evaluated as
\begin{align}
\mathfrak{Z}_{\textrm{NS}}^{\textrm{2d $(0,2)$ $SU(2)$ $+\Phi_{\alpha}$}}(s,z)
&=-\zeta(s,4R_x+z)-\zeta(s,1-4R_x+z)
\nonumber\\
&\quad +6\left[\zeta(s,2R_x+z)+\zeta(s,1-2R_x+z)\right]. 
\end{align}
It has a simple pole at $s=1$ with residue
\begin{align}
\label{Res1_2d02su2+4cm}
\mathrm{Res}_{s=1} \mathfrak{Z}_{\textrm{NS}}^{\textrm{2d $(0,2)$ $SU(2)$ $+\Phi_{\alpha}$}}(s,z)
&=10. 
\end{align}
The Zeta-index vanishes
\begin{align}
\mathfrak{Z}_{\textrm{NS}}^{\textrm{2d $(0,2)$ $SU(2)$ $+\Phi_{\alpha}$}}(0,0)&=0. 
\end{align}
For $s=-1$ and $z=0$ the supersymmetric zeta value is evaluated as
\begin{align}
\label{-1_2d02su2+4cm}
\mathfrak{Z}_{\textrm{NS}}^{\textrm{2d $(0,2)$ $SU(2)$ $+\Phi_{\alpha}$}}(-1,0)
&=-\left(8R_x^2-8R_x+\frac56\right). 
\end{align}
From (\ref{Res1_2d02su2+4cm}) and (\ref{-1_2d02su2+4cm}) we obtain the trial central charges
\begin{align}
\label{trialCL_2d02su2+4cm}
c_L&=96R_x^2-96R_x+10, \\
\label{trialCR_2d02su2+4cm}
c_R&=96R_x^2-96R_x+15. 
\end{align}
Applying the $c$-extremization \cite{Benini:2012cz,Benini:2013cda}, one finds $R_x=1/2$, which leads to $c_L=-14$ and $c_R=-9$. 
As the negative values violate the unitarity bound, they indicate the failure of the naive application of the $c$-extremization. 
It was proposed by Sacchi \cite{Sacchi:2020pet} 
that the correct central charges can be obtained by choosing the vanishing mixing coefficients $R_x=0$ 
for the associated flavor currents arising from non-compact directions in the moduli space. 
Then one finds 
\begin{align}
\label{CL_2d02su2+4cm}
c_L&=10, \\
\label{CR_2d02su2+4cm}
c_R&=15. 
\end{align}

We have the supersymmetric determinant
\begin{align}
\mathfrak{D}_{\textrm{NS}}^{\textrm{2d $(0,2)$ $SU(2)$ $+\Phi_{\alpha}$}}(z)
&=\frac{\Gamma(z)^5\Gamma(1+z)^5}{(2\pi)^5}. 
\end{align}
The vacuum exponent is 
\begin{align}
\mathfrak{D}_{\textrm{NS}}^{\textrm{2d $(0,2)$ $SU(2)$ $+\Phi_{\alpha}$}}(0)
&=\infty. 
\end{align}
The divergence of the vacuum exponent can be traced back to the existence of gauge-invariant chiral operators with vanishing R-charge. 
As is immediately seen from the matter content, there exist composite operators whose total R-charge is exactly zero. 
In two-dimensional SCFTs, such operators parametrize flat directions in the moduli space, corresponding to non-compact directions of the target space geometry. 
These directions give rise to a continuous spectrum rather than a discrete normalizable set of states, leading to a divergence in the vacuum exponent. 
In this sense, the divergence reflects the intrinsic non-compactness of the moduli space associated with the chiral operators with vanishing R-charge. 

\subsubsection{2d $\mathcal{N}=(0,2)$ $SU(N)$ with $2N$ chirals $(\Phi_{\alpha},\bar{\Phi}_{\beta})$}
Consider the $SU(N)$ gauge theory with $N$ fundamental chiral multiplets $\Phi_{\alpha}$ and $N$ antifundamental chiral multiplets $\bar{\Phi}_{\beta}$. 
It is argued \cite{Putrov:2015jpa,Gadde:2015wta} that the theory is dual to the LG theory of $N^2$ chiral multiplets $\Phi'$, 
$2$ chiral multiplets $(B',\bar{B}')$ and the Fermi multiplet $\Psi'$ with a superpotential
\begin{align}
\mathcal{W}&=\Psi'(B'\bar{B}'+\det \Phi'). 
\end{align}
The field content and charges are given by 
\begin{align}
\begin{array}{c|cccccc}
&SU(N)&SU(N)_x&SU(N)_{y}&U(1)_x&U(1)_y&U(1)_R \\ \hline 
\Phi_{\alpha}&\mathbf{N}&\overline{\mathbf{N}}&\mathbf{1}&+1&0&0 \\
\bar{\Phi}_{\beta}&\overline{\mathbf{N}}&\mathbf{1}&\mathbf{N}&0&+1&0 \\ \hline 
\Phi'&\mathbf{1}&\overline{\mathbf{N}}&\mathbf{N}&+1&+1&0 \\
B'&\mathbf{1}&\mathbf{1}&\mathbf{1}&+N&0&0 \\
\bar{B}'&\mathbf{1}&\mathbf{1}&\mathbf{1}&0&+N&0 \\
\Psi'&\mathbf{1}&\mathbf{1}&\mathbf{1}&-N&-N&+1 \\
\end{array}
\end{align}

The elliptic genus of the $SU(N)$ gauge theory reads
\begin{align}
&
\mathcal{I}_{\textrm{NS}}^{\textrm{2d $(0,2)$ $SU(N)$ $+(\Phi_{\alpha},\bar{\Phi}_{\beta})$}}(x_{\alpha},y_{\beta},x,y;q)
\nonumber\\
&=\frac{1}{N!}(q)_{\infty}^{N-1}\oint_{\textrm{JK}} 
\prod_{i=1}^{N-1}
\frac{ds_i}{2\pi is_i}
\prod_{i<j}(s_i^{\pm}s_j^{\mp};q)_{\infty}(qs_i^{\pm}s_j^{\mp};q)_{\infty}
\nonumber\\
&\times 
\prod_{i=1}^{N}
\prod_{\alpha=1}^{N}
\frac{1}{(s_ix_{\alpha}^{-1}x;q)_{\infty} (qs_i^{-1}x_{\alpha}x^{-1};q)_{\infty}}
\prod_{i=1}^{N}
\prod_{\beta=1}^{N}
\frac{1}{(s_i^{-1}y_{\beta}y;q)_{\infty} (qs_iy_{\beta}^{-1}y^{-1};q)_{\infty}}, 
\end{align}
with $\prod_{i=1}^{N}s_i=1$, $\prod_{\alpha}x_{\alpha}=1$ and $\prod_{\beta}y_{\beta}=1$. 
As discussed in \cite{Putrov:2015jpa}, 
the expression agrees with 
\begin{align}
&
\mathcal{I}_{\textrm{NS}}^{\textrm{2d $(0,2)$ $\Phi'+(B'\bar{B}')+\Psi'$}}(x_{\alpha},y_{\beta},x,y;q)
\nonumber\\
&=\frac{(qx^{-N}y^{-N};q)_{\infty}(x^Ny^N;q)_{\infty}}
{(x^N;q)_{\infty}(qx^{-N};q)_{\infty}(y^N;q)_{\infty}(qy^{-N};q)_{\infty}
\prod_{\alpha=1}^{N}\prod_{\beta=1}^{N}
(x_{\alpha}^{-1}y_{\beta}xy;q)_{\infty}
(qx_{\alpha}y_{\beta}^{-1}x^{-1}y^{-1};q)_{\infty}
}, 
\end{align}
which is the elliptic genus of the dual LG theory. 
The single-particle index is
\begin{align}
&
i_{\textrm{NS}}^{\textrm{2d $(0,2)$ $SU(N)$ $+(\Phi_{\alpha},\bar{\Phi}_{\beta})$}}(x_{\alpha},y_{\beta},x,y;q)
\nonumber\\
&=-\frac{x^{-N}y^{-N}q}{1-q}-\frac{x^{N}y^{N}}{1-q}
+\frac{x^N}{1-q}+\frac{x^{-N}q}{1-q}+\frac{y^N}{1-q}+\frac{y^{-N}q}{1-q}
\nonumber\\
&\quad +\sum_{\alpha=1}^N \sum_{\beta=1}^N 
\left[
\frac{x_{\alpha}^{-1}y_{\beta}xy}{1-q}+\frac{x_{\alpha}y_{\beta}^{-1}x^{-1}y^{-1}q}{1-q}
\right]. 
\end{align}
Setting $x_{\alpha}=y_{\beta}=1$, $x=q^{R_x}$ and $y=q^{R_y}$, 
it reduces to
\begin{align}
&
i_{\textrm{NS}}^{\textrm{2d $(0,2)$ $SU(N)$ $+(\Phi_{\alpha},\bar{\Phi}_{\beta})$}}(x_{\alpha}=1,y_{\beta}=1,x=q^{R_x},y=q^{R_y};q)
\nonumber\\
&=-\frac{q^{1-N(R_x+R_y)}}{1-q}-\frac{q^{N(R_x+R_y)}}{1-q}
+\frac{q^{NR_x}}{1-q}+\frac{q^{1-NR_x}}{1-q}
+\frac{q^{NR_y}}{1-q}+\frac{q^{1-NR_y}}{1-q}
\nonumber\\
&+N^2\left(\frac{q^{R_x+R_y}}{1-q}
+\frac{q^{1-R_x-R_y}}{1-q}
\right). 
\end{align}

The supersymmetric zeta function is evaluated as
\begin{align}
&
\mathfrak{Z}_{\textrm{NS}}^{\textrm{2d $(0,2)$ $SU(N)$ $+(\Phi_{\alpha},\bar{\Phi}_{\beta})$}}(s,z)
\nonumber\\
&=-\zeta(s,1-N(R_x+R_y)+z)-\zeta(s,N(R_x+R_y)+z)
\nonumber\\
&+\zeta(s,NR_x+z)+\zeta(s,1-NR_x+z)+\zeta(s,NR_y+z)+\zeta(s,1-NR_y+z)
\nonumber\\
&+N^2\left[
\zeta(s,R_x+R_y+z)+\zeta(s,1-R_x-R_y+z)
\right]. 
\end{align}
The residue at a simple pole $s=1$ is 
\begin{align}
\mathrm{Res}_{s=1}\mathfrak{Z}_{\textrm{NS}}^{\textrm{2d $(0,2)$ $SU(N)$ $+(\Phi_{\alpha},\bar{\Phi}_{\beta})$}}(s,z)
&=2(N^2+1). 
\end{align}
The special supersymmetric zeta value for $s=0$ and $z=0$ vanishes
\begin{align}
\mathfrak{Z}_{\textrm{NS}}^{\textrm{2d $(0,2)$ $SU(N)$ $+(\Phi_{\alpha},\bar{\Phi}_{\beta})$}}(0,0)&=0. 
\end{align}
For $s=-1$ and $z=0$ the supersymmetric zeta value is
\begin{align}
\mathfrak{Z}_{\textrm{NS}}^{\textrm{2d $(0,2)$ $SU(N)$ $+(\Phi_{\alpha},\bar{\Phi}_{\beta})$}}(-1,0)
&=-N^2(R_x^2-R_x)-N^2(R_y^2-R_y)-\frac{N^2}{6}-\frac16. 
\end{align}
Hence we get the trial central charges
\begin{align}
\label{trialCL_2d02suN+2Ncm}
c_L&=12(N^2R_x^2-N^2R_x+N^2R_y^2-N^2R_y)+2(N^2+1), \\
\label{trialCR_2d02suN+2Ncm}
c_R&=12(N^2R_x^2-N^2R_x+N^2R_y^2-N^2R_y)+3(N^2+1). 
\end{align}
Upon the $c$-extremization \cite{Benini:2012cz,Benini:2013cda}, we get $R_x=R_y=1/2$, 
which leads to the negative central charges $c_L$ $=$ $2-4N^2$ and $c_R$ $=$ $3-3N^2$. 
Again the correct central charges will be obtained by requiring no mixing anomaly coefficients \cite{Sacchi:2020pet}. 
Then we get
\begin{align}
\label{CL_2d02suN+2Ncm}
c_L&=2(N^2+1), \\
\label{CR_2d02suN+2Ncm}
c_R&=3(N^2+1).
\end{align}
When $N=2$, they reproduce the central charges (\ref{CL_2d02su2+4cm}) and (\ref{CR_2d02su2+4cm}) 
for the $SU(2)$ gauge theory with $4$ fundamental chirals, as expected. 

The supersymmetric determinant is
\begin{align}
\mathfrak{D}_{\textrm{NS}}^{\textrm{2d $(0,2)$ $SU(N)$ $+(\Phi_{\alpha},\bar{\Phi}_{\beta})$}}(z)
&=\frac{\Gamma(z)^{N^2+1}\Gamma(1+z)^{N^2+1}}{(2\pi)^{N^2+1}}. 
\end{align}
The vacuum exponent is 
\begin{align}
\mathfrak{D}_{\textrm{NS}}^{\textrm{2d $(0,2)$ $SU(N)$ $+(\Phi_{\alpha},\bar{\Phi}_{\beta})$}}(0)
&=\infty. 
\end{align}
Again, this would imply that the theory has the non-compact free bosons. 

\subsubsection{2d $\mathcal{N}=(0,2)$ $USp(2N)$ with $2N+2$ chirals $\Phi_{\alpha}$}
Next consider the $USp(2N)$ gauge theory with $2N+2$ fundamental chiral multiplets 
that can be viewed as an alternative higher rank generalization of the $SU(2)$ gauge theory with $4$ fundamental chiral multiplets. 
It is conjectured \cite{Gadde:2015wta,Sacchi:2020pet} that the theory is dual to 
the LG model of a single Fermi multiplet $\Psi'$ and $(N+1)(2N+1)$ chiral multiplets $\Phi'$ with a superpotential 
\begin{align}
\mathcal{W}&=\Psi'\mathrm{Pf}\Phi'. 
\end{align}
The field content and the charges are given by
\begin{align}
\begin{array}{c|cccc}
&USp(2N)&SU(2N+2)&U(1)_x&U(1)_R \\ \hline 
\Phi_{\alpha}&\mathbf{2N}&\mathbf{2N+2}&+1&0 \\ \hline 
\Phi'&\mathbf{1}&\mathbf{(N+1)(2N+1)}&+2&0 \\
\Psi'&\mathbf{1}&\mathbf{1}&-2(N+1)&+1 \\
\end{array}
\end{align}
The duality between these two theories results from the dimensional reduction of the Intriligator-Pouliot duality \cite{Intriligator:1995ne} for 4d $\mathcal{N}=1$ gauge theories. 
When $N=1$, the theory reduces to the $SU(2)$ gauge theory with $4$ fundamental chirals.\footnote{In addition, it is conjectured in \cite{Jiang:2024ifv} that these theories are dual to 
the $SU(2N)$ gauge theory with an antisymmetric chiral, $2N+2$ fundamental chirals and a neutral Fermi interacting with a J-term superpotential. }

The elliptic genus of the $USp(2N)$ gauge theory reads
\begin{align}
\label{ind_2d02usp2N+2N+2cm}
&
\mathcal{I}_{\textrm{NS}}^{\textrm{2d $(0,2)$ $USp(2N)$ $+\Phi_{\alpha}$}}(x_{\alpha},x;q)
\nonumber\\
&=\frac{1}{2^N N!}
(q)_{\infty}^{2N}
\oint_{\textrm{JK}} \prod_{i=1}^{N} \frac{ds_i}{2\pi is_i}
(s_i^{\pm 2};q)_{\infty}(qs_i^{\pm2};q)_{\infty}
\nonumber\\
&\times \prod_{i<j}
(s_i^{\pm}s_j^{\mp};q)_{\infty}(s_i^{\pm}s_j^{\pm};q)_{\infty}
(qs_i^{\pm}s_j^{\mp};q)_{\infty}(qs_i^{\pm}s_j^{\pm};q)_{\infty}
\nonumber\\
&\times 
\prod_{i=1}^{N}
\prod_{\alpha=1}^{2N+2}
\frac{1}{(s_ix_{\alpha}x;q)_{\infty}(qs_i^{-1}x_{\alpha}^{-1}x^{-1};q)_{\infty}
(s_i^{-1}x_{\alpha}x;q)_{\infty}(qs_ix_{\alpha}^{-1}x^{-1};q)_{\infty}}. 
\end{align}
It coincides with the elliptic genus for the dual LG model of the form
\begin{align}
\mathcal{I}_{\textrm{NS}}^{\textrm{2d $(0,2)$ $\Phi'+\Psi'$}}(x_{\alpha},x;q)
&=\frac{(qx^{-2(N+1)};q)_{\infty}(x^{2(N+1)};q)_{\infty}}
{\prod_{\alpha<\beta}^{2N+2}(x_{\alpha}x_{\beta}x^2;q)_{\infty}(qx_{\alpha}^{-1}x_{\beta}^{-1}x^{-2};q)_{\infty}}. 
\end{align}
The single-particle index takes the form
\begin{align}
&
i_{\textrm{NS}}^{\textrm{2d $(0,2)$ $USp(2N)$ $+\Phi_{\alpha}$}}(x_{\alpha},x;q)
\nonumber\\
&=-\frac{x^{-2(N+1)}q}{1-q}-\frac{x^{2(N+1)}}{1-q}
+\sum_{\alpha<\beta}^{2N+2}
\left[
\frac{x_{\alpha}x_{\beta}x^2}{1-q}
+\frac{x_{\alpha}^{-1}x_{\beta}^{-1}x^{-2}q}{1-q}
\right]. 
\end{align}
With $x_{\alpha}=1$ and $x=q^{R_x}$, it becomes
\begin{align}
&
i_{\textrm{NS}}^{\textrm{2d $(0,2)$ $USp(2N)$ $+\Phi_{\alpha}$}}(x_{\alpha}=1,x=q^{R_x};q)
\nonumber\\
&=-\frac{q^{1-2(N+1)R_x}}{1-q}-\frac{q^{2(N+1)R_x}}{1-q}
+(2N+1)(N+1)
\left[
\frac{q^{2R_x}}{1-q}
+\frac{q^{1-2R_x}}{1-q}
\right]. 
\end{align}

One finds the supersymmetric zeta function
\begin{align}
&
\mathfrak{Z}_{\textrm{NS}}^{\textrm{2d $(0,2)$ $USp(2N)$ $+\Phi_{\alpha}$}}(s,z)
\nonumber\\
&=-\zeta(s,1-2(N+1)R_x+z)-\zeta(s,2(N+1)R_x+z)
\nonumber\\
&+(2N+1)(N+1)\left[
\zeta(s,2R_x+z)+\zeta(s,1-2R_x+z)
\right]. 
\end{align}
The residue at a simple pole $s=1$ is 
\begin{align}
\mathrm{Res}_{s=1}\mathfrak{Z}_{\textrm{NS}}^{\textrm{2d $(0,2)$ $USp(2N)$ $+\Phi_{\alpha}$}}(s,z)
&=2N(2N+3). 
\end{align}
The special supersymmetric zeta value with $s=0$ and $z=0$ vanishes
\begin{align}
\mathfrak{Z}_{\textrm{NS}}^{\textrm{2d $(0,2)$ $USp(2N)$ $+\Phi_{\alpha}$}}(0,0)&=0. 
\end{align}
The special supersymmetric zeta value with $s=-1$ and $z=0$ is
\begin{align}
\mathfrak{Z}_{\textrm{NS}}^{\textrm{2d $(0,2)$ $USp(2N)$ $+\Phi_{\alpha}$}}(-1,0)
&=-4N(N+1)R_x^2+4N(N+1)R_x-\frac16N(2N+3). 
\end{align}
Accordingly, we obtain the trial central charges
\begin{align}
\label{cLtrial_2d02usp2N+2N+2cm}
c_L&=48N(N+1)R_x^2-48N(N+1)R_x+2N(2N+3), \\
\label{cRtrial_2d02usp2N+2N+2cm}
c_R&=48N(N+1)R_x^2-48N(N+1)R_x+3N(2N+3). 
\end{align}
By extremizing the trial central charges according to the $c$-extremization \cite{Benini:2012cz,Benini:2013cda}, we get $R_x=1/2$. 
However, for this value the trial central charges become negative, $c_L$ $=$ $-2N(4N+3)$ and $c_R$ $=$ $-3N(2N+1)$. 
Following the proposal in \cite{Sacchi:2020pet}, the correct central charges will be obtained by simply setting the mixing coefficient $R_x$ to zero. 
Then the expressions (\ref{cLtrial_2d02usp2N+2N+2cm}) and (\ref{cRtrial_2d02usp2N+2N+2cm}) reduce to 
\begin{align}
c_L&=2N(2N+3), \\
c_R&=3N(2N+3). 
\end{align}

The supersymmetric determinant is
\begin{align}
\mathfrak{D}_{\textrm{NS}}^{\textrm{2d $(0,2)$ $USp(2N)$ $+\Phi_{\alpha}$}}(z)
&=\frac{\Gamma(z)^{N(2N+3)}\Gamma(1+z)^{N(2N+3)}}{(2\pi)^{N(2N+3)}}. 
\end{align}
As the theory has the non-compact free bosons the vacuum exponent is divergent 
\begin{align}
\mathfrak{D}_{\textrm{NS}}^{\textrm{2d $(0,2)$ $USp(2N)$ $+\Phi_{\alpha}$}}(0)
&=\infty. 
\end{align}

\subsubsection{2d $\mathcal{N}=(0,2)$ $USp(2N)$ with chirals $(\Phi_{\alpha},A)$ and Fermis $\Gamma_a$}
Another type of higher rank $USp(2N)$ gauge theories which generalizes the $SU(2)$ gauge theory with $4$ fundamental chiral multiplets 
can be obtained by introducing the chiral multiplet $A$ transforming in the antisymmetric representation. 
When the theory has $4$ fundamental chiral multiplets $\Phi_{\alpha}$, an antisymmetric chiral multiplet $A$ and $N$ neutral Fermi multiplets $\Gamma_a$ with a superpotential 
\begin{align}
\mathcal{W}&=\sum_{a=1}^{N}\Gamma_a \Tr A^a, 
\end{align}
where the contractions are performed using the $USp(2N)$ invariant antisymmetric tensor $J=i\sigma_2\otimes \mathbb{I}_{N}$, 
it is conjectured to be dual to the LG model of $6N$ chiral multiplets ${\Phi'}_a^{\alpha\beta}$ and $N$ Fermi multiplets $\Psi'_a$ with a superpotential \cite{Sacchi:2020pet}
\begin{align}
\mathcal{W}&=\sum_{a,b,c=1}^N \sum_{\alpha,\beta,\gamma,\delta=1}^{4}
\epsilon_{\alpha\beta\gamma\delta}\Psi'_a {\Phi'}^{\alpha\beta}_{b} {\Phi'}^{\gamma\delta}_{c}\delta_{a+b+c,2N+1}. 
\end{align}
The field content and charges are summarized as
\begin{align}
\begin{array}{c|ccccc}
&USp(2N)&SU(4)&U(1)_w&U(1)_x&U(1)_R\\ \hline 
\Phi_{\alpha}&\bm{2N}&\bm{4}&\frac{1-N}{3}&+1&0 \\
A&\bm{N(2N-1)}&\bm{1}&+1&0&0 \\
\Gamma_a&\bm{1}&\bm{1}&-a&0&+1 \\ \hline
\Phi_a&\bm{1}&\bm{6}&a-\frac{2N+1}{3}&+2&0 \\
\Psi_a&\bm{1}&\bm{1}&a-\frac{2N+1}{3}&-4&+1 \\
\end{array}
\end{align}
with $a=1,\cdots, N$. 

The elliptic genus of the $USp(2N)$ gauge theory reads
\begin{align}
&\mathcal{I}_{\textrm{NS}}^{\textrm{2d $(0,2)$ $USp(2N)$ $\Phi_{\alpha}+A+\Gamma_a$}}(x_{\alpha},w,x;q)
\nonumber\\
&=\frac{1}{2^N N!}(q)_{\infty}^{2N}
\oint_{\textrm{JK}}\prod_{i=1}^N \frac{ds_i}{2\pi is_i}
(s_i^{\pm 2};q)_{\infty}(qs_i^{\pm2};q)_{\infty}
\nonumber\\
&\times \prod_{i<j}
(s_i^{\pm}s_j^{\mp};q)_{\infty}(s_i^{\pm}s_j^{\pm};q)_{\infty}
(qs_i^{\pm}s_j^{\mp};q)_{\infty}(qs_i^{\pm}s_j^{\pm};q)_{\infty}
\nonumber\\
&\times 
\prod_{i=1}^{N}
\prod_{\alpha=1}^{4}
\frac{1}{(s_ix_{\alpha}w^{\frac{1-N}{3}}x;q)_{\infty}(qs_i^{-1}x_{\alpha}^{-1}w^{\frac{N-1}{3}}x^{-1};q)_{\infty}
(s_i^{-1}x_{\alpha}w^{\frac{1-N}{3}}x;q)_{\infty}(qs_ix_{\alpha}^{-1}w^{\frac{N-1}{3}}x^{-1};q)_{\infty}}
\nonumber\\
&\times 
\prod_{i,j=1}^N
\frac{1}{(s_is_j^{-1}w;q)_{\infty}(qs_i^{-1}s_jw;q)_{\infty}}
\prod_{i<j}\frac{1}{(s_i^{\pm}s_j^{\pm}w;q)_{\infty}(qs_i^{\mp}s_j^{\mp}w^{-1};q)_{\infty}}
\nonumber\\
&\times 
\prod_{a=1}^{N}(qw^{-a};q)_{\infty}(w^{a};q)_{\infty}. 
\end{align}
It agrees with 
\begin{align}
&
\mathcal{I}_{\textrm{NS}}^{\textrm{2d $(0,2)$ $\Phi'_a+\Psi'_a$}}(x_{\alpha},w,x;q)
\nonumber\\
&=\prod_{i=1}^N \frac{(x^4w^{-i+\frac{2N+1}{3}};q)_{\infty}(qx^{-4}w^{i-\frac{2N+1}{3}};q)_{\infty}}
{\prod_{\alpha<\beta}^4 
(x^2w^{i-\frac{2N+1}{3}}x_{\alpha}x_{\beta};q)_{\infty}
(qx^{-2}w^{-i+\frac{2N+1}{3}}x_{\alpha}^{-1}x_{\beta}^{-1};q)_{\infty}}. 
\end{align}
With $x_{\alpha}=1$, $w=q^{R_w}$ and $x=q^{R_x}$, the single-particle index reads
\begin{align}
&i_{\textrm{NS}}^{\textrm{2d $(0,2)$ $USp(2N)$ $\Phi_{\alpha}+A+\Gamma_a$}}
(x_{\alpha}=1,w=q^{R_w},x=q^{R_x};q)
\nonumber\\
=&\sum_{i=1}^N 
\Biggl[
-\frac{q^{4R_x-(i-\frac{2N+1}{3})R_w}}{1-q}
-\frac{q^{1-4R_x+(i-\frac{2N+1}{3})R_w}}{1-q}
\nonumber\\
&\qquad +\frac{6q^{2R_x+(i-\frac{2N+1}{3})R_w}}{1-q}
+\frac{6q^{1-2R_x-(i-\frac{2N+1}{3})R_w}}{1-q}
\Biggr]. 
\end{align}

The supersymmetric zeta function is evaluated as
\begin{align}
&\mathfrak{Z}_{\textrm{NS}}^{\textrm{2d $(0,2)$ $USp(2N)$ $\Phi_{\alpha}+A+\Gamma_a$}}(s,z)
\nonumber\\
&=\sum_{i=1}^{N} 
\Biggl[
-\zeta\left(s,4R_x-(i-\frac{2N+1}{3})R_w+z\right)
-\zeta\left(s,1-4R_x+(i-\frac{2N+1}{3})R_w+z\right)
\nonumber\\
&+6\zeta\left(s,2R_x+(i-\frac{2N+1}{3})R_w+z\right)
+6\zeta\left(s,1-2R_x-(i-\frac{2N+1}{3})R_w+z\right)
\Biggr]. 
\end{align}
The residue at a simple pole $s=1$ is
\begin{align}
\mathrm{Res}_{s=1} \mathfrak{Z}_{\textrm{NS}}^{\textrm{2d $(0,2)$ $USp(2N)$ $\Phi_{\alpha}+A+\Gamma_a$}}(s,z)&=10N. 
\end{align}
For $s=0$ and $z=0$ the supersymmetric zeta value vanishes
\begin{align}
\mathfrak{Z}_{\textrm{NS}}^{\textrm{2d $(0,2)$ $USp(2N)$ $\Phi_{\alpha}+A+\Gamma_a$}}(0,0)&=0.
\end{align}
The supersymmetric zeta value with $s=-1$ and $z=0$ is given by
\begin{align}
&
\mathfrak{Z}_{\textrm{NS}}^{\textrm{2d $(0,2)$ $USp(2N)$ $\Phi_{\alpha}+A+\Gamma_a$}}(-1,0)
\nonumber\\
&=\sum_{i=1}^N 
\Biggl[
-\frac59 (2N-3i+1)^2 R_w^2-8R_x^2+\frac{32}{3}(2N-3i+1)R_wR_x
\nonumber\\
&-\frac73(2N-3i+1)R_w+8R_x-\frac56
\Biggr]
\nonumber\\
&=-\frac{5}{18}N(2N^2-N-1)R_w^2-8NR_x^2+\frac{16}{3}N(N-1)R_wR_x
\nonumber\\
&\quad -\frac{7}{6}N(N-1)R_w+8NR_x-\frac56 N. 
\end{align}
Hence the trial central charges are 
\begin{align}
c_L&=\frac{10}{3}N(2N^2-N-1)R_w^2+96NR_x^2-64N(N-1)R_wR_x
\nonumber\\
&+14N(N-1)R_w-96NR_x+10N, \\
c_R&=\frac{10}{3}N(2N^2-N-1)R_w^2+96NR_x^2-64N(N-1)R_wR_x
\nonumber\\
&+14N(N-1)R_w-96NR_x+15N. 
\end{align}
When $N=1$, they agree with (\ref{trialCL_2d02su2+4cm}) and (\ref{trialCR_2d02su2+4cm}). 
Applying the $c$-extremization, one finds that the mixing coefficients are given by
\begin{align}
R_w&=-\frac{9}{2(2N-7)}, \qquad 
R_x=-\frac{N+4}{2(2N-7)}. 
\end{align}
Then the values of the central charges are 
\begin{align}
c_L&=\frac{5N(5N+23)}{2(2N-7)}, \\
c_R&=\frac{45N(N+1)}{2(2N-7)}. 
\end{align}
Note that they are negative for $N=1,2$ and $3$. 
Again the prescription for the failure of the $c$-extremization will be given by the vanishing two mixing coefficients $R_w$ and $R_x$ \cite{Sacchi:2020pet}.  
Consequently, we arrive at 
\begin{align}
c_L&=10N, \\
c_R&=15N. 
\end{align}

The supersymmetric determinant is
\begin{align}
\mathfrak{D}_{\textrm{NS}}^{\textrm{2d $(0,2)$ $USp(2N)$ $\Phi_{\alpha}+A+\Gamma_a$}}(z)
&=\frac{\Gamma(z)^{5N}\Gamma(1+z)^{5N}}{(2\pi)^{5N}}. 
\end{align}
We find the divergent vacuum exponent 
\begin{align}
\mathfrak{D}_{\textrm{NS}}^{\textrm{2d $(0,2)$ $USp(2N)$ $\Phi_{\alpha}+A+\Gamma_a$}}(0)
&=\infty. 
\end{align}

\subsection{Zeta functions for 2d $\mathcal{N}=(2,2)$ elliptic genera}

The 2d $\mathcal{N}=(2,2)$ supersymmetric field theories have a left-moving $U(1)$ R-symmetry $J_L$ and a right-moving $U(1)$ R-symmetry $J_R$. 
They are related to the vectorial ($R_V$) and axial ($R_A$) R-symmetries as follows:
\begin{align}
R_V&=J_L+J_R,&
R_A&=J_L-J_R. 
\end{align}

The elliptic genera for 2d $\mathcal{N}=(2,2)$ supersymmetric field theories can be defined in a similar way:
\begin{align}
\mathcal{I}_{\textrm{R}}^{\textrm{2d $(2,2)$}}(y,x;q)&={\Tr}_{\textrm{R}} (-1)^F q^{H_L} \overline{q}^{H_R}y^{J_L}\prod_{\alpha}x_{\alpha}^{f_{\alpha}}, \\
\mathcal{I}_{\textrm{NS}}^{\textrm{2d $(2,2)$}}(y,x;q)&={\Tr}_{\textrm{NS}} (-1)^F q^{H_L} \overline{q}^{H_R-\frac{J_R}{2}}y^{J_L}\prod_{\alpha}x_{\alpha}^{f_{\alpha}}. 
\end{align}
The elliptic genus in the RR sector and that in the NSNS sector are related by spectral flow:
\begin{align}
\mathcal{I}_{\textrm{R}}^{\textrm{2d $(2,2)$}}(y,x;q)&=y^{-\frac{c}{6}}\mathcal{I}_{\textrm{NS}}^{\textrm{2d $(2,2)$}}(q^{-\frac12}y,x;q), 
\end{align}
where $c=c_L=c_R$. 

Similarly, we define the supersymmetric zeta functions $\mathfrak{Z}_{\textrm{NS}}^{\textrm{2d $(2,2)$}}(s,z)$ 
by the Mellin transform(\ref{Mellin_transf1}) of the plethystic logarithms of the NSNS sector elliptic genera with $y=x=1$, 
and the associated supersymmetric determinants $\mathfrak{D}_{\textrm{NS}}^{\textrm{2d $(2,2)$}}(z)$ by (\ref{sDet_DEF}).

\subsubsection{2d $\mathcal{N}=(2,2)$ chiral multiplet}

Consider the $\mathcal{N}=(2,2)$ chiral multiplet of the vectorial R-charge $2r$ and of the vanishing axial R-charge. 
It has the left-moving R-charge $r$. 
The elliptic genus in the NSNS sector is given by
\begin{align}
\label{ind_2d(2,2)cm}
\mathcal{I}_{\textrm{NS}}^{\textrm{2d $(2,2)$ chiral}_r}(y,x;q)
&=\mathcal{I}_{\textrm{NS}}^{\textrm{2d $(0,2)$ chiral}_r}(y^{r}x;q)
\times \mathcal{I}_{\textrm{NS}}^{\textrm{2d $(0,2)$ Fermi}_r}(y^{r-1}x;q)
\nonumber\\
&=\frac{(q^{\frac12+\frac{r}{2}}y^{r-1}x;q)_{\infty}(q^{\frac12-\frac{r}{2}}y^{1-r}x^{-1};q)_{\infty}}
{(q^{\frac{r}{2}}y^{r}x;q)_{\infty}(q^{1-\frac{r}{2}}y^{-r}x^{-1};q)_{\infty}}. 
\end{align}
The single-particle index is
\begin{align}
\label{sind_2d(2,2)cm}
i_{\textrm{NS}}^{\textrm{2d $(2,2)$ chiral}_r}(y,x;q)
&=i_{\textrm{NS}}^{\textrm{2d $(0,2)$ chiral}_r}(y^{r}x;q)+i_{\textrm{NS}}^{\textrm{2d $(0,2)$ Fermi}_r}(y^{r-1}x;q)
\nonumber\\
&=\frac{q^{\frac{r}{2}}y^{r}x}{1-q}+\frac{q^{1-\frac{r}{2}}y^{-r}x^{-1}}{1-q}
-\frac{q^{\frac{1+r}{2}}y^{r-1}x}{1-q}-\frac{q^{\frac{1-r}{2}}y^{-r+1}x^{-1}}{1-q}. 
\end{align}

The supersymmetric zeta function is evaluated as 
\begin{align}
\label{zeta_2d(2,2)cm}
&
\mathfrak{Z}_{\textrm{NS}}^{\textrm{2d $(2,2)$ chiral}_r}(s,z)
=\mathfrak{Z}_{\textrm{NS}}^{\textrm{$({0},2)$ chiral}_r}(s,z)+\mathfrak{Z}_{\textrm{NS}}^{\textrm{$(0,2)$ Fermi}_r}(s,z)
\nonumber\\
&=\zeta\left(s,\frac{r}{2}+z\right)+\zeta\left(s,1-\frac{r}{2}+z\right)
-\zeta\left(s,\frac{1+r}{2}+z\right)-\zeta\left(s,\frac{1-r}{2}+z\right). 
\end{align}
While each term has a simple pole at $s=1$, the total sum of the residues vanishes
\begin{align}
\label{Res1_2d(2,2)cm}
\mathrm{Res}_{s=1}\mathfrak{Z}_{\textrm{NS}}^{\textrm{2d $(2,2)$ chiral}_r}(s,z)&=0. 
\end{align}
Also the Zeta-index vanishes
\begin{align}
\label{0_2d(2,2)cm}
\mathfrak{Z}_{\textrm{NS}}^{\textrm{2d $(2,2)$ chiral}_r}(0,0)&=0. 
\end{align}
For $s=-1$ and $z=0$ the supersymmetric zeta value is 
\begin{align}
\label{-1_2d(2,2)cm}
\mathfrak{Z}_{\textrm{NS}}^{\textrm{2d $(2,2)$ chiral}_r}(-1,0)
&=\frac{2r-1}{4}. 
\end{align}
Plugging (\ref{Res1_2d(2,2)cm}) and (\ref{-1_2d(2,2)cm}) into (\ref{02k_zeta}) and (\ref{02kRR_zeta}), 
we find the gravitational and 't Hooft anomalies
\begin{align}
k&=0, \\
k_{RR}&=1-2r. 
\end{align}
From the formulae (\ref{cL_zeta}) and (\ref{cR_zeta}), we get 
\begin{align}
\label{c_2d(2,2)cm}
c&=c_L=c_R=3(1-2r). 
\end{align}
This is the expected anomaly contribution from the $\mathcal{N}=(2,2)$ chiral multiplet \cite{Benini:2013xpa} for generic R-symmetry assignment. 

\begin{figure}
\begin{center}
\includegraphics[width=10cm]{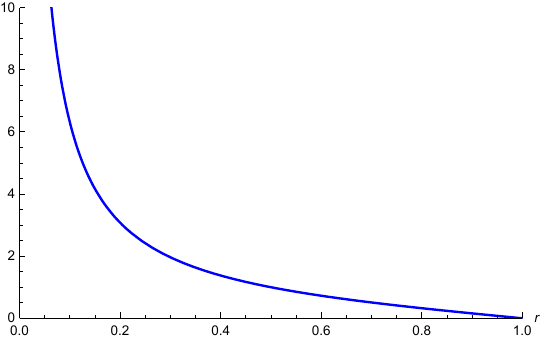}
\caption{The vacuum exponent $\mathfrak{D}_{\textrm{NS}}^{\textrm{2d $(2,2)$ chiral}_r}(0)$.}
\label{fig_vacexp_2d22cm}
\end{center}
\end{figure}

The supersymmetric determinant is
\begin{align}
\mathfrak{D}_{\textrm{NS}}^{\textrm{2d $(2,2)$ chiral}_r}(z)
&=\frac{\Gamma\left(\frac{r}{2}+z\right)\Gamma\left(1-\frac{r}{2}+z\right)}
{\Gamma\left(\frac{1+r}{2}+z\right)\Gamma\left(\frac{1-r}{2}+z\right)}. 
\end{align}
The vacuum exponent is
\begin{align}
\mathfrak{D}_{\textrm{NS}}^{\textrm{2d $(2,2)$ chiral}_r}(0)
&=\frac{\Gamma\left(\frac{r}{2}\right)\Gamma\left(1-\frac{r}{2}\right)}
{\Gamma\left(\frac{1+r}{2}\right)\Gamma\left(\frac{1-r}{2}\right)}
\nonumber\\
&=\cot(\frac{\pi r}{2}), 
\end{align}
which is shown in Figure \ref{fig_vacexp_2d22cm}. 

Similarly, we obtain from the expression (\ref{ind_2d(2,2)cm}) the elliptic genus of the $\mathcal{N}=(2,2)$ twisted chiral multiplet 
by replacing $y$ with $y^{-1}$
\begin{align}
\mathcal{I}_{\textrm{NS}}^{\textrm{2d $(2,2)$ tchiral}_r}(y,x;q)
&=\mathcal{I}_{\textrm{NS}}^{\textrm{2d $(2,2)$ chiral}_r}(y^{-1},x;q). 
\end{align} 
Hence the supersymmetric zeta function associated with the unflavored index takes the same form as (\ref{zeta_2d(2,2)cm}) 
and therefore the anomaly contribution is given by (\ref{c_2d(2,2)cm}). 

\subsubsection{2d $\mathcal{N}=(2,2)$ vector multiplet}
The $\mathcal{N}=(2,2)$ $U(1)$ vector multiplet can be viewed as the twisted chiral multiplet of the left-moving R-charge $r=+1$. 

The elliptic genus in the NSNS sector is given 
by combining the indices of the $\mathcal{N}=(0,2)$ vector multiplet and the $\mathcal{N}=(0,2)$ chiral multiplet of R-charge $+1$
\begin{align}
\label{ind_2d(2,2)vm}
\mathcal{I}_{\textrm{NS}}^{\textrm{2d $(2,2)$ vec}}(y;q)
&=\frac{(q)_{\infty}^2}{(q^{\frac12}y;q)_{\infty}(q^{\frac12}y^{-1};q)_{\infty}}. 
\end{align}
The single-particle index of the $\mathcal{N}=(2,2)$ $U(1)$ vector multiplet 
\begin{align}
\label{sind_2d(2,2)vm}
i_{\textrm{NS}}^{\textrm{2d $(2,2)$ vec}}(y;q)
&=-\frac{2q}{1-q}+\frac{(y+y^{-1})q^{\frac12}}{1-q}. 
\end{align}

The supersymmetric zeta function is given by
\begin{align}
\label{zeta_2d(2,2)vm}
\mathfrak{Z}_{\textrm{NS}}^{\textrm{2d $(2,2)$ vec}}(s,z)
&=-2\zeta(s,1+z)+2\zeta\left(s,\frac12+z\right). 
\end{align}
The residue at $s=1$ is
\begin{align}
\label{Res1_2d(2,2)vm}
\mathrm{Res}_{s=1}\mathfrak{Z}_{\textrm{NS}}^{\textrm{2d $(2,2)$ vec}}(s,z)&=0. 
\end{align}
The Zeta-index is given by
\begin{align}
\label{0_2d(2,2)vm}
\mathfrak{Z}_{\textrm{NS}}^{\textrm{2d $(2,2)$ vec}}(0,0)&=1. 
\end{align}
The supersymmetric zeta value with $s=-1$ and $z=0$ is 
\begin{align}
\label{-1_2d(2,2)vm}
\mathfrak{Z}_{\textrm{NS}}^{\textrm{2d $(2,2)$ vec}}(-1,0)&=\frac14. 
\end{align}
Substituting (\ref{Res1_2d(2,2)vm}) and (\ref{-1_2d(2,2)vm}) into (\ref{02k_zeta}) and (\ref{02kRR_zeta}), 
we get the anomalies 
\begin{align}
k&=0, \\
k_{RR}&=-1. 
\end{align}
Also we obtain from (\ref{cL_zeta}) and (\ref{cR_zeta})
\begin{align}
c&=c_L=c_R=-3. 
\end{align}
This is the expected anomaly contribution from the $\mathcal{N}=(2,2)$ $U(1)$ vector multiplet \cite{Benini:2013xpa}. 

We find the supersymmetric determinant
\begin{align}
\mathfrak{D}_{\textrm{NS}}^{\textrm{2d $(2,2)$ vec}}(z)
&=\frac{\Gamma\left(\frac12+z\right)^2}{\Gamma(1+z)^2}. 
\end{align}
The vacuum exponent is 
\begin{align}
\mathfrak{D}_{\textrm{NS}}^{\textrm{2d $(2,2)$ vec}}(0)
&=\pi. 
\end{align}

\subsubsection{2d $\mathcal{N}=(2,2)$ pure SYM theories}
Now consider the $\mathcal{N}=(2,2)$ pure SYM theory of gauge group $G$. 
It is conjectured \cite{Aharony:2016jki} that when $G$ is semisimple and simply-connected, the theory flows to a free theory of 
$\mathrm{rank}(G)$ twisted chiral multiplets with the left-moving R-charges being the degrees $d_i$, $i=1,\cdots, \mathrm{rank}(G)$ of the Casimir operators at low energy 
(see e.g. \cite{Gu:2018fpm,Chen:2018wep,Gu:2020ivl,Eager:2020rra} for further discussions). 
The degrees of the Casimirs are given by (see e.g. \cite{MR511189})
\begin{align}
\begin{array}{c|c}
G&d_i\\ \hline 
SU(N)&2,3,\cdots, N \\ 
SO(2N+1)&2,4,\cdots, 2N \\ 
USp(2N)&2,4,\cdots, 2N \\ 
SO(2N)&2,4,\cdots, 2(N-1), N \\ 
G_2&2,6 \\
F_4&2,6,8,12 \\
E_6&2,5,6,8,9,12 \\
E_7&2,6,8,10,12,14,18 \\
E_8&2,8,12,14,18,20,24,30 \\
\end{array}
\end{align}
For example, for $G=SU(N)$ it is dual to the theory of $(N-1)$ free twisted chiral multiplets of the left-moving R-charges $n$ with $n=2,\cdots N$. 

The elliptic genus in the NSNS sector can be evaluated by the matrix integral
\begin{align}
\label{ind_2d(2,2)vmG}
&
\mathcal{I}_{\textrm{NS}}^{\textrm{2d $(2,2)$ $G$}}(y;q)
\nonumber\\
&=\frac{1}{|\mathrm{Wey}(G)|}
\frac{(q)_{\infty}^{2\mathrm{rank}(G)}}
{(q^{\frac12}y^{\pm};q)_{\infty}^{\mathrm{rank}(G)}}
\oint_{\textrm{JK}} \prod_{\alpha\in \mathrm{root}(G)}
\frac{ds}{2\pi is}
\frac{(s^{\alpha};q)_{\infty}(qs^{\alpha};q)_{\infty}}
{(q^{\frac12}s^{\alpha}y;q)_{\infty}(q^{\frac12}s^{-\alpha}y^{-1};q)_{\infty}}, 
\end{align}
where $|\mathrm{Weyl}(G)|$ is the order of the Weyl group of $G$. 
According to the dualities, it is equal to
\begin{align}
\label{ind_2d(2,2)G_closed}
\mathcal{I}_{\textrm{NS}}^{\textrm{2d $(2,2)$ $G$}}(y;q)
&=\prod_{i=1}^{\mathrm{rank}(G)}
\frac{(q^{\frac{1+d_i}{2}}y^{1-d_i};q)_{\infty}(q^{\frac{1-d_i}{2}}y^{-1+d_i};q)_{\infty}}
{(q^{\frac{d_i}{2}}y^{-d_i};q)_{\infty}(q^{1-\frac{d_i}{2}}y^{d_i};q)_{\infty}}
\end{align}
and the single-particle index is given by
\begin{align}
&
i_{\textrm{NS}}^{\textrm{2d $(2,2)$ $G$}}(y;q)
\nonumber\\
&=\sum_{i=1}^{\mathrm{rank}(G)}
\Biggl[
-\frac{y^{1-d_i}q^{\frac{1+d_i}{2}}}{1-q}-\frac{y^{-1+d_i}q^{\frac{1-d_i}{2}}}{1-q}
+\frac{y^{-d_i}q^{\frac{d_i}{2}}}{1-q}+\frac{y^{d_i}q^{1-\frac{d_i}{2}}}{1-q}
\Biggr]. 
\end{align}

The supersymmetric zeta function is evaluated as
\begin{align}
&
\mathfrak{Z}_{\textrm{NS}}^{\textrm{2d $(2,2)$ $G$}}(s,z)
\nonumber\\
&=\sum_{i=1}^{\mathrm{rank}(G)}
\Biggl[
-\zeta\left(s,\frac{1+d_i}{2}+z\right)
-\zeta\left(s,\frac{1-d_i}{2}+z\right)
+\zeta\left(s,\frac{d_i}{2}+z\right)
+\zeta\left(s,1-\frac{d_i}{2}+z\right)
\Biggr]. 
\end{align}
It follows that
\begin{align}
\mathrm{Res}_{s=1}\mathfrak{Z}_{\textrm{NS}}^{\textrm{2d $(2,2)$ $G$}}(s,z)&=0, \\
\mathfrak{Z}_{\textrm{NS}}^{\textrm{2d $(2,2)$ $G$}}(0,0)&=0, \\
\mathfrak{Z}_{\textrm{NS}}^{\textrm{2d $(2,2)$ $G$}}(-1,0)&=\frac{\dim G}{4}, 
\end{align}
where we have used the fact that 
\begin{align}
\label{d_Casimir_relation1}
\dim G&=\sum_{i=1}^{\mathrm{rank}(G)}(2d_i-1). 
\end{align}
According to the formulae (\ref{02k_zeta}), (\ref{02kRR_zeta}), (\ref{cL_zeta}) and (\ref{cR_zeta}), we find the anomalies 
\begin{align}
k&=0, \\
k_{RR}&=-\dim G
\end{align}
and 
\begin{align}
\label{c_2d(2,2)vmG}
c=c_L=c_R=-3\dim G. 
\end{align}
While the values (\ref{c_2d(2,2)vmG}) should not be understood as the ordinary central charges, 
they are the anomaly contributions of the $\mathcal{N}=(2,2)$ vector multiplet of gauge group $G$ \cite{Benini:2013xpa}. 
They are also called the effective central charges in \cite{Eager:2020rra}. 
The pure SYM theory of gauge group $G$ can  be viewed as a sigma model on the algebraic stack $BG$ $=$ $[\textrm{point}/G]$ \cite{MR1818418,Pantev:2005rh,Pantev:2005zs,Pantev:2005wj}. 
The dimension of the quotient algebraic stack $[X/G]$ is given by \cite{MR1818418}
\begin{align}
\dim[X/G]&=\dim X-\dim G, 
\end{align}
which can be negative. 
For the stack $BG$ $=$ $[\textrm{point}/G]$, the dimension is $-\dim G$, 
which matches with $c/3$.  

The supersymmetric determinant is
\begin{align}
\mathfrak{D}_{\textrm{NS}}^{\textrm{2d $(2,2)$ $G$}}(z)
&=\prod_{i=1}^{\mathrm{rank}(G)}\frac{\Gamma\left(\frac{d_i}{2}+z\right)\Gamma\left(1-\frac{d_i}{2}+z\right)}
{\Gamma\left(\frac{1+d_i}{2}+z\right)\Gamma\left(\frac{1-d_i}{2}+z\right)}. 
\end{align}
The vacuum exponent takes the form 
\begin{align}
\mathfrak{D}_{\textrm{NS}}^{\textrm{2d $(2,2)$ $G$}}(0)
&=\prod_{i=1}^{\mathrm{rank}(G)}\cot\left(\frac{\pi d_i}{2}\right). 
\end{align}
It contains divergent factors associated with the degrees $d_i$ of the Casimirs being even integers. 
This indicates the non-compact degrees of freedom.\footnote{For $SU(N)$ with $N$ odd, $SO(2N)$ with $N$ odd, and $E_6$, the degrees $d_i$ of the Casimirs contain odd integers corresponding to zeros so that the vacuum exponents are indeterminate.}
\subsubsection{2d $\mathcal{N}=(2,2)$ minimal models}
Let us consider the $\mathcal{N}=(2,2)$ minimal models with $c<3$ 
which can be described by the LG theories of twisted chiral multiplets $\widetilde{\Phi}_i$ with a quasi-homogeneous twisted superpotential $\widetilde{\mathcal{W}}$. 
They are the rational conformal field theories with an ADE classification of the twisted superpotentials given by
\begin{align}
\begin{array}{c|cc}
\textrm{minimal models}&\widetilde{\mathcal{W}} \\ \hline 
A_{k+1} &\widetilde{\Phi}^{k+2}&k\ge0\\
D_{k+2} &\widetilde{\Phi}_1^{k+1}+\widetilde{\Phi}_1\widetilde{\Phi}_2^2&k\ge2 \\
E_6&\widetilde{\Phi}_1^3+\widetilde{\Phi}_2^4\\
E_7&\widetilde{\Phi}_1^3+\widetilde{\Phi}_1\widetilde{\Phi}_2^3\\
E_8&\widetilde{\Phi}_1^3+\widetilde{\Phi}_2^5\\
\end{array}
\end{align}
The twisted superpotential has the left-moving R-charge $+1$ and the right-moving R-charge $-1$. 

The elliptic genera of the $\mathcal{N}=(2,2)$ minimal models in the NSNS sector are given by 
\cite{Witten:1993jg,DiFrancesco:1993ty}
\begin{align}
\mathcal{I}_{\textrm{NS}}^{\textrm{2d $(2,2)$ min $A_{k+1}$}}(y;q)
&=\frac{(q^{\frac12+\frac{1}{2(k+2)}}y^{\frac{k+1}{k+2}};q)_{\infty} (q^{\frac12-\frac{1}{2(k+2)}}y^{-\frac{k+1}{k+2}};q)_{\infty}}
{(q^{\frac{1}{2(k+2)}}y^{-\frac{1}{k+2}};q)_{\infty} (q^{1-\frac{1}{2(k+2)}}y^{\frac{1}{k+2}};q)_{\infty}}, \\
\mathcal{I}_{\textrm{NS}}^{\textrm{2d $(2,2)$ min $D_{k+2}$}}(y;q)
&=\frac{(q^{\frac12+\frac{1}{2(k+1)}}y^{-\frac{1}{k+1}+1};q)_{\infty}(q^{\frac12-\frac{1}{2(k+1)}}y^{\frac{1}{k+1}-1};q)_{\infty}}
{(q^{\frac{1}{2(k+1)}}y^{-\frac{1}{k+1}};q)_{\infty}(q^{1-\frac{1}{2(k+1)}}y^{\frac{1}{k+1}};q)_{\infty}}
\nonumber\\
& \times 
\frac{(q^{\frac12+\frac{k}{4(k+1)}}y^{-\frac{k}{2(k+1)}+1};q)_{\infty}
(q^{\frac12-\frac{k}{4(k+1)}}y^{\frac{k}{2(k+1)}-1};q)_{\infty}}
{(q^{\frac{k}{4(k+1)}}y^{-\frac{k}{2(k+1)}};q)_{\infty}
(q^{1-\frac{k}{4(k+1)}}y^{\frac{k}{2(k+1)}};q)_{\infty}}
, \\
\mathcal{I}_{\textrm{NS}}^{\textrm{2d $(2,2)$ min $E_{6}$}}(y;q)
&=\frac{(q^{\frac23}y^{\frac23};q)_{\infty}(q^{\frac13}y^{-\frac23};q)_{\infty}}
{(q^{\frac16}y^{-\frac13};q)_{\infty}(q^{\frac56}y^{\frac13};q)_{\infty}}
\frac{(q^{\frac58}y^{\frac34};q)_{\infty}(q^{\frac38}y^{-\frac34};q)_{\infty}}
{(q^{\frac18}y^{-\frac14};q)_{\infty}(q^{\frac78}y^{\frac14};q)_{\infty}}, \\
\mathcal{I}_{\textrm{NS}}^{\textrm{2d $(2,2)$ min $E_{7}$}}(y;q)
&=\frac{(q^{\frac23}y^{\frac23};q)_{\infty}(q^{\frac13}y^{-\frac23};q)_{\infty}}
{(q^{\frac16}y^{-\frac13};q)_{\infty}(q^{\frac56}y^{\frac13};q)_{\infty}}
\frac{(q^{\frac{11}{18}}y^{\frac79};q)_{\infty}(q^{\frac{7}{18}}y^{-\frac79};q)_{\infty}}
{(q^{\frac{1}{9}}y^{-\frac29};q)_{\infty}(q^{\frac{8}{9}}y^{\frac29};q)_{\infty}}, \\
\mathcal{I}_{\textrm{NS}}^{\textrm{2d $(2,2)$ min $E_{8}$}}(y;q)
&=\frac{(q^{\frac23}y^{\frac23};q)_{\infty}(q^{\frac13}y^{-\frac23};q)_{\infty}}
{(q^{\frac16}y^{-\frac13};q)_{\infty}(q^{\frac56}y^{\frac13};q)_{\infty}}
\frac{(q^{\frac{3}{5}}y^{\frac45};q)_{\infty}(q^{\frac{2}{5}}y^{-\frac45};q)_{\infty}}
{(q^{\frac{1}{10}}y^{-\frac15};q)_{\infty}(q^{\frac{9}{10}}y^{\frac15};q)_{\infty}}. 
\end{align}
The single-particle indices are
\begin{align}
i_{\textrm{NS}}^{\textrm{2d $(2,2)$ min $A_{k+1}$}}(q)
&=-\frac{q^{\frac12+\frac{1}{2(k+2)}}y^{\frac{k+1}{k+2}}+q^{\frac12-\frac{1}{2(k+2)}}y^{-\frac{k+1}{k+2}}}
{1-q}
\nonumber\\
&\quad +\frac{q^{\frac{1}{2(k+2)}}y^{-\frac{1}{k+2}}+q^{1-\frac{1}{2(k+2)}}y^{\frac{1}{k+2}}}
{1-q}
, \\
i_{\textrm{NS}}^{\textrm{2d $(2,2)$ min $D_{k+2}$}}(q)
&=-\frac{q^{\frac12+\frac{1}{2(k+1)}}y^{-\frac{1}{k+1}+1}+q^{\frac12-\frac{1}{2(k+1)}}y^{\frac{1}{k+1}-1}}{1-q}
+\frac{q^{\frac{1}{2(k+1)}}y^{-\frac{1}{k+1}}+q^{1-\frac{1}{2(k+1)}}y^{\frac{1}{k+1}}}{1-q}
\nonumber\\
&\quad -\frac{q^{\frac12+\frac{k}{4(k+1)}}y^{-\frac{k}{2(k+1)}+1}+q^{\frac12-\frac{k}{4(k+1)}}y^{\frac{k}{2(k+1)}-1}}{1-q}
+\frac{q^{\frac{k}{4(k+1)}}y^{-\frac{k}{2(k+1)}}+q^{1-\frac{k}{4(k+1)}}y^{\frac{k}{2(k+1)}}}{1-q}, \\
i_{\textrm{NS}}^{\textrm{2d $(2,2)$ min $E_{6}$}}(q)
&=-\frac{q^{\frac23}y^{\frac23}+q^{\frac13}y^{-\frac23}}{1-q}
+\frac{q^{\frac16}y^{-\frac13}+q^{\frac56}y^{\frac13}}{1-q}
-\frac{q^{\frac58}y^{\frac34}+q^{\frac38}y^{-\frac34}}{1-q}
+\frac{q^{\frac18}y^{-\frac14}+q^{\frac78}y^{\frac14}}{1-q}, \\
i_{\textrm{NS}}^{\textrm{2d $(2,2)$ min $E_{7}$}}(q)
&=-\frac{q^{\frac23}y^{\frac23}+q^{\frac13}y^{-\frac23}}{1-q}
+\frac{q^{\frac16}y^{-\frac13}+q^{\frac56}y^{\frac13}}{1-q}
-\frac{q^{\frac{11}{18}}y^{\frac79}+q^{\frac{7}{18}}y^{-\frac79}}{1-q}
+\frac{q^{\frac{1}{9}}y^{-\frac29}+q^{\frac{8}{9}}y^{\frac29}}{1-q}, \\
i_{\textrm{NS}}^{\textrm{2d $(2,2)$ min $E_{8}$}}(q)
&=-\frac{q^{\frac23}y^{\frac23}+q^{\frac13}y^{-\frac23}}{1-q}
+\frac{q^{\frac16}y^{-\frac13}+q^{\frac56}y^{\frac13}}{1-q}
-\frac{q^{\frac{3}{5}}y^{\frac45}+q^{\frac{2}{5}}y^{-\frac45}}{1-q}
+\frac{q^{\frac{1}{10}}y^{-\frac15}+q^{\frac{9}{10}}y^{\frac15}}{1-q}. 
\end{align}

The supersymmetric zeta functions for the $\mathcal{N}=(2,2)$ minimal models are evaluated as
\begin{align}
\mathfrak{Z}_{\textrm{NS}}^{\textrm{2d $(2,2)$ min $A_{k+1}$}}(s,z)
&=-\zeta\left(s,\frac{k+3}{2(k+2)}+z\right)-\zeta\left(s,\frac{k+1}{2(k+2)}+z\right)
\nonumber\\
&\quad +\zeta\left(s,\frac{1}{2(k+2)}+z\right)+\zeta\left(s,\frac{2k+3}{2(k+2)}+z\right), \\
\mathfrak{Z}_{\textrm{NS}}^{\textrm{2d $(2,2)$ min $D_{k+2}$}}(s,z)
&=-\zeta\left(s,\frac{k+2}{2(k+1)}+z \right)-\zeta\left(s,\frac{k}{2(k+1)}+z \right)
\nonumber\\
&\quad +\zeta\left(s,\frac{1}{2(k+1)}+z \right)+\zeta\left(s,\frac{2k+1}{2(k+1)}+z \right)
\nonumber\\
&\quad -\zeta\left(s,\frac{3k+2}{4(k+1)}+z \right)-\zeta\left(s,\frac{k+2}{4(k+1)}+z \right)
\nonumber\\
&\quad +\zeta\left(s,\frac{k}{4(k+1)}+z \right)+\zeta\left(s,\frac{3k+4}{4(k+1)}+z \right), \\
\mathfrak{Z}_{\textrm{NS}}^{\textrm{2d $(2,2)$ min $E_{6}$}}(s,z)
&=-\zeta\left(s,\frac13+z\right)-\zeta\left(s,\frac23+z\right)
+\zeta\left(s,\frac16+z\right)+\zeta\left(s,\frac56+z\right)
\nonumber\\
&\quad -\zeta\left(s,\frac58+z\right)-\zeta\left(s,\frac38+z\right)
+\zeta\left(s,\frac18+z\right)+\zeta\left(s,\frac78+z\right), \\
\mathfrak{Z}_{\textrm{NS}}^{\textrm{2d $(2,2)$ min $E_{7}$}}(s,z)
&=-\zeta\left(s,\frac13+z\right)-\zeta\left(s,\frac23+z\right)
+\zeta\left(s,\frac16+z\right)+\zeta\left(s,\frac56+z\right)
\nonumber\\
&\quad -\zeta\left(s,\frac{11}{18}+z\right)-\zeta\left(s,\frac{7}{18}+z\right)
+\zeta\left(s,\frac19+z\right)+\zeta\left(s,\frac89+z\right), \\
\mathfrak{Z}_{\textrm{NS}}^{\textrm{2d $(2,2)$ min $E_{8}$}}(s,z)
&=-\zeta\left(s,\frac13+z\right)-\zeta\left(s,\frac23+z\right)
+\zeta\left(s,\frac16+z\right)+\zeta\left(s,\frac56+z\right)
\nonumber\\
&\quad-\zeta\left(s,\frac{3}{5}+z\right)-\zeta\left(s,\frac{2}{5}+z\right)
+\zeta\left(s,\frac{1}{10}+z\right)+\zeta\left(s,\frac{9}{10}+z\right). 
\end{align}
For any series of the $\mathcal{N}=(2,2)$ minimal models, we have
\begin{align}
\mathrm{Res}_{s=1} \mathfrak{Z}_{\textrm{NS}}^{\textrm{2d $(2,2)$ min $G$}}(s,z)&=0
\end{align}
and 
\begin{align}
\mathfrak{Z}_{\textrm{NS}}^{\textrm{2d $(2,2)$ min $G$}}(0,0)&=0. 
\end{align}
The special supersymmetric zeta values with $s=-1$ and $z=0$ are given by
\begin{align}
\mathfrak{Z}_{\textrm{NS}}^{\textrm{2d $(2,2)$ min $A_{k+1}$}}(-1,0)&=-\frac{k}{4(k+2)}, \\
\mathfrak{Z}_{\textrm{NS}}^{\textrm{2d $(2,2)$ min $D_{k+2}$}}(-1,0)&=-\frac{k}{4(k+1)}, \\
\mathfrak{Z}_{\textrm{NS}}^{\textrm{2d $(2,2)$ min $E_6$}}(-1,0)&=-\frac{5}{24}, \\
\mathfrak{Z}_{\textrm{NS}}^{\textrm{2d $(2,2)$ min $E_7$}}(-1,0)&=-\frac{2}{9}, \\
\mathfrak{Z}_{\textrm{NS}}^{\textrm{2d $(2,2)$ min $E_8$}}(-1,0)&=-\frac{7}{30}.
\end{align}
According to the formulae (\ref{cL_zeta}) and (\ref{cR_zeta}), 
we get the central charges \cite{Vafa:1988uu}
\begin{align}
c&=c_L=c_R=3-\frac{6}{h^{\vee}}
\end{align}
of the $\mathcal{N}=(2,2)$ minimal models, 
where $h^{\vee}$ is the dual Coxeter number of the relevant Lie algebra 
\begin{align}
\begin{array}{c|cc}
\textrm{minimal modes}&c&h^{\vee} \\ \hline 
A_{k+1}&\frac{3k}{k+2}&k+2\\
D_{k+2}&\frac{3k}{k+1}&2k+2\\
E_6&\frac52&12\\
E_7&\frac83&18\\
E_8&\frac{14}{5}&30\\
\end{array}. 
\end{align}

The supersymmetric determinants are given by
\begin{align}
\mathfrak{D}_{\textrm{NS}}^{\textrm{2d $(2,2)$ min $A_{k+1}$}}(z)
&=\frac{\Gamma\left(\frac{1}{2(k+2)}+z\right) \Gamma\left(\frac{2k+3}{2(k+2)}+z\right)}
{\Gamma\left(\frac{k+3}{2(k+2)}+z\right)\Gamma\left(\frac{k+1}{2(k+2)}+z\right)}, \\
\mathfrak{D}_{\textrm{NS}}^{\textrm{2d $(2,2)$ min $D_{k+2}$}}(z)
&=\frac{\Gamma\left(\frac{1}{2(k+1)}+z\right)\Gamma\left(\frac{2k+1}{2(k+1)}+z\right)
\Gamma\left(\frac{k}{4(k+1)}+z\right)\Gamma\left(\frac{3k+4}{4(k+1)}+z\right)}
{\Gamma\left(\frac{k+2}{2(k+1)}+z\right)\Gamma\left(\frac{k}{2(k+1)}+z\right)
\Gamma\left(\frac{3k+2}{4(k+1)}+z\right)\Gamma\left(\frac{k+2}{4(k+1)}+z\right)}, \\
\mathfrak{D}_{\textrm{NS}}^{\textrm{2d $(2,2)$ min $E_{6}$}}(z)
&=\frac{\Gamma\left(\frac16+z\right)\Gamma\left(\frac56+z\right)
\Gamma\left(\frac18+z\right)\Gamma\left(\frac78+z\right)}
{\Gamma\left(\frac13+z\right)\Gamma\left(\frac23+z\right)
\Gamma\left(\frac58+z\right)\Gamma\left(\frac38+z\right)}, \\
\mathfrak{D}_{\textrm{NS}}^{\textrm{2d $(2,2)$ min $E_{7}$}}(z)
&=\frac{\Gamma\left(\frac16+z\right)\Gamma\left(\frac56+z\right)
\Gamma\left(\frac19+z\right)\Gamma\left(\frac89+z\right)}
{\Gamma\left(\frac13+z\right)\Gamma\left(\frac23+z\right)
\Gamma\left(\frac{11}{18}+z\right)\Gamma\left(\frac{7}{18}+z\right)}, \\
\mathfrak{D}_{\textrm{NS}}^{\textrm{2d $(2,2)$ min $E_{8}$}}(z)
&=\frac{\Gamma\left(\frac16+z\right)\Gamma\left(\frac56+z\right)
\Gamma\left(\frac{1}{10}+z\right)\Gamma\left(\frac{9}{10}+z\right)}
{\Gamma\left(\frac13+z\right)\Gamma\left(\frac23+z\right)
\Gamma\left(\frac{3}{5}+z\right)\Gamma\left(\frac{2}{5}+z\right)}. 
\end{align}
The vacuum exponents are
\begin{align}
\mathfrak{D}_{\textrm{NS}}^{\textrm{2d $(2,2)$ min $A_{k+1}$}}(0)
&=\cot\left(\frac{\pi}{2k+4}\right), \\
\mathfrak{D}_{\textrm{NS}}^{\textrm{2d $(2,2)$ min $D_{k+2}$}}(0)
&=\cot\left(\frac{\pi}{2(k+1)}\right)\cot\left(\frac{\pi k}{4(k+1)}\right), \\
\mathfrak{D}_{\textrm{NS}}^{\textrm{2d $(2,2)$ min $E_{6}$}}(0)
&=\sqrt{3}\cot\left(\frac{\pi}{8}\right)=4.18154\ldots, \\
\mathfrak{D}_{\textrm{NS}}^{\textrm{2d $(2,2)$ min $E_{7}$}}(0)
&=\sqrt{3}\cot\left(\frac{\pi}{9}\right)=4.75877\ldots, \\
\mathfrak{D}_{\textrm{NS}}^{\textrm{2d $(2,2)$ min $E_{8}$}}(0)
&=\sqrt{3}\cot\left(\frac{\pi}{10}\right)=\sqrt{15+6\sqrt{5}}=5.3307\ldots. 
\end{align}

\subsubsection{2d $\mathcal{N}=(2,2)$ Kazama-Suzuki models}
Kazama-Suzuki models \cite{Kazama:1988qp,Kazama:1988uz} 
are the rational $\mathcal{N}=(2,2)$ SCFTs that are described by the coset 
\begin{align}
\frac{SU(M+1)_k\times SO(2M)_1}
{SU(M)_{k+1}\times U(1)_{M(M+1)(M+k+1)}}. 
\end{align}
It is expected to be described by the LG theory of $M$ twisted chiral multiplets $\widetilde{\Phi}_i$ 
with a superpotential \cite{Gepner:1990gr}
\begin{align}
\widetilde{\mathcal{W}}
&=\sum_{i=1}^{M}\frac{1}{M+k+1}\widetilde{\phi}_i^{M+k+1}, 
\end{align} 
where
\begin{align}
\sum_{i=0}^{M}\widetilde{\Phi}_it^i
&=\prod_{i=1}^{M}(1+\widetilde{\phi}_i t). 
\end{align}
For $M=1$ they are identical to the $\mathcal{N}=(2,2)$ $A_{k+1}$ minimal model. 

The elliptic genus of the $A_{k+1}$ Kazama-Suzuki model in the NSNS sector is given by \cite{DiFrancesco:1993ty}
\begin{align}
\mathcal{I}_{\textrm{NS}}^{\textrm{2d $(2,2)$ KS}_{M,k}}(y;q)
&=\prod_{j=1}^{M}
\frac{(q^{\frac{k+j}{2(M+k+1)}}y^{-\frac{k+j}{M+k+1}};q)_{\infty}(q^{1-\frac{k+j}{2(M+k+1)}}y^{\frac{k+j}{M+k+1}};q)_{\infty}}
{(q^{\frac{j}{2(M+k+1)}}y^{-\frac{j}{M+k+1}};q)_{\infty}(q^{1-\frac{j}{2(M+k+1)}}y^{\frac{j}{M+k+1}};q)_{\infty}}. 
\end{align}
Then the single-particle index reads
\begin{align}
&
i_{\textrm{NS}}^{\textrm{2d $(2,2)$ KS}_{M,k}}(y;q)
\nonumber\\
&=\sum_{j=1}^M 
\Biggl[
-\frac{q^{\frac{k+j}{2(M+k+1)}}y^{-\frac{k+j}{M+k+1}}}{1-q}-\frac{q^{1-\frac{k+j}{2(M+k+1)}}y^{\frac{k+j}{M+k+1}}}{1-q}
+\frac{q^{\frac{j}{2(M+k+1)}}y^{-\frac{j}{M+k+1}}}{1-q}+\frac{q^{1-\frac{j}{2(M+k+1)}}y^{\frac{j}{M+k+1}}}{1-q}
\Biggr]. 
\end{align}

The supersymmetric zeta function is 
\begin{align}
&
\mathfrak{Z}_{\textrm{NS}}^{\textrm{2d $(2,2)$ KS}_{M,k}}(s,z)
\nonumber\\
&=\sum_{j=1}^M 
\Biggl[
-\zeta\left(s,\frac{k+j}{2(M+k+1)}+z\right)
-\zeta\left(s,1-\frac{k+j}{2(M+k+1)}+z\right)
\nonumber\\
&\qquad +\zeta\left(s,\frac{j}{2(M+k+1)}+z\right)
+\zeta\left(s,1-\frac{j}{2(M+k+1)}+z\right)
\Biggr]. 
\end{align}
Although each of the Hurwitz zeta function has a simple pole at $s=1$, 
the total sum of the residues vanishes
\begin{align}
\mathrm{Res}_{s=1}\mathfrak{Z}_{\textrm{NS}}^{\textrm{2d $(2,2)$ KS}_{M,k}}(s,z)=0. 
\end{align}
Also the special supersymmetric zeta value with $s=0$ and $z=0$ vanishes
\begin{align}
\mathfrak{Z}_{\textrm{NS}}^{\textrm{2d $(2,2)$ KS}_{M,k}}(0,0)&=0. 
\end{align}
The special supersymmetric zeta value with $s=-1$ and $z=0$ is given by
\begin{align}
\mathfrak{Z}_{\textrm{NS}}^{\textrm{2d $(2,2)$ KS}_{M,k}}(-1,0)
&=-\sum_{j=1}^M \frac{k(2M+k-2j+2)}{4(M+k+1)^2}
\nonumber\\
&=-\frac{Mk}{4(M+k+1)}. 
\end{align}
Therefore we get from (\ref{cL_zeta}) and (\ref{cR_zeta}) the central charge 
\begin{align}
c&=c_L=c_R=\frac{3Mk}{M+k+1}. 
\end{align}

The supersymmetric determinant is
\begin{align}
\mathfrak{D}_{\textrm{NS}}^{\textrm{2d $(2,2)$ KS}_{M,k}}(z)
&=\prod_{j=1}^M \frac{\Gamma\left(\frac{j}{2(M+k+1)}+z\right)\Gamma\left(1-\frac{j}{2(M+k+1)}+z\right)}
{\Gamma\left(\frac{k+j}{2(M+k+1)}+z\right)\Gamma\left(1-\frac{k+j}{2(M+k+1)}+z\right)}. 
\end{align}
The vacuum exponent is evaluated as
\begin{align}
\mathfrak{D}_{\textrm{NS}}^{\textrm{2d $(2,2)$ KS}_{M,k}}(0)
&=\prod_{i=1}^{\lfloor\frac{M+1}{2}\rfloor}
\cot\left(\frac{\pi (2i-1)}{2(k+M+1)}\right)
\prod_{j=1}^{\lfloor \frac{M}{2} \rfloor}
\cot\left(\frac{\pi j}{k+M+1}\right), 
\end{align}
where $\lfloor x\rfloor$ $:=$ $\max\{m\in \mathbb{Z}|m\le x\}$ is the floor function. 

\subsubsection{2d $\mathcal{N}=(2,2)$ $\mathbb{CP}^{N-1}$ models}
The $\mathcal{N}=(2,2)$ supersymmetric $\mathbb{CP}^{N-1}$ model is described by a $U(1)$ vector multiplet with $N$ chiral multiplets of charge $+1$ 
(See e.g. \cite{DAdda:1978vbw,DAdda:1978dle,Witten:1978bc,Witten:1993yc} for earlier works on the model). 
It is known as an asymptotic free theory with $N$ vacua with mass gap. 
So the theory is not conformal in the IR. 
It is mirror to the $A_{N-1}$ affine Toda theory that is an LG theory of $N-1$ periodic chiral multiplet fields $\Phi_{\alpha}$ 
with a superpotential of the form
\begin{align}
\mathcal{W}&=\Lambda\left(
\sum_{\alpha=1}^{N-1}e^{\Phi_{\alpha}}
+\prod_{\alpha=1}^{N-1}e^{-\Phi_{\alpha}}
\right), 
\end{align}
with $\Lambda$ being some constant, corresponding to the dynamical scale generated in the $\mathbb{CP}^{N-1}$ model. 

The elliptic genus is given by \cite{Benini:2013nda}
\begin{align}
\label{ind_2d22_CPN-1}
\mathcal{I}_{\textrm{NS}}^{\textrm{2d $(2,2)$ $\mathbb{CP}^{N-1}$}}(y,x_{\alpha};q)
&=\frac{(q)_{\infty}^2}{(q^{\frac12}y^{\pm};q)_{\infty}}
\oint_{\textrm{JK}} \frac{ds}{2\pi is}
\prod_{\alpha=1}^{N}
\frac{(q^{\frac12}y^{\mp}s^{\pm}x_{\alpha}^{\mp};q)_{\infty}}
{(sx_{\alpha}^{-1};q)_{\infty}(qs^{-1}x_{\alpha};q)_{\infty}}, 
\end{align}
where $y^N=1$ and $\prod_{\alpha}x_{\alpha}=1$. 
Picking up the JK residues at $N$ simple poles $s=x_{\beta}$ with $\beta=1,\cdots, N$, we obtain
\begin{align}
\mathcal{I}_{\textrm{NS}}^{\textrm{2d $(2,2)$ $\mathbb{CP}^{N-1}$}}(y,x_{\alpha};q)
&=\sum_{\beta=1}^{N}
\mathcal{I}_{\textrm{NS},\beta}^{\textrm{2d $(2,2)$ $\mathbb{CP}^{N-1}$}}(y,x_{\alpha};q), 
\end{align}
where 
\begin{align}
\label{ind_2d22_CPN-1sub}
\mathcal{I}_{\textrm{NS},\beta}^{\textrm{2d $(2,2)$ $\mathbb{CP}^{N-1}$}}(y,x_{\alpha};q)
&=\prod_{\gamma\neq \beta}
\frac{(q^{\frac12}y^{-1}\frac{x_{\beta}}{x_{\gamma}};q)_{\infty} (q^{\frac12}y\frac{x_{\gamma}}{x_{\beta}};q)_{\infty}}
{(\frac{x_{\beta}}{x_{\gamma}};q)_{\infty} (q\frac{x_{\gamma}}{x_{\beta}};q)_{\infty}}. 
\end{align}
In the unflavored limit each residue (\ref{ind_2d22_CPN-1sub}) takes the same form albeit with a singularity 
so that the full index is simply given by multiplying it by $N$. 
When we take the plethystic logarithm of the full index, the multiplicative factor $N$ simply gives rise to the constant term $\log N$, 
which is irrelevant upon the Mellin transform. 
So it is enough to consider the single-particle index as the plethystic logarithm of the contribution (\ref{ind_2d22_CPN-1sub})
\begin{align}
\label{sind_2d22_CPN-1sub}
i_{\textrm{NS},\beta}^{\textrm{2d $(2,2)$ $\mathbb{CP}^{N-1}$}}(y,x_{\alpha};q)
&=\sum_{\gamma\neq \beta}
\Biggl[
-\frac{q^{\frac12}y^{-1}\frac{x_{\beta}}{x_{\gamma}}}{1-q}
-\frac{q^{\frac12}y\frac{x_{\gamma}}{x_{\beta}}}{1-q}
+\frac{\frac{x_{\beta}}{x_{\gamma}}}{1-q}
+\frac{q\frac{x_{\gamma}}{x_{\beta}}}{1-q}
\Biggr]. 
\end{align}

Then the supersymmetric zeta function associated with (\ref{sind_2d22_CPN-1sub}) takes the form
\begin{align}
&
\mathfrak{Z}_{\textrm{NS},\beta}^{\textrm{2d $(2,2)$ $\mathbb{CP}^{N-1}$}}(s,z)
\nonumber\\
&=-2(N-1)\zeta\left(s,\frac12+z\right)+(N-1)\zeta\left(s,z\right)+(N-1)\zeta\left(s,1+z\right). 
\end{align}
We have
\begin{align}
\mathrm{Res}_{s=1}\mathfrak{Z}_{\textrm{NS},\beta}^{\textrm{2d $(2,2)$ $\mathbb{CP}^{N-1}$}}(s,z)&=0
\end{align}
and 
\begin{align}
\mathfrak{Z}_{\textrm{NS},\beta}^{\textrm{2d $(2,2)$ $\mathbb{CP}^{N-1}$}}(0,0)&=0. 
\end{align}
The supersymmetric zeta value with $s=-1$ and $z=0$ is given by
\begin{align}
\mathfrak{Z}_{\textrm{NS},\beta}^{\textrm{2d $(2,2)$ $\mathbb{CP}^{N-1}$}}(-1,0)&=-\frac{N-1}{4}. 
\end{align}
From (\ref{cL_zeta}) and (\ref{cR_zeta}) we obtain 
\begin{align}
\label{c_2d22_CPN-1}
c&=c_L=c_R=3(N-1). 
\end{align}
This is the 't Hooft anomaly rather than the central charge. The value $\frac{c}{3}$ is equal to the dimension $N-1$ of the $\mathbb{CP}^{N-1}$. 

The supersymmetric determinant is
\begin{align}
\mathfrak{D}_{\textrm{NS},\beta}^{\textrm{2d $(2,2)$ $\mathbb{CP}^{N-1}$}}(z)
&=\frac{\Gamma(z)^{N-1}\Gamma(1+z)^{N-1}}{\Gamma(\frac12+z)^{2(N-1)}}. 
\end{align}
The vacuum exponent is divergent
\begin{align}
\mathfrak{D}_{\textrm{NS},\beta}^{\textrm{2d $(2,2)$ $\mathbb{CP}^{N-1}$}}(0)
&=\infty. 
\end{align}

\section{4d supersymmetric field theories}
\label{sec_4d}

\subsection{Zeta functions for 4d supersymmetric indices}

The supersymmetric indices for 4d $\mathcal{N}=1$ supersymmetric field theories can be defined as follows
\cite{Kinney:2005ej,Romelsberger:2005eg}:
\begin{align}
\label{4d_ind_DEF}
\mathcal{I}^{\textrm{4d $\mathcal{N}=1$}}(x;p;q)
&=\Tr (-1)^F p^{j_{3}+\overline{j}_3-\frac{R}{2}} q^{j_{3}-\overline{j}_3-\frac{R}{2}} \prod_{\alpha} x_{\alpha}^{f_{\alpha}}, 
\end{align}
where $j_3$ and $\overline{j}_{3}$ are the Cartan generators of the rotation group $SU(2)_1\times SU(2)_{2}$, 
$R$ is the $U(1)_R$ R-charge, and $f_{\alpha}$ are the Cartan generators of the other global symmetries. 
The supersymmetric indices of 4d $\mathcal{N}=1$ gauge theories can be evaluated as the elliptic hypergeometric integrals \cite{Dolan:2008qi}. 

The $6$-form anomaly polynomial takes the form 
\begin{align}
\label{4d_anomaly}
\mathcal{A}_6&=\frac{1}{6}k_{RRR} c_1(R)^3-\frac{1}{24}k_R c_1(R)p_1(T), 
\end{align}
where $c_1(R)$ is the first Chern class for the $U(1)_R$ R-symmetry background gauge field 
and $p_1(T)$ is the first Pontryagin class of the tangent bundle $T$ of the four-dimensional spacetime. 
The 't Hooft anomaly coefficient $k_{RRR}$ for the $U(1)$ R-symmetry 
and the mixed $U(1)_R$-gravitational anomaly coefficient $k_R$ are given by
\begin{align}
k_{RRR}&=\Tr U(1)_R^3,& 
k_{R}&=\Tr U(1)_R. 
\end{align}
For the superconformal field theories, the $c$-anomaly $c_{4d}$ and the $a$-anomaly $a_{4d}$ are linearly related to the 't Hooft anomaly coefficient for the $U(1)_R$ R-symmetry and the mixed $U(1)$-gravitational anomaly coefficient \cite{Anselmi:1997am}
\begin{align}
c_{4d}&=\frac{9}{32}k_{RRR}-\frac{5}{32}k_R, \\
a_{4d}&=\frac{9}{32}k_{RRR}-\frac{3}{32}k_R. 
\end{align}

We define the multiple supersymmetric zeta functions  
$\mathfrak{Z}^{\textrm{4d $\mathcal{N}=1$}}(s,z;\omega_1,\omega_2)$ for the 4d $\mathcal{N}=1$ supersymmetric field theories by taking the Mellin transform (\ref{Mellin_transf1}) or equivalently (\ref{Mellin_transf2}) of the plethystic logarithms of the 4d supersymmetric indices (\ref{4d_ind_DEF}) with the fugacities $p$ and $q$ being replaced with $p = q^{\omega_1}$ and $q = q^{\omega_2}$.
Subsequently, the supersymmetric determinants $\mathfrak{D}^{\textrm{4d $\mathcal{N}=1$}}(z;\omega_1,\omega_2)$ 
associated with the supersymmetric zeta functions $\mathfrak{Z}^{\textrm{4d $\mathcal{N}=1$}}$ $(s,z;\omega_1,\omega_2)$ are defined by (\ref{sDet_DEF}). 
These supersymmetric zeta functions can be expressed in terms of the Barnes double zeta functions \cite{barnes1904theory,MR1576719}, which have simple poles at $s=2$ and $s=1$. 
However, the supersymmetric zeta functions 
$\mathfrak{Z}^{\textrm{4d $\mathcal{N}=1$}}$ $(s,z;\omega_1,\omega_2)$ may only have a simple pole at $s=1$. 

From the asymptotic formula (\ref{Cardy_lim0}), we have 
\begin{align}
&
\log \mathcal{I}^{\textrm{4d $\mathcal{N}=1$}}(1;q^{\omega_1};q^{\omega_2})
\nonumber\\
&\sim \frac{\pi^2 \mathrm{Res}_{s=1}\mathfrak{Z}^{\textrm{4d $\mathcal{N}=1$}}(s,0;\omega_1,\omega_2)}
{6\beta}
-\mathfrak{Z}^{\textrm{4d $\mathcal{N}=1$}}(0,0;\omega_1,\omega_2)\log\beta
+\log \mathfrak{D}^{\textrm{4d $\mathcal{N}=1$}}(0;\omega_1,\omega_2). 
\end{align}
When we expand the 4d $\mathcal{N}=1$ supersymmetric index with $p=q$ as 
\begin{align}
\mathcal{I}^{\textrm{4d $\mathcal{N}=1$}}(1;q;q)
&=\sum_{n} d^{\textrm{4d $\mathcal{N}=1$}}(n) q^n, 
\end{align}
the analysis of the asymptotic behavior of the expansion coefficients $d^{\textrm{4d $\mathcal{N}=1$}}(n)$ 
needs delicate and subtle care since the expansion coefficients $d^{\textrm{4d $\mathcal{N}=1$}}(n)$ generally do not grow exponentially. 

Di Pietro and Komargodski argued \cite{DiPietro:2014bca} that 
the leading terms in the Cardy-like limit of the 4d supersymmetric indices are universally captured by the mixed $U(1)$-gravitational anomalies  
and they can be expressed in terms of the 4d central charges $c_{4d}$ and $a_{4d}$ for superconformal theories. 
This implies that the residues of the supersymmetric zeta functions at a simple pole $s=1$ are given by
\begin{align}
\label{Cardy_4dSCI}
\mathrm{Res}_{s=1}\mathfrak{Z}^{\textrm{4d $\mathcal{N}=1$}}(s,z;\omega_1,\omega_2)
&=16(c_{4d}-a_{4d}) \frac{\omega_1+\omega_2}{\omega_1\omega_2}. 
\end{align}

From the various examples we have examined, the special supersymmetric zeta values with $s=-1$ and $z=0$ can be determined by the $c$- and $a$-anomalies
\begin{align}
\label{Cas_4dSCI}
&
\mathfrak{Z}^{\textrm{4d $\mathcal{N}=1$}}(-1,0;\omega_1,\omega_2)
\nonumber\\
&=\frac{4(\omega_1^3+\omega_2^3)}{9\omega_1\omega_2}c_{4d}
-\frac{4(2\omega_1^3-3\omega_1^2\omega_2-3\omega_1\omega_2^2+2\omega_2^3)}{27\omega_1\omega_2}a_{4d}. 
\end{align}
Comparison with the supersymmetric Casimir energies for 4d $\mathcal{N}=1$ supersymmetric field theories \cite{Kim:2012ava,Assel:2014paa} 
(also see \cite{Assel:2015nca,Assel:2014tba,Cassani:2014zwa,ArabiArdehali:2015iow,Bobev:2015kza}) shows that our proposed relation (\ref{zeta-1_Cas}) holds.

For superconformal field theories, from (\ref{Cardy_4dSCI}) and (\ref{Cas_4dSCI}) the $c$- and $a$-anomalies can be expressed in terms of the supersymmetric zeta functions 
\begin{align}
\label{c4d_zeta}
c_{4d}&=\frac{1}{128}\mathrm{Res}_{s=1}\mathfrak{Z}^{\textrm{4d $\mathcal{N}=1$}}(s,0;1,1)
+\frac{27}{32}\mathfrak{Z}^{\textrm{4d $\mathcal{N}=1$}}(-1,0;1,1), \\
\label{a4d_zeta}
a_{4d}&=-\frac{3}{128}\mathrm{Res}_{s=1}\mathfrak{Z}^{\textrm{4d $\mathcal{N}=1$}}(s,0;1,1)
+\frac{27}{32}\mathfrak{Z}^{\textrm{4d $\mathcal{N}=1$}}(-1,0;1,1). 
\end{align}
Also, the $6$-form anomaly coefficients are given by
\begin{align}
\label{kRRR_zeta}
k_{RRR}&=-\frac{1}{4}\mathrm{Res}_{s=1}\mathfrak{Z}^{\textrm{4d $\mathcal{N}=1$}}(s,0;1,1)
+3\mathfrak{Z}^{\textrm{4d $\mathcal{N}=1$}}(-1,0;1,1), \\
\label{kR_zeta}
k_{R}&=-\frac{1}{2}\mathrm{Res}_{s=1}\mathfrak{Z}^{\textrm{4d $\mathcal{N}=1$}}(s,0;1,1). 
\end{align}

\subsubsection{4d $\mathcal{N}=1$ chiral multiplet}
The supersymmetric index of the 4d $\mathcal{N}=1$ chiral multiplet with R-charge $r$ takes the form \cite{Romelsberger:2005eg}
\begin{align}
\label{ind_4dchiral}
\mathcal{I}^{\textrm{4d $\mathcal{N}=1$ chiral}_r}(x;p;q)
&=\Gamma((pq)^{\frac{r}{2}}x;p,q), 
\end{align}
where
\begin{align}
\Gamma(x;p,q)
&=\prod_{m=0}^{\infty}\prod_{n=0}^{\infty}
\frac{1-p^{m+1}q^{n+1}x^{-1}}{1-p^{m}q^nx}
\end{align}
is the elliptic gamma function. 
The single-particle index is given by
\begin{align}
\label{sind_4dchiral}
i^{\textrm{4d $\mathcal{N}=1$ chiral}_r}(x;p;q)&=
\frac{(pq)^{\frac{r}{2}}x-(pq)^{\frac{2-r}{2}}x^{-1}}
{(1-p)(1-q)}. 
\end{align}
For $p=q$ it becomes
\begin{align}
\label{sind_4dchiral1}
i^{\textrm{4d $\mathcal{N}=1$ chiral}_r}(x;q;q)&=
\frac{q^rx-q^{2-r}x^{-1}}{(1-q)^2}
\nonumber\\
&=\sum_{n=1}^{\infty}(nq^{n-1+r}x-nq^{n+1-r}x^{-1}). 
\end{align}

The supersymmetric zeta function is evaluated as
\begin{align}
\label{zeta_4dchiral}
&
\mathfrak{Z}^{\textrm{4d $\mathcal{N}=1$ chiral}_r}(s,z;\omega_1,\omega_2)
\nonumber\\
&=\zeta_2\left(s,\frac{r(\omega_1+\omega_2)}{2}+z;\omega_1,\omega_2\right)
-\zeta_2\left(s,\frac{(2-r)(\omega_1+\omega_2)}{2}+z;\omega_1,\omega_2\right), 
\end{align}
where
\begin{align}
\label{Barnes_double_Z}
\zeta_2(s,z;a,b)
&=\sum_{m,n\ge0}(am+bn+z)^{-s}
\end{align}
is the Barnes double zeta function \cite{barnes1904theory,MR1576719} defined by the double series. 
When $\omega_1=\omega_2=1$, it reduces to
\begin{align}
\label{zeta_4dchiral1}
\mathfrak{Z}^{\textrm{4d $\mathcal{N}=1$ chiral}_r}(s,z;1,1)
&=\zeta(s-1,r+z)-\zeta(s-1,2-r+z)
\nonumber\\
&+(1-r-z)\zeta(s,r+z)-(-1+r-z)\zeta(s,2-r+z). 
\end{align}
The Barnes double zeta function (\ref{Barnes_double_Z}) has an analytic continuation 
to the complex $s$-plane with simple poles at $s=1$ and $s=2$. 
The residues at these poles are given by \cite{MR2528052}
\begin{align}
\mathrm{Res}_{s=2}\zeta_2(s,z;a,b)&=\frac{1}{ab}, \\
\mathrm{Res}_{s=1}\zeta_2(s,z;a,b)&=\frac12\left(\frac{1}{a}+\frac{1}{b} \right)-\frac{z}{ab}. 
\end{align}
Hence the residues at simple poles at $s=2$ and $s=1$ for the supersymmetric zeta function (\ref{zeta_4dchiral}) are evaluated as
\begin{align}
\label{Res2_4dchiral}
\mathrm{Res}_{s=2}\mathfrak{Z}^{\textrm{4d $\mathcal{N}=1$ chiral}_r}(s,z;\omega_1,\omega_2)
&=0, \\
\label{Res1_4dchiral}
\mathrm{Res}_{s=1}\mathfrak{Z}^{\textrm{4d $\mathcal{N}=1$ chiral}_r}(s,z;\omega_1,\omega_2)
&=\frac{(1-r)(\omega_1+\omega_2)}{\omega_1\omega_2}. 
\end{align}
The special supersymmetric zeta values can be evaluated by making use of the formula (\ref{values_doubleBzeta}). 
The Zeta-index is 
\begin{align}
\mathfrak{Z}^{\textrm{4d $\mathcal{N}=1$ chiral}_r}(0,0;\omega_1,\omega_2)&=0. 
\end{align}
The supersymmetric zeta value with $s=-1$ and $z=0$ is
\begin{align}
\label{-1_4dchiral}
&
\mathfrak{Z}^{\textrm{4d $\mathcal{N}=1$ chiral}_r}(-1,0;\omega_1,\omega_2)
\nonumber\\
&=\frac{\omega_1+\omega_2}{24\omega_1\omega_2}
(r-1)
\left[
(\omega_1+\omega_2)^2r^2-2(\omega_1+\omega_2)^2r+2\omega_1\omega_2
\right]. 
\end{align}
Substituting (\ref{Res1_4dchiral}) and (\ref{-1_4dchiral}) into the formulae (\ref{c4d_zeta}), (\ref{a4d_zeta}), 
(\ref{kRRR_zeta}) and (\ref{kR_zeta}), we find the anomalies 
\begin{align}
\label{c_4dchiral}
c_{4d}&=\frac{9}{32}k_{RRR}-\frac{5}{32}k_R
\nonumber\\
&=\frac{9}{32}r^3-\frac{27}{32}r^2+\frac{11}{16}r-\frac18, \\
\label{a_4dchiral}
a_{4d}
&=\frac{9}{32}k_{RRR}-\frac{3}{32}k_R
\nonumber\\
&=\frac{9}{32}r^3-\frac{27}{32}r^2+\frac{3}{4}r-\frac{3}{16}, \\
k_{RRR}&=(r-1)^3, \\
k_R&=r-1. 
\end{align}

For a free chiral multiplet the R-charge is $r=2/3$ 
the expressions (\ref{c_4dchiral}) and (\ref{a_4dchiral}) reduce to the central charges
\begin{align}
\label{c_4dchiral2/3}
c_{4d}&=\frac{1}{24}, \\
\label{a_4dchiral2/3}
a_{4d}&=\frac{1}{48}. 
\end{align}
In this case, the value $a_{4d}/c_{4d}$ $=$ $1/2$ hits the lower bound for $\mathcal{N}=1$ SCFT \cite{Hofman:2008ar}. 

\begin{figure}
\begin{center}
\includegraphics[width=10cm]{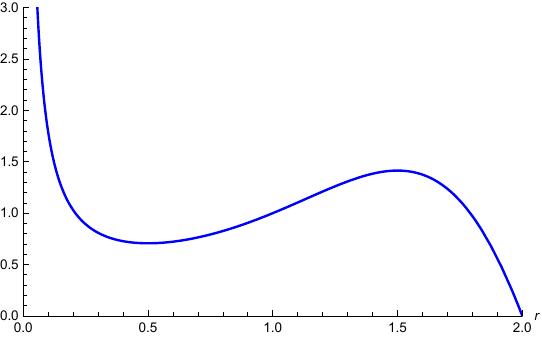}
\caption{The vacuum exponent $\mathfrak{D}^{\textrm{4d $\mathcal{N}=1$ chiral}_r}(0;1,1)$.}
\label{fig_vacexp_4dcm}
\end{center}
\end{figure}

The supersymmetric determinant is 
\begin{align}
\mathfrak{D}^{\textrm{4d $\mathcal{N}=1$ chiral}_r}(z;\omega_1,\omega_2)
&=\frac{\Gamma_2\left(\frac{r(\omega_1+\omega_2)}{2}+z;\omega_1,\omega_2\right)}
{\Gamma_2\left(\frac{(2-r)(\omega_1+\omega_2)}{2}+z;\omega_1,\omega_2\right)}, 
\end{align}
where 
\begin{align}
\Gamma_2(z;\omega_1,\omega_2)
&:=\exp\left[\frac{\partial}{\partial s} \zeta_2(s,z;\omega_1,\omega_2)\Biggl|_{s=0}\right]
\end{align}
is the Barnes double gamma function \cite{barnes1904theory}. 
The vacuum exponent with $\omega_1=\omega_2$ $=$ $1$ is given by
\begin{align}
\label{vexp_4dchiral}
\mathfrak{D}^{\textrm{4d $\mathcal{N}=1$ chiral}_r}(0;1,1)
&=e^{\zeta'(-1,r)-\zeta'(-1,2-r)}
\left[
\frac{\Gamma(r)\Gamma(2-r)}{2\pi}
\right]^{1-r}. 
\end{align}
We show a plot of the vacuum exponent (\ref{vexp_4dchiral}) in Figure \ref{fig_vacexp_4dcm}. 
%
%
%
%
%

When one introduces a mass term $\mathcal{W}=\lambda \Phi^2$, the chiral multiplet has R-charge $r=1$. 
The theory has a single supersymmetric vacuum. 
The supersymmetric index (\ref{ind_4dchiral}) is equal to $1$ 
and the supersymmetric zeta function (\ref{zeta_4dchiral}) vanishes. 
The vacuum exponent $\mathfrak{D}^{\textrm{4d $\mathcal{N}=1$ chiral}_r}(0;1,1)$ is $1$. 

For the theory of a chiral multiplet with a linear superpotential $\mathcal{W}=\lambda \Phi$, the Polonyi model \cite{Polonyi:1977pj}, 
the chiral multiplet has R-charge $r=2$. 
There is no supersymmetric vacuum so that supersymmetry is spontaneously broken. 
The supersymmetric index (\ref{ind_4dchiral}) vanishes. 
It stems from the fermionic contribution $-1$ in the single-particle index (\ref{sind_4dchiral}), corresponding to the Goldstino. 
In this case the vacuum exponent $\mathfrak{D}^{\textrm{4d $\mathcal{N}=1$ chiral}_r}(0;1,1)$ vanishes, as discussed in (\ref{vacexp_criteria}). 

\subsubsection{4d $\mathcal{N}=1$ vector multiplet}

The supersymmetric index of the 4d $\mathcal{N}=1$ $U(1)$ vector multiplet reads \cite{Romelsberger:2005eg}
\begin{align}
\label{ind_4dvec}
\mathcal{I}^{\textrm{4d $\mathcal{N}=1$ vec}}(p;q)
&=(p;p)_{\infty}(q;q)_{\infty}. 
\end{align}
The single-particle index of the vector multiplet is 
\begin{align}
\label{sind_4dvec}
i^{\textrm{4d $\mathcal{N}=1$ vec}}(p;q)
&=-\frac{p}{1-p}-\frac{q}{1-q}. 
\end{align}
In the limit $p=q$, it reduces to
\begin{align}
i^{\textrm{4d $\mathcal{N}=1$ vec}}(q)
&=\frac{-2q}{1-q}
=-2\sum_{n=1}^{\infty}q^{n}. 
\end{align}

The supersymmetric zeta function is given by
\begin{align}
\label{zeta_4dvec}
\mathfrak{Z}^{\textrm{4d $\mathcal{N}=1$ vec}}(s,z;\omega_1,\omega_2)
&=-\omega_1^{-s}\zeta\left(s,\frac{\omega_1+z}{\omega_1}\right)
-\omega_2^{-s}\zeta\left(s,\frac{\omega_2+z}{\omega_2}\right). 
\end{align}
It has a simple pole at $s=1$ with residue
\begin{align}
\label{Res1_4dvec}
\mathrm{Res}_{s=1}\mathfrak{Z}^{\textrm{4d $\mathcal{N}=1$ vec}}(s,z;\omega_1,\omega_2)&=-\frac{1}{\omega_1}-\frac{1}{\omega_2}. 
\end{align}
The Zeta-index is 
\begin{align}
\label{0_4dvec}
\mathfrak{Z}^{\textrm{4d $\mathcal{N}=1$ vec}}(0,0;\omega_1,\omega_2)&=1. 
\end{align}
The supersymmetric zeta value with $s=-1$ and $z=0$ is
\begin{align}
\label{-1_4dvec}
\mathfrak{Z}^{\textrm{4d $\mathcal{N}=1$ vec}}(-1,0;\omega_1,\omega_2)
&=\frac{\omega_1+\omega_2}{12}. 
\end{align}
Plugging (\ref{Res1_4dvec}) and (\ref{-1_4dvec}) into the formulae (\ref{kRRR_zeta}) and (\ref{kR_zeta}), we get the anomalies 
\begin{align}
k_{RRR}&=1, \\
k_{R}&=1. 
\end{align}
Also from (\ref{c4d_zeta}) and (\ref{a4d_zeta}), we get 
\begin{align}
\label{c_4dvec}
c_{4d}&=\frac18, \\ 
\label{a_4dvec}
 a_{4d}&=\frac{3}{16}, 
\end{align}
which are equal to the central charges contributed from the 4d $\mathcal{N}=1$ $U(1)$ vector multiplet. 
The value $a_{4d}/c_{4d}$ $=$ $3/2$ hits the upper bound for $\mathcal{N}=1$ SCFT \cite{Hofman:2008ar}. 

The supersymmetric determinant is
\begin{align}
\mathfrak{D}^{\textrm{4d $\mathcal{N}=1$ vec}}(z;\omega_1,\omega_2)
&=\frac{2\pi \omega_1^{\frac12-\frac{\omega_1+z}{\omega_1}}\omega_2^{\frac12-\frac{\omega_2+z}{\omega_2}} }
{\Gamma\left(\frac{\omega_1+z}{\omega_1}\right)\Gamma\left(\frac{\omega_2+z}{\omega_2}\right)}. 
\end{align}
The vacuum exponent is 
\begin{align}
\mathfrak{D}^{\textrm{4d $\mathcal{N}=1$ vec}}(0;\omega_1,\omega_2)
&=\frac{2\pi}{\sqrt{\omega_1\omega_2}}. 
\end{align}

\subsubsection{4d $\mathcal{N}=1$ $SU(2)$ with $6$ chirals $(\Phi_{\alpha},\bar{\Phi}_{\alpha})$}
Let us consider the 4d $\mathcal{N}=1$ $SU(2)$ gauge theory with $3$ fundamental chiral multiplets $\Phi_{\alpha}$ and $3$ antifundamental chiral multiplets $\bar{\Phi}_{\alpha}$. 
The theory is dual to a free theory of $15$ chiral multiplets $(\Phi',B',\bar{B}')$ interacting with a superpotential \cite{Seiberg:1994pq}
\begin{align}
\mathcal{W}&=\Phi'B'\bar{B}'-\det \Phi'. 
\end{align}
The field content is summarized as
\begin{align}
\begin{array}{c|ccccc}
&SU(2)&SU(3)_x&SU(3)_y&U(1)_B&U(1)_R \\ \hline
\Phi_{\alpha}&\mathbf{2}&\mathbf{3}&\mathbf{1}&+1&\frac13 \\ 
\bar{\Phi}_{\alpha}&\overline{\mathbf{2}}&\mathbf{1}&\mathbf{3}&-1&\frac13 \\ \hline
\Phi'&\mathbf{1}&\mathbf{3}&\mathbf{3}&0&\frac23 \\ 
B'&\mathbf{1}&\overline{\mathbf{3}}&\mathbf{1}&+2&\frac23 \\ 
\bar{B}'&\mathbf{1}&\mathbf{1}&\overline{\mathbf{3}}&-2&\frac23 \\ 
\end{array}
\end{align}

The supersymmetric index is given by
\begin{align}
\label{ind_4dsuN3cm}
&
\mathcal{I}^{\textrm{4d $\mathcal{N}=1$ $SU(2)$ $\Phi_{\alpha}$}}(b,x_{\alpha},\tilde{x}_{\alpha};p;q)
\nonumber\\
&=\frac{(p;p)_{\infty}(q;q)_{\infty}}{2}
\oint \frac{ds}{2\pi is}
\frac{1}{\Gamma(s^{\pm 2};p,q)}
\prod_{\alpha=1}^3
\Gamma((pq)^{\frac16}bx_{\alpha}s^{\mp};p,q)
\Gamma((pq)^{\frac16}b^{-1}\tilde{x}_{\alpha}s^{\pm};p,q), 
\end{align}
with $\prod_{\alpha}x_{\alpha}$ $=$ $\prod_{\alpha}\tilde{x}_{\alpha}$ $=$ $1$.  
It is shown to be equal to \cite{MR1846786}
\begin{align}
\prod_{\alpha=1}^3 \prod_{\beta=1}^3
\Gamma((pq)^{\frac13}x_{\alpha}\tilde{x}_{\beta};p,q)
\prod_{\alpha=1}^{3}\Gamma((pq)^{\frac13}b^2x_{\alpha}^{-1};p,q)
\prod_{\alpha=1}^{3}\Gamma((pq)^{\frac13}b^{-2}\tilde{x}_{\alpha}^{-1};p,q), 
\end{align}
which is identified with the index of the dual theory. 
The single-particle index reads
\begin{align}
\label{sind_4dsuN3cm}
&
i^{\textrm{4d $\mathcal{N}=1$ $SU(2)$ $\Phi_{\alpha}$}}(b,x_{\alpha},\tilde{x}_{\alpha};p;q)
\nonumber\\
&=\sum_{\alpha=1}^3\sum_{\beta=1}^3 
\frac{(pq)^{\frac13}x_{\alpha}\tilde{x}_{\beta}-(pq)^{\frac23}\tilde{x}_{\alpha}^{-1}x_{\beta}^{-1}}
{(1-p)(1-q)}
\nonumber\\
&+\sum_{\alpha=1}^3 
\frac{(pq)^{\frac13}b^2x_{\alpha}^{-1}-(pq)^{\frac23}b^{-2}x_{\alpha}}
{(1-p)(1-q)}
+\sum_{\alpha=1}^3 
\frac{(pq)^{\frac13}b^{-2}\tilde{x}_{\alpha}^{-1}-(pq)^{\frac23}b^{2}\tilde{x}_{\alpha}}
{(1-p)(1-q)}. 
\end{align}
With $p=q^{\omega_1}$, $q=q^{\omega_2}$, $x_{\alpha}=\tilde{x}_{\alpha}=1$ and $b=q^{B}$, 
it becomes
\begin{align}
&
i^{\textrm{4d $\mathcal{N}=1$ $SU(2)$ $\Phi_{\alpha}$}}(q^{B},1,1;q^{\omega_1};q^{\omega_2})
\nonumber\\
&=9\frac{q^{\frac{\omega_1+\omega_2}{3}}-q^{\frac{2(\omega_1+\omega_2)}{3}}}
{(1-q^{\omega_1})(1-q^{\omega_2})}
+3\frac{q^{\frac{\omega_1+\omega_2}{3}+2B}-q^{\frac{2(\omega_1+\omega_2)}{3}-2B}}
{(1-q^{\omega_1})(1-q^{\omega_2})}
+3\frac{q^{\frac{\omega_1+\omega_2}{3}-2B}-q^{\frac{2(\omega_1+\omega_2)}{3}+2B}}
{(1-q^{\omega_1})(1-q^{\omega_2})}. 
\end{align}

The supersymmetric zeta function is evaluated as
\begin{align}
\label{zeta_4dsuN3cm}
&
\mathfrak{Z}^{\textrm{4d $\mathcal{N}=1$ $SU(2)$ $\Phi_{\alpha}$}}(s,z;\omega_1,\omega_2;B)
\nonumber\\
&=9\left[\zeta_2\left(s,\frac{\omega_1+\omega_2}{3}+z;\omega_1,\omega_2\right)
-\zeta_2\left(s,\frac{2(\omega_1+\omega_2)}{3}+z;\omega_1,\omega_2\right)\right]
\nonumber\\
&+3\left[\zeta_2\left(s,\frac{\omega_1+\omega_2}{3}+2B+z;\omega_1,\omega_2\right)
-\zeta_2\left(s,\frac{2(\omega_1+\omega_2)}{3}-2B+z;\omega_1,\omega_2\right)\right]
\nonumber\\
&+3\left[\zeta_2\left(s,\frac{\omega_1+\omega_2}{3}-2B+z;\omega_1,\omega_2\right)
-\zeta_2\left(s,\frac{2(\omega_1+\omega_2)}{3}+2B+z;\omega_1,\omega_2\right)\right]. 
\end{align}
While the Barnes double zeta function has simple poles at $s=2$ and $1$, 
the total residues at these poles are 
\begin{align}
\label{Res2_4dsuN3cm}
\mathrm{Res}_{s=2}\mathfrak{Z}^{\textrm{4d $\mathcal{N}=1$ $SU(2)$ $\Phi_{\alpha}$}}(s,z;\omega_1,\omega_2;B)
&=0, \\
\label{Res1_4dsuN3cm}
\mathrm{Res}_{s=1}\mathfrak{Z}^{\textrm{4d $\mathcal{N}=1$ $SU(2)$ $\Phi_{\alpha}$}}(s,z;\omega_1,\omega_2;B)
&=\frac{5(\omega_1+\omega_2)}{\omega_1\omega_2}. 
\end{align}
Also we have 
\begin{align}
\label{0_4dsuN3cm}
\mathfrak{Z}^{\textrm{4d $\mathcal{N}=1$ $SU(2)$ $\Phi_{\alpha}$}}(0,0;\omega_1,\omega_2;B)&=0
\end{align}
and
\begin{align}
\label{-1_4dsuN3cm}
\mathfrak{Z}^{\textrm{4d $\mathcal{N}=1$ $SU(2)$ $\Phi_{\alpha}$}}(-1,0;\omega_1,\omega_2;B)
&=
-4B^2\frac{\omega_1+\omega_2}{\omega_1\omega_2}
+\frac{5}{36}(\omega_1+\omega_2)
+\frac{5}{27}\frac{\omega_1^3+\omega_2^3}{\omega_1\omega_2}. 
\end{align}
From (\ref{Res1_4dsuN3cm}) and (\ref{-1_4dsuN3cm}) as well as the formulae (\ref{c4d_zeta}) and (\ref{a4d_zeta}), 
we find the trial central charges
\begin{align}
\label{ctrial_4dsuN3cm}
c_{4d}&=\frac58-\frac{27B^2}{(\omega_1+\omega_2)^2}, \\ 
\label{atrial_4dsuN3cm}
 a_{4d}&=\frac{5}{16}-\frac{27B^2}{(\omega_1+\omega_2)^2}. 
\end{align}
According to the $a$-maximization \cite{Intriligator:2003jj}, 
the mixed anomaly coefficient $B$ is required to vanish. 
Then we get
\begin{align}
\label{c_4dsuN3cm}
c_{4d}&=\frac58, \\ 
\label{a_4dsuN3cm}
 a_{4d}&=\frac{5}{16}, 
\end{align}
which agree with the central charges. 
In this case, $a_{4d}/c_{4d}=1/2$ hits the lower bound for $\mathcal{N}=1$ SCFT \cite{Hofman:2008ar}. 

The supersymmetric determinant is 
\begin{align}
\mathfrak{D}^{\textrm{4d $\mathcal{N}=1$ $SU(2)$ $\Phi_{\alpha}$}}(z;\omega_1,\omega_2)
&=\frac{\Gamma_2\left(\frac{\omega_1+\omega_2}{3}+z;\omega_1,\omega_2\right)^{15}}
{\Gamma_2\left(\frac{2(\omega_1+\omega_2)}{3}+z;\omega_1,\omega_2\right)^{15}}. 
\end{align}
Using Mathematica, the vacuum exponent with $\omega_1=\omega_2$ $=$ $1$ is evaluated as
\begin{align}
\mathfrak{D}^{\textrm{4d $\mathcal{N}=1$ $SU(2)$ $\Phi_{\alpha}$}}(0;1,1)
&=\frac{1}{9\sqrt{3}}
e^{\frac{5\left(2\pi^2-3\psi^{(1)}(\frac13)\right)}{6\sqrt{3}\pi}}
=0.0127546\ldots
\end{align}

\subsubsection{4d $\mathcal{N}=1$ $SU(N)$ with $2(N+1)$ chirals $(\Phi_{\alpha},\bar{\Phi}_{\alpha})$}
Consider the 4d $\mathcal{N}=1$ $SU(N)$ gauge theory 
with $N+1$ fundamental chiral multiplets $\Phi_{\alpha}$ and $N+1$ antifundamental chiral multiplets $\bar{\Phi}_{\alpha}$. 
It is dual to a free theory consisting of $(N+1)^2+(N+1)+(N+1)$ chiral multiplets 
$(\Phi',B',\bar{B}')$ interacting with a superpotential \cite{Seiberg:1994pq}
\begin{align}
\mathcal{W}&=\Phi'B'\bar{B}'-\det \Phi'. 
\end{align}
The field content and charges are 
\begin{align}
\begin{array}{c|ccccc}
&SU(N)&SU(N+1)_x&SU(N+1)_y&U(1)_B&U(1)_R \\ \hline
\Phi_{\alpha}&\mathbf{N}&\mathbf{N+1}&\mathbf{1}&+1&\frac{1}{N+1} \\ 
\bar{\Phi}_{\alpha}&\overline{\mathbf{N}}&\mathbf{1}&\mathbf{N+1}&-1&\frac{1}{N+1} \\ \hline
\Phi'&\mathbf{1}&\mathbf{N+1}&\mathbf{N+1}&0&\frac{2}{N+1} \\ 
B'&\mathbf{1}&\overline{\mathbf{N+1}}&\mathbf{1}&+N&\frac{N}{N+1} \\ 
\bar{B}'&\mathbf{1}&\mathbf{1}&\overline{\mathbf{N+1}}&-N&\frac{N}{N+1} \\ 
\end{array}
\end{align}

The supersymmetric index is given by
\begin{align}
\label{ind_4dsuNnfN+1cm}
&
\mathcal{I}^{\textrm{4d $\mathcal{N}=1$ $SU(N)$ $\Phi_{\alpha}$}}(b,x_{\alpha},\tilde{x}_{\alpha};p;q)
\nonumber\\
&=\frac{(p;p)_{\infty}^{N-1}(q;q)_{\infty}^{N-1}}{N!}
\oint \prod_{i=1}^{N-1} \frac{ds_i}{2\pi is_i}
\prod_{i<j}
\frac{1}{\Gamma(s_i^{\pm}s_j^{\mp};p,q)}
\nonumber\\
&\times 
\prod_{i=1}^{N}
\prod_{\alpha=1}^{N+1}
\Gamma((pq)^{\frac{1}{2(N+1)}}bx_{\alpha}s_i^{\mp};p,q)
\Gamma((pq)^{\frac{1}{2(N+1)}}b^{-1}\tilde{x}_{\alpha}s_i^{\pm};p,q), 
\end{align}
with $\prod_{\alpha}x_{\alpha}$ $=$ $\prod_{\alpha}\tilde{x}_{\alpha}$ $=$ $1$. 
It agrees with the index 
\begin{align}
\prod_{\alpha=1}^{N+1}
\prod_{\beta=1}^{N+1}
\Gamma((pq)^{\frac{1}{N+1}}x_{\alpha}\tilde{x}_{\beta};p,q)
\prod_{\alpha=1}^{N+1}
\Gamma((pq)^{\frac{N}{2(N+1)}}b^N x_{\alpha}^{-1};p,q)
\Gamma((pq)^{\frac{N}{2(N+1)}}b^{-N} \tilde{x}_{\alpha}^{-1};p,q)
\end{align}
of the dual theory. 
The single-particle index reads
\begin{align}
\label{sind_4dsuNnfN+1cm}
&
i^{\textrm{4d $\mathcal{N}=1$ $SU(N)$ $\Phi_{\alpha}$}}(b,x_{\alpha},\tilde{x}_{\alpha};p;q)
\nonumber\\
&=\sum_{\alpha=1}^{N+1}\sum_{\beta=1}^{N+1}
\frac{(pq)^{\frac{1}{N+1}}x_{\alpha}\tilde{x}_{\beta}-(pq)^{1-\frac{1}{N+1}}x_{\alpha}^{-1}\tilde{x}_{\beta}^{-1}}
{(1-p)(1-q)}
\nonumber\\
&+\sum_{\alpha=1}^{N+1}
\frac{(pq)^{\frac{N}{2(N+1)}}b^Nx_{\alpha}^{-1}-(pq)^{1-\frac{N}{2(N+1)}}b^{-N}x_{\alpha}}
{(1-p)(1-q)}
+\sum_{\alpha=1}^{N+1}
\frac{(pq)^{\frac{N}{2(N+1)}}b^{-N}\tilde{x}_{\alpha}^{-1}-(pq)^{1-\frac{N}{2(N+1)}}b^{N}\tilde{x}_{\alpha}}
{(1-p)(1-q)}. 
\end{align}
If we set $p=q^{\omega_1}$, $q=q^{\omega_2}$ and $b=q^B$ 
and turn off the flavor fugacities $x_{\alpha}$, $\tilde{x}_{\alpha}$, 
it reduces to
\begin{align}
&
i^{\textrm{4d $\mathcal{N}=1$ $SU(N)$ $\Phi_{\alpha}$}}(q^{B},x_{\alpha},\tilde{x}_{\alpha};q^{\omega_1};q^{\omega_2})
\nonumber\\
&=(N+1)^2\frac{q^{\frac{\omega_1+\omega_2}{N+1}}-q^{(1-\frac{1}{N+1})(\omega_1+\omega_2)}}
{(1-q^{\omega_1})(1-q^{\omega_2})}
\nonumber\\
&+(N+1)\frac{q^{\frac{N}{2(N+1)}(\omega_1+\omega_2)+BN}-q^{1-\frac{N}{2(N+1)}(\omega_1+\omega_2)-BN}}
{(1-q^{\omega_1})(1-q^{\omega_2})}
\nonumber\\
&+(N+1)\frac{q^{\frac{N}{2(N+1)}(\omega_1+\omega_2)-BN}-q^{1-\frac{N}{2(N+1)}(\omega_1+\omega_2)+BN}}
{(1-q^{\omega_1})(1-q^{\omega_2})}. 
\end{align}

The supersymmetric zeta function is
\begin{align}
\label{zeta_4dsuNnfN+1cm}
&
\mathfrak{Z}^{\textrm{4d $\mathcal{N}=1$ $SU(N)$ $\Phi_{\alpha}$}}(s,z;\omega_1,\omega_2)
\nonumber\\
&=(N+1)^2 \Biggl[
\zeta_2\left(s,\frac{\omega_1+\omega_2}{N+1}+z;\omega_1,\omega_2\right)
-\zeta_2\left(s,\frac{N(\omega_1+\omega_2)}{N+1}+z;\omega_1,\omega_2\right)
\Biggr]
\nonumber\\
&+(N+1)\Biggl[
\zeta_2\left(s,\frac{N(\omega_1+\omega_2)}{2(N+1)}+BN+z;\omega_1,\omega_2\right)
-\zeta_2\left(s,\frac{(N+2)(\omega_1+\omega_2)}{2(N+1)}-BN+z;\omega_1,\omega_2\right)
\Biggr]
\nonumber\\
&+(N+1)\Biggl[
\zeta_2\left(s,\frac{N(\omega_1+\omega_2)}{2(N+1)}-BN+z;\omega_1,\omega_2\right)
-\zeta_2\left(s,\frac{(N+2)(\omega_1+\omega_2)}{2(N+1)}+BN+z;\omega_1,\omega_2\right)
\Biggr]. 
\end{align}
The residues at simple poles $s=2$ and $1$ are evaluated as
\begin{align}
\label{Res2_4dsuNnfN+1cm}
\mathrm{Res}_{s=2}\mathfrak{Z}^{\textrm{4d $\mathcal{N}=1$ $SU(N)$ $\Phi_{\alpha}$}}(s,z;\omega_1,\omega_2)
&=0, \\
\label{Res1_4dsuNnfN+1cm}
\mathrm{Res}_{s=1}\mathfrak{Z}^{\textrm{4d $\mathcal{N}=1$ $SU(N)$ $\Phi_{\alpha}$}}(s,z;\omega_1,\omega_2)
&=(N^2+1)\frac{\omega_1+\omega_2}{\omega_1\omega_2}. 
\end{align}
For $s=0$ and $z=0$ the supersymmetric zeta value vanishes
\begin{align}
\label{0_4dsuNnfN+1cm}
\mathfrak{Z}^{\textrm{4d $\mathcal{N}=1$ $SU(N)$ $\Phi_{\alpha}$}}(0,0;\omega_1,\omega_2)&=0. 
\end{align}
For $s=-1$ and $z=0$ we find
\begin{align}
&
\label{-1_4dsuNnfN+1cm}
\mathfrak{Z}^{\textrm{4d $\mathcal{N}=1$ $SU(N)$ $\Phi_{\alpha}$}}(-1,0;\omega_1,\omega_2)
\nonumber\\
&=-B^2N^2\frac{\omega_1+\omega_2}{\omega_1\omega_2}
-\frac{N^4-4N^3-N^2+2N+1}{12(N+1)^2}(\omega_1+\omega_2)
+\frac{N^2(2N+1)(\omega_1^3+\omega_2^3)}{12(N+1)^2\omega_1\omega_2}. 
\end{align}
According to the formulae (\ref{c4d_zeta}) and (\ref{a4d_zeta}), 
we find 
\begin{align}
\label{ctrial_4dsuNnfN+1cm}
c_{4d}&=\frac{1}{16}\Biggl[
18N-\left(2+\frac{108B^2}{(\omega_1+\omega_2)^2}\right)N^2-\frac{N(29N+22)+2}{(N+1)^2}
\Biggr], \\
\label{atrial_4dsuNnfN+1cm}
a_{4d}&=\frac{3}{16}\Biggl[
6N-\left(1+\frac{36B^2}{(\omega_1+\omega_2)^2}\right)N^2-\frac{2N(5N+4)+1}{(N+1)^2}
\Biggr]. 
\end{align}
The value $a_{4d}$ is maximum at $B=0$, i.e. the vanishing mixed anomaly coefficient. 
We then arrive at
\begin{align}
\label{c_4dsuNnfN+1cm}
c_{4d}&=-\frac{9N^4}{16(N+1)^2}+\frac{7}{16}N^2-\frac{1}{8}, \\
\label{a_4dsuNnfN+1cm}
a_{4d}&=-\frac{9N^4}{16(N+1)^2}+\frac{3}{8}N^2-\frac{3}{16}. 
\end{align}
For $N=2$ they reduce to the central charges (\ref{c_4dsuN3cm}) and (\ref{a_4dsuN3cm}). 
For $N>2$ they do not obey the universal bounds on $c$ and $a$ for $\mathcal{N}=1$ SCFTs, 
wheres they match the expected anomalies of the 4d $\mathcal{N}=1$ $SU(N)$ SQCD with $2(N+1)$ chirals.\footnote{
For 4d $\mathcal{N}=1$ SQCD with $N_f$ flavors the central charges are 
\begin{align}
c_{4d}&=-\frac{9N^4}{16N_f^2}+\frac{7}{16}N^2-\frac18, \\
a_{4d}&=-\frac{9N^4}{16N_f^2}+\frac{3}{8}N^2-\frac{3}{16}. 
\end{align}
}

The supersymmetric determinant is
\begin{align}
&
\mathfrak{D}^{\textrm{4d $\mathcal{N}=1$ $SU(N)$ $\Phi_{\alpha}$}}(z;\omega_1,\omega_2)
\nonumber\\
&=\frac{
\Gamma_2\left(\frac{\omega_1+\omega_2}{N+1}+z;\omega_1,\omega_2\right)^{(N+1)^2}
\Gamma_2\left(\frac{N(\omega_1+\omega_2)}{2(N+1)}+z;\omega_1,\omega_2\right)^{2(N+1)}
}
{
\Gamma_2\left(\frac{N(\omega_1+\omega_2)}{N+1}+z;\omega_1,\omega_2\right)^{(N+1)^2}
\Gamma_2\left(\frac{(N+2)(\omega_1+\omega_2)}{2(N+1)}+z;\omega_1,\omega_2\right)^{2(N+1)}
}.
\end{align}
It turns out that 
the vacuum exponents with $\omega_1=\omega_2$ $=$ $1$ are finite values for finite $N\ge2$. 
For example, for $N=3$ we find the finite vacuum exponent with $\omega_1=\omega_2$ $=$ $1$ given by
\begin{align}
\mathfrak{D}^{\textrm{4d $\mathcal{N}=1$ $SU(3)$ $\Phi_{\alpha}$}}(0;1,1)
&=\frac{1}{512}e^{-\frac{4G}{\pi}}
=0.000608467\ldots, 
\end{align}
where
\begin{align}
G&=\sum_{k=0}^{\infty}\frac{(-1)^k}{(2k+1)^2}=0.915965\ldots
\end{align}
is Catalan's constant. 
For $N=2,\cdots, 10$ the vacuum exponents with $\omega_1=\omega_2$ $=$ $1$ are numerically computed as follows: 
\begin{align}
\begin{array}{c|c}
N&\mathfrak{D}^{\textrm{4d $\mathcal{N}=1$ $SU(N)$ $\Phi_{\alpha}$}}(0;1,1)\\ \hline 
2&0.0127546\\
3&0.000608467 \\
4&0.0000520141 \\
5&0.0000130731 \\
6&0.0000142974 \\
7&0.0000941049 \\
8&0.00491762 \\
9&2.599 \\
10&17226.5 \\
\end{array}
\end{align}

\subsubsection{4d $\mathcal{N}=1$ $USp(2N)$ with $2N+4$ chirals $\Phi_{\alpha}$}
Another type of higher rank generalizations of the $SU(2)$ gauge theory with $6$ chiral multiplets 
is the $\mathcal{N}=1$ $USp(2N)$ gauge theory with $2N+4$ fundamental chiral multiplets $\Phi_{\alpha}$. 
It is dual to the free theory of $(N+2)(2N+3)$ chiral multiplets $\Phi'$ with a superpotential 
\begin{align}
\mathcal{W}&=\mathrm{Pf}\Phi'. 
\end{align}
The duality belongs to the Intriligator-Pouliot duality \cite{Intriligator:1995ne}. 
The field content is summarized as 
\begin{align}
\begin{array}{c|ccc}
&USp(2N)&SU(2N+4)&U(1)_R\\ \hline 
\Phi_{\alpha}&\mathbf{2N}&\mathbf{2N+4}&\frac{1}{N+2} \\ \hline 
\Phi'&\mathbf{1}&\mathbf{(N+2)(2N+3)}&\frac{2}{N+2} \\
\end{array}
\end{align}

The supersymmetric index of the $USp(2N)$ gauge theory is 
\begin{align}
\label{ind_4dusp2Nnf2N+4cm}
&
\mathcal{I}^{\textrm{4d $\mathcal{N}=1$ $USp(2N)$ $\Phi_{\alpha}$}}(x_{\alpha};p;q)
\nonumber\\
&=\frac{(p;p)_{\infty}^N(q;q)_{\infty}^{N}}{2^N N!}
\oint \prod_{i=1}^N \frac{ds_i}{2\pi is_i}
\frac{1}{\Gamma(s_i^{\pm 2};p,q)}
\prod_{i<j}\frac{1}{\Gamma(s_i^{\pm}s_j^{\mp};p,q)\Gamma(s_i^{\pm}s_j^{\pm};p,q)}
\nonumber\\
&\times 
\prod_{i=1}^N\prod_{\alpha=1}^{2N+4}
\Gamma((pq)^{\frac{1}{2(N+2)}}s_i^{\pm}x_{\alpha};p,q). 
\end{align}
It agrees with
\begin{align}
\prod_{\alpha<\beta}^{2N+4}
\Gamma((pq)^{\frac{1}{N+2}}x_{\alpha}x_{\beta};p,q), 
\end{align}
which is the index of the dual free theory. 
The single-particle index is
\begin{align}
\label{sind_4dusp2Nnf2N+4cm}
&
i^{\textrm{4d $\mathcal{N}=1$ $USp(2N)$ $\Phi_{\alpha}$}}(x_{\alpha};p;q)
\nonumber\\
&=\sum_{\alpha<\beta}^{2N+4}
\frac{(pq)^{\frac{1}{N+2}}x_{\alpha}x_{\beta}-(pq)^{1-\frac{1}{N+2}}x_{\alpha}^{-1}x_{\beta}^{-1}}{(1-p)(1-q)}. 
\end{align}

The supersymmetric zeta function is
\begin{align}
\label{zeta_4dusp2Nnf2N+4cm}
&
\mathfrak{Z}^{\textrm{4d $\mathcal{N}=1$ $USp(2N)$ $\Phi_{\alpha}$}}(s,z;\omega_1,\omega_2)
\nonumber\\
&=(N+2)(2N+3)
\Biggl[
\zeta_2\left(s,\frac{\omega_1+\omega_2}{N+2}+z;\omega_1,\omega_2\right)
-\zeta_2\left(s,\frac{(N+1)(\omega_1+\omega_2)}{N+2}+z;\omega_1,\omega_2\right)
\Biggr]. 
\end{align}
The residues at simple poles $s=2$ and $1$ are
\begin{align}
\label{Res2_4dusp2Nnf2N+4cm}
\mathrm{Res}_{s=2}\mathfrak{Z}^{\textrm{4d $\mathcal{N}=1$ $USp(2N)$ $\Phi_{\alpha}$}}(s,z;\omega_1,\omega_2)&=0, \\
\label{Res1_4dusp2Nnf2N+4cm}
\mathrm{Res}_{s=1}\mathfrak{Z}^{\textrm{4d $\mathcal{N}=1$ $USp(2N)$ $\Phi_{\alpha}$}}(s,z;\omega_1,\omega_2)&=\frac{N(2N+3)(\omega_1+\omega_2)}{\omega_1\omega_2}. 
\end{align}
The supersymmetric zeta value with $s=0$ and $z=0$ vanishes
\begin{align}
\mathfrak{Z}^{\textrm{4d $\mathcal{N}=1$ $USp(2N)$ $\Phi_{\alpha}$}}(0,0;\omega_1,\omega_2)&=0. 
\end{align}
For $s=-1$ and $z=0$ the supersymmetric zeta value is given by
\begin{align}
\label{-1_4dusp2Nnf2N+4cm}
&
\mathfrak{Z}^{\textrm{4d $\mathcal{N}=1$ $USp(2N)$ $\Phi_{\alpha}$}}(-1,0;\omega_1,\omega_2)
\nonumber\\
&=\frac{N(N+1)(2N+3)}{6(N+2)^2}\frac{\omega_1^3+\omega_2^3}{\omega_1\omega_2}
-\frac{N(2N+3)(N^2-2N-2)}{12(N+2)^2}(\omega_1+\omega_2). 
\end{align}
Substituting (\ref{Res1_4dusp2Nnf2N+4cm}) and (\ref{-1_4dusp2Nnf2N+4cm}) into formulae (\ref{c4d_zeta}) and (\ref{a4d_zeta}), we obtain 
\begin{align}
\label{c_4dusp2Nnf2N+4cm}
c_{4d}&=-\frac{N(2N+3)(N^2-5N-5)}{8(N+2)^2} \\
\label{a_4dusp2Nnf2N+4cm}
a_{4d}&=-\frac{3N(2N^3-N^2-10N-6)}{16(N+2)^2}. 
\end{align}
When $N=1$ they reduce to (\ref{c_4dsuN3cm}) and (\ref{a_4dsuN3cm}). 
For $N>2$ they do not obey the universal bounds on $a$ and $c$, 
while the values (\ref{c_4dusp2Nnf2N+4cm}) and (\ref{a_4dusp2Nnf2N+4cm}) are the contributions to the $c$- and $a$-anomalies. 
As before, the reason they do not obey the universal bound is for $N>3$, the term in the superpotential $\Pf\Phi'$ is not renormalizable, and the true ``IR fixed point" is located at the zero value of the coefficient, leading to an enhanced symmetry.

The supersymmetric determinant is
\begin{align}
\mathfrak{D}^{\textrm{4d $\mathcal{N}=1$ $USp(2N)$ $\Phi_{\alpha}$}}(z;\omega_1,\omega_2)
&=\frac{\Gamma_2\left(\frac{\omega_1+\omega_2}{N+2}+z;\omega_1,\omega_2\right)^{(N+2)(2N+3)}}
{\Gamma_2\left(\frac{(N+1)(\omega_1+\omega_2)}{N+2}+z;\omega_1,\omega_2\right)^{(N+2)(2N+3)}}. 
\end{align}
We find that the vacuum exponents with $\omega_1=\omega_2$ $=$ $1$ are finite values for finite $N$. 
For $N=1,\cdots, 8$ the vacuum exponents are numerically computed as follows: 
\begin{align}
\begin{array}{c|c}
N&\mathfrak{D}^{\textrm{4d $\mathcal{N}=1$ $USp(2N)$ $\Phi_{\alpha}$}}(0;1,1)\\ \hline 
1&0.0127546 \\
2&0.0000610352 \\
3&6.06434\times 10^{-7} \\
4&3.89098\times 10^{-8} \\
5&3.79592\times 10^{-8} \\
6&1.12473\times 10^{-6} \\
7&0.00181271 \\
8&263.197 \\
\end{array}
\end{align}

\subsubsection{4d $\mathcal{N}=2$ hypermultiplet}
The 4d $\mathcal{N}=2$ hypermultiplet can be constructed by a pair of two $\mathcal{N}=1$ chiral multiplets of $r=2/3$. 

The supersymmetric index is given by
\begin{align}
\label{ind_4dN2hyp}
&
\mathcal{I}^{\textrm{4d $\mathcal{N}=2$ hyp}}(y,z;p;q)
\nonumber\\
&=\mathcal{I}^{\textrm{4d $\mathcal{N}=1$ chiral}_{r=\frac23}}(yz;p;q)
\times \mathcal{I}^{\textrm{4d $\mathcal{N}=1$ chiral}_{r=\frac23}}(yz^{-1};p;q)
\nonumber\\
&=\Gamma((pq)^{\frac{1}{3}}yz;p,q)\Gamma((pq)^{\frac{1}{3}}yz^{-1};p,q), 
\end{align}
where $y$ is the fugacity for the $SU(2) \times U(1)$ R-symmetry. 
The single-particle index is
\begin{align}
\label{sind_4dN2hyp}
i^{\textrm{4d $\mathcal{N}=2$ hyp}}(y,z;p;q)
&=\frac{(pq)^{\frac13}y(z+z^{-1})-(pq)^{\frac23}y^{-1}(z+z^{-1})}{(1-p)(1-q)}. 
\end{align}

The supersymmetric zeta function is
\begin{align}
\label{zeta_4dN2hyp}
&
\mathfrak{Z}^{\textrm{4d $\mathcal{N}=2$ hyp}}(s,z;\omega_1,\omega_2)
\nonumber\\
&=2\Biggl[
\zeta_2\left(s,\frac{\omega_1+\omega_2}{3}+z,\omega_1,\omega_2\right)
-\zeta_2\left(s,\frac{2(\omega_1+\omega_2)}{3}+z,\omega_1,\omega_2\right)
\Biggr]. 
\end{align}
The residues at simple poles $s=2$ and $1$ are
\begin{align}
\label{Res2_4dN2hyp}
\mathrm{Res}_{s=2}\mathfrak{Z}^{\textrm{4d $\mathcal{N}=2$ hyp}}(s,z;\omega_1,\omega_2)&=0, \\
\label{Res1_4dN2hyp}
\mathrm{Res}_{s=1}\mathfrak{Z}^{\textrm{4d $\mathcal{N}=2$ hyp}}(s,z;\omega_1,\omega_2)&=\frac23\frac{\omega_1+\omega_2}{\omega_1\omega_2}. 
\end{align}
For $s=0$ and $z=0$ it vanishes
\begin{align}
\label{0_4dN2hyp}
\mathfrak{Z}^{\textrm{4d $\mathcal{N}=2$ hyp}}(0,0;\omega_1,\omega_2)&=0. 
\end{align}
The special value with $s=-1$ and $z=0$ is
\begin{align}
\label{-1_4dN2hyp}
\mathfrak{Z}^{\textrm{4d $\mathcal{N}=2$ hyp}}(-1,0;\omega_1,\omega_2)
&=\frac{1}{54}(\omega_1+\omega_2)+\frac{2}{81}\frac{\omega_1^3+\omega_2^3}{\omega_1\omega_2}. 
\end{align}
From (\ref{Res1_4dN2hyp}) and (\ref{-1_4dN2hyp}) we get
\begin{align}
\label{c_4dN2hyp}
c_{4d}&=\frac{1}{12}, \\
\label{a_4dN2hyp}
a_{4d}&=\frac{1}{24}. 
\end{align}
They are the 4d central charges of the $\mathcal{N}=2$ hypermultiplet. 
The value $a_{4d}/c_{4d}$ $=$ $1/2$ saturates the lower bound for $\mathcal{N}=2$ SCFT \cite{Shapere:2008zf,Hofman:2008ar}. 

The supersymmetric determinant is
\begin{align}
\mathfrak{D}^{\textrm{4d $\mathcal{N}=2$ hyp}}(z;\omega_1,\omega_2)
&=\frac{\Gamma_2\left(\frac{(\omega_1+\omega_2)}{3}+z;\omega_1,\omega_2\right)^2}
{\Gamma_2\left(\frac{2(\omega_1+\omega_2)}{3}+z;\omega_1,\omega_2\right)^2}
\end{align}
The vacuum exponent with $\omega_1=\omega_2$ $=$ $1$ is given by
\begin{align}
\label{vexp_4dN2hyp}
\mathfrak{D}^{\textrm{4d $\mathcal{N}=2$ hyp}}(0;1,1)
&=\frac{1}{3^{\frac13}}e^{\frac{2\pi^2-3\psi^{(1)}(\frac13)}{9\sqrt{3}\pi}}
=0.559014\ldots
\end{align}

\subsubsection{4d $\mathcal{N}=2$ vector multiplet}
The 4d $\mathcal{N}=2$ $U(1)$ vector multiplet can be realized as a combination of 
the $\mathcal{N}=1$ $U(1)$ vector multiplet and the $\mathcal{N}=1$ chiral multiplet with $r=2/3$. 

The supersymmetric index is given by
\begin{align}
\label{ind_4dN2vec}
&
\mathcal{I}^{\textrm{4d $\mathcal{N}=2$ vec}}(y;p;q)
\nonumber\\
&=\mathcal{I}^{\textrm{4d $\mathcal{N}=1$ vec}}(p;q)\times 
\mathcal{I}^{\textrm{4d $\mathcal{N}=1$ chiral}_{r=\frac23}}(y^{-2};p;q)
\nonumber\\
&=(p;p)_{\infty}(q;q)_{\infty}\Gamma((pq)^{\frac13}y^{-2};p,q), 
\end{align}
where $y$ is the fugacity for the $SU(2) \times U(1)$ R-symmetry.
The single-particle index reads
\begin{align}
i^{\textrm{4d $\mathcal{N}=2$ vec}}(y;p;q)
&=-\frac{p}{1-p}-\frac{q}{1-q}
+\frac{(pq)^{\frac13}y^{-2}-(pq)^{\frac23}y^2}{(1-p)(1-q)}. 
\end{align}

The supersymmetric zeta function is
\begin{align}
\label{zeta_4dN2vec}
&
\mathfrak{Z}^{\textrm{4d $\mathcal{N}=2$ vec}}(s,z;\omega_1,\omega_2)
\nonumber\\
&=-\omega_1^{-s}\zeta\left(s,\frac{\omega_1+z}{\omega_1}\right)
-\omega_2^{-s}\zeta\left(s,\frac{\omega_2+z}{\omega_2}\right)
\nonumber\\
&\quad 
+\zeta_2\left(s,\frac{\omega_1+\omega_2}{3}+z;\omega_1,\omega_2\right)
-\zeta_2\left(s,\frac{2(\omega_1+\omega_2)}{3}+z;\omega_1,\omega_2\right). 
\end{align}
The residues at simple poles $s=2$ and $1$ are
\begin{align}
\label{Res2_4dN2vec}
\mathrm{Res}_{s=2}\mathfrak{Z}^{\textrm{4d $\mathcal{N}=2$ vec}}(s,z;\omega_1,\omega_2)
&=0, \\
\label{Res1_4dN2vec}
\mathrm{Res}_{s=1}\mathfrak{Z}^{\textrm{4d $\mathcal{N}=2$ vec}}(s,z;\omega_1,\omega_2)
&=-\frac23\frac{\omega_1+\omega_2}{\omega_1\omega_2}. 
\end{align}
We have the non-trivial Zeta-index 
\begin{align}
\label{0_4dN2vec}
\mathfrak{Z}^{\textrm{4d $\mathcal{N}=2$ vec}}(0,0;\omega_1,\omega_2)&=1. 
\end{align}
For $s=-1$ and $z=0$ one finds
\begin{align}
\label{-1_4dN2vec}
\mathfrak{Z}^{\textrm{4d $\mathcal{N}=2$ vec}}(-1,0;\omega_1,\omega_2)
&=\frac{5}{54}(\omega_1+\omega_2)+\frac{1}{81}\frac{\omega_1^3+\omega_2^3}{\omega_1\omega_2}. 
\end{align}
We obtain from (\ref{Res1_4dN2vec}) and (\ref{-1_4dN2vec})
\begin{align}
\label{c_4dN2vec}
c_{4d}&=\frac16,\\ 
\label{a_4dN2vec}
a_{4d}&=\frac{5}{24}. 
\end{align}
They agree with the 4d central charges of the $\mathcal{N}=2$ $U(1)$ vector multiplet. 
The value $a_{4d}/c_{4d}$ $=$ $5/4$ hits the upper bound for $\mathcal{N}=2$ SCFT \cite{Shapere:2008zf,Hofman:2008ar}. 

The supersymmetric determinant is
\begin{align}
\mathfrak{D}^{\textrm{4d $\mathcal{N}=2$ vec}}(z;\omega_1,\omega_2)
&=
\frac{2\pi 
\omega_1^{\frac12-\frac{\omega_1+z}{\omega_1}}
\omega_2^{\frac12-\frac{\omega_2+z}{\omega_2}}
}{
\Gamma\left(\frac{\omega_1+z}{\omega_1}\right)
\Gamma\left(\frac{\omega_2+z}{\omega_2}\right)
}
\frac{\Gamma_2\left(\frac{(\omega_1+\omega_2)}{3}+z;\omega_1,\omega_2\right)}
{\Gamma_2\left(\frac{2(\omega_1+\omega_2)}{3}+z;\omega_1,\omega_2\right)}
\end{align}
The vacuum exponent with $\omega_1=\omega_2$ $=$ $1$ is
\begin{align}
\label{vexp_4dN2vec}
\mathfrak{D}^{\textrm{4d $\mathcal{N}=2$ vec}}(z;1,1)
&=\frac{2\pi}{3^{\frac16}}
e^{\frac{2\pi^2-3\psi^{(1)}(\frac13)}{18\sqrt{3}\pi}}=4.69776\ldots
\end{align}

\subsubsection{4d $\mathcal{N}=4$ vector multiplet}
The $\mathcal{N}=4$ $U(1)$ vector multiplet comprises  
the $\mathcal{N}=2$ $U(1)$ vector multiplet and the $\mathcal{N}=2$ hypermultiplet. 

The supersymmetric index reads
\begin{align}
\label{ind_4dN4vec}
&
\mathcal{I}^{\textrm{4d $\mathcal{N}=4$ vec}}(v,w;p;q)
\nonumber\\
&=\mathcal{I}^{\textrm{4d $\mathcal{N}=2$ vec}}\left(y=\sqrt{\frac{v}{w}};p;q\right)\times 
\mathcal{I}^{\textrm{4d $\mathcal{N}=2$ hyp}}\left(y=\sqrt{\frac{v}{w}},z=\sqrt{vw};p;q\right)
\nonumber\\
&=(p;p)_{\infty}(q;q)_{\infty}
\Gamma((pq)^{\frac13}v;p,q)\Gamma((pq)^{\frac13}w^{-1};p,q)\Gamma((pq)^{\frac13}wv^{-1};p,q). 
\end{align}
The single-particle index is 
\begin{align}
\label{sind_4dN4vec}
i^{\textrm{4d $\mathcal{N}=4$ vec}}(v,w;p;q)
&=\frac{(pq)^{\frac13}\left(v+\frac{1}{w}+\frac{w}{v}\right)-(pq)^{\frac23}\left(w+\frac{1}{v}+\frac{v}{w}\right) -p-q+2pq}{(1-p)(1-q)}. 
\end{align}

The supersymmetric zeta function is 
\begin{align}
\label{zeta_4dN4vec}
&
\mathfrak{Z}^{\textrm{4d $\mathcal{N}=4$ vec}}(s,z;\omega_1,\omega_2)
\nonumber\\
&=-\omega_1^{-s}\zeta\left(s,\frac{\omega_1+z}{\omega_1}\right)
-\omega_2^{-s}\zeta\left(s,\frac{\omega_2+z}{\omega_2}\right)
\nonumber\\
&\quad 
+3\Biggl[\zeta_2\left(s,\frac{\omega_1+\omega_2}{3}+z;\omega_1,\omega_2\right)
-\zeta_2\left(s,\frac{2(\omega_1+\omega_2)}{3}+z;\omega_1,\omega_2\right)\Biggr]. 
\end{align}
The residues at simple poles $s=2$ and $1$ vanish
\begin{align}
\label{Res2_4dN4vec}
\mathrm{Res}_{s=2}\mathfrak{Z}^{\textrm{4d $\mathcal{N}=4$ vec}}(s,z;\omega_1,\omega_2)
&=0, \\
\label{Res1_4dN4vec}
\mathrm{Res}_{s=1}\mathfrak{Z}^{\textrm{4d $\mathcal{N}=4$ vec}}(s,z;\omega_1,\omega_2)
&=0. 
\end{align}
The supersymmetric zeta value with $s=0$ and $z=0$ is
\begin{align}
\label{0_4dN4vec}
\mathfrak{Z}^{\textrm{4d $\mathcal{N}=4$ vec}}(0,0;\omega_1,\omega_2)&=1. 
\end{align}
For $s=-1$ and $z=0$ we have 
\begin{align}
\label{-1_4dN4vec}
\mathfrak{Z}^{\textrm{4d $\mathcal{N}=4$ vec}}(-1,0;\omega_1,\omega_2)
&=\frac{1}{9}(\omega_1+\omega_2)+\frac{1}{27}\frac{\omega_1^3+\omega_2^3}{\omega_1\omega_2}. 
\end{align}
Plugging  (\ref{Res1_4dN4vec}) and (\ref{-1_4dN4vec}) into the formulae (\ref{c4d_zeta}) and (\ref{a4d_zeta}), 
we get the central charges
\begin{align}
\label{ca_4dN4vec}
c_{4d}&=a_{4d}=\frac14. 
\end{align}

The supersymmetric determinant is
\begin{align}
\mathfrak{D}^{\textrm{4d $\mathcal{N}=4$ vec}}(z;\omega_1,\omega_2)
&=
\frac{2\pi 
\omega_1^{\frac12-\frac{\omega_1+z}{\omega_1}}
\omega_2^{\frac12-\frac{\omega_2+z}{\omega_2}}
}{
\Gamma\left(\frac{\omega_1+z}{\omega_1}\right)
\Gamma\left(\frac{\omega_2+z}{\omega_2}\right)
}
\frac{\Gamma_2\left(\frac{(\omega_1+\omega_2)}{3}+z;\omega_1,\omega_2\right)^3}
{\Gamma_2\left(\frac{2(\omega_1+\omega_2)}{3}+z;\omega_1,\omega_2\right)^3}
\end{align}
The vacuum exponent with $\omega_1=\omega_2$ $=$ $1$ is
\begin{align}
\label{vexp_4dN4vec}
\mathfrak{D}^{\textrm{4d $\mathcal{N}=4$ vec}}(0;1,1)
&=\frac{2\pi}{\sqrt{3}}
e^{\frac{2\pi^2-3\psi^{(1)}(\frac13)}{6\sqrt{3}\pi}}=2.62611\ldots
\end{align}

\subsection{Zeta functions for Schur and Macdonald indices}

For 4d $\mathcal{N}\ge 2$ supersymmetric field theories, 
one can define special types of supersymmetric indices by taking certain fugacity limits of the supersymmetric indices 
so that only certain short multiplets contribute to the indices \cite{Gadde:2011ik,Gadde:2011uv}. 
The $\mathcal{N}=2$ supersymmetric index is given by
\begin{align}
\label{4dN2_ind_DEF}
\mathcal{I}^{\textrm{4d $\mathcal{N}=2$}}(x;y;p;q)
&=\Tr(-1)^F p^{j_3+\overline{j}_3+\frac13(2l-\frac{r}{2})}
q^{j_3-\overline{j}_3+\frac13(2l-\frac{r}{2})}
y^{2l+r}\prod_{\alpha}x_{\alpha}^{f_{\alpha}}, 
\end{align}
where $l$ and $r$ stand for the Cartan generators of $SU(2)_R$ and $U(1)_r$ R-symmetry. 
These are related to the $\mathcal{N}=1$ R-charge $R$ as $R=\frac23(-2l+\frac{r}{2})$. 
Replacement $y\rightarrow (pq)^{-\frac13}\tilde{q}^{\frac12}$ leads to
\begin{align}
\label{4dN2_ind_DEF2}
\mathcal{I}^{\textrm{4d $\mathcal{N}=2$}}(x;\tilde{q};p;q)
&=\Tr(-1)^F p^{j_3+\overline{j}_3-\frac{r}{2}}
q^{j_3-\overline{j}_3-\frac{r}{2}}
\tilde{q}^{l+\frac{r}{2}}\prod_{\alpha}x_{\alpha}^{f_{\alpha}}. 
\end{align}
If we take the limit $p$ $\rightarrow$ $0$ of the index (\ref{4dN2_ind_DEF2}), we obtain the Macdonald index \cite{Gadde:2011uv}\footnote{The fugacity $t$ in \cite{Gadde:2011uv} corresponds to $\tilde{q}$. } 
\begin{align}
\label{Macdonald_lim}
\mathcal{I}_{\textrm{Mac}}(x;\tilde{q};q)=
\mathcal{I}^{\textrm{4d $\mathcal{N}=2$}}(x;\tilde{q};0;q). 
\end{align}
Furthermore, if we set $\tilde{q}$ to $q$, the Macdonald index reduces to the Schur index 
\begin{align}
\label{Schur_lim}
\mathcal{I}_{\textrm{Schur}}(x;q)&=\mathcal{I}_{\textrm{Mac}}(x;q;q). 
\end{align}
While the Schur index is typically associated with the Higgs branch of the theory, 
it also encodes some properties of the Coulomb branch \cite{Cordova:2015nma,Cordova:2016uwk,Gaiotto:2024ioj,Deb:2025ypl}. 
In \cite{Deb:2025ypl} Deb and Razamat propose that 
the generalized Schur partition function may encode the Schur indices of the theories 
that are obtained from a seed theory by tuning to singular loci of its Coulomb branch. 
The generalized Schur partition function, which we denote by $\mathcal{I}_{\textrm{Schur}_{\alpha}}(x;q)$ is obtained from (\ref{4dN2_ind_DEF2}) by setting 
\begin{align}
\tilde{q}=(qp)^{1+\frac{\alpha}{\log q}\epsilon}
\end{align}
and replacing $p$ with $1-\epsilon$ and then taking $\epsilon\rightarrow 0$. 
Here $\alpha$ is non-negative real value.\footnote{The limit for integer value of $\alpha$ was discussed in \cite{Cecotti:2015lab}. } 
When $\alpha=1$, it reduces to the ordinary Schur index. 

We define the supersymmetric zeta functions $\mathfrak{Z}_{\textrm{Schur}}(s,z)$ (resp. $\mathfrak{Z}_{\textrm{Schur}_{\alpha}}(s,z)$)
for the 4d $\mathcal{N}=2$ supersymmetric field theories by taking the Mellin transform of the plethystic logarithms of the unflavored Schur indices 
(resp the unflavored generalized Schur partition functions). 
Similarly, we define $\mathfrak{Z}_{\textrm{Mac}}(s,z;\omega_1,\omega_2)$ as the Mellin transformed plethystic logarithms of the Macdonald indices  
with $q=q^{\omega_1}$ and $\tilde{q}=q^{\omega_2}$. 
Thus we have $\mathfrak{Z}_{\textrm{Mac}}(s,z;1,1)$ $=$ $\mathfrak{Z}_{\textrm{Schur}}(s,z)$. 
These supersymmetric zeta functions have simple pole at $s=1$. 
In addition, for the Schur index, the generalized Schur partition function and Macdonald index 
we denote the associated supersymmetric determinants 
by $\mathfrak{D}_{\textrm{Schur}}(z)$, $\mathfrak{D}_{\textrm{Schur}_{\alpha}}(z)$ and $\mathfrak{D}_{\textrm{Mac}}(z;\omega_1,\omega_2)$ respectively. 

Again the leading behaviors in the Cardy-like limits of the Schur and Macdonald indices are captured by the supersymmetric zeta functions. 
It follows from (\ref{Cardy_lim0}) and (\ref{Asymptotic_Dege}) that 
\begin{align}
&
\log \mathcal{I}_{\textrm{Mac}}(1;q^{\omega_2};q^{\omega_1})
\nonumber\\
&\sim \frac{\pi^2 \mathrm{Res}_{s=1} \mathfrak{Z}_{\textrm{Mac}}(s,0;\omega_1,\omega_2)}
{6\beta}
-\mathfrak{Z}_{\textrm{Mac}}(0,0;\omega_1,\omega_2)\log\beta
+\log \mathfrak{D}_{\textrm{Mac}}(0;\omega_1,\omega_2), \\
&d_{\textrm{Schur}}(n)
\nonumber\\
&\sim \left(\frac{\mathrm{Res}_{s=1} \mathfrak{Z}_{\textrm{Schur}}(s,0)}{96n^3}\right)^{\frac14}
\left(\frac{6n}{\pi^2 \mathrm{Res}_{s=1}\mathfrak{Z}_{\textrm{Schur}}(s,0)}\right)^{\frac{\mathfrak{Z}_{\textrm{Schur}}(0,0)}{2}}
\nonumber\\
&\quad \times  
\mathfrak{D}_{\textrm{Schur}}(0)\exp\left[
2\pi \sqrt{\frac{\mathrm{Res}_{s=1}\mathfrak{Z}_{\textrm{Schur}}(s,0)}{6}} n^{\frac12}
\right]. 
\end{align}
For the superconformal field theories 
the leading Cardy-like asymptotics of the Macdonald and Schur indices are universally controlled by the anomalies. 
In other words, we have 
\begin{align}
\label{Cardy_Schur}
\mathrm{Res}_{s=1}\mathfrak{Z}_{\textrm{Schur}}(s,0)&=48 (c_{\textrm{4d}}-a_{\textrm{4d}}), \\
\label{Cardy_Mac}
\mathrm{Res}_{s=1}\mathfrak{Z}_{\textrm{Mac}}(s,0;\omega_1,\omega_2)&=\frac{48}{\omega_1} (c_{\textrm{4d}}-a_{\textrm{4d}}). 
\end{align}
The relation (\ref{Cardy_Schur}) follows from the universal formula (\ref{Cardy_lim0}) and 
the leading terms in the Cardy-like limit of the Schur indices, which are discussed in \cite{Buican:2015ina,ArabiArdehali:2015ybk,Cecotti:2015lab,Beem:2017ooy,Xie:2019zlb,Xie:2019vzr,Kaidi:2021tgr,Rastelli:2023sfk}.\footnote{We remark that deviations from the universal Cardy-like growth of the Schur index have been reported in \cite{ArabiArdehali:2023bpq}. } 
On the other hand, we propose the universal relation (\ref{Cardy_Mac}) associated with the Macdonald indices.\footnote{Also see \cite{Choi:2018hmj} for the study of the Cardy-like limits of the Macdonald indices. }

Also we propose that the supersymmetric zeta values with $s=-1$ and $z=0$ are universally given by the anomalies 
\begin{align}
\label{Cas_Schur}
\mathfrak{Z}_{\textrm{Schur}}(-1,0)&=c_{\textrm{4d}},\\
\label{Cas_Mac}
\mathfrak{Z}_{\textrm{Mac}}(-1,0;\omega_1,\omega_2)&=-4\omega_1(c_{4d}-a_{4d})
+12\omega_2(c_{4d}-a_{4d})-\frac{\omega_2^2}{\omega_1}(7c_{4d}-8a_{4d}). 
\end{align}
From (\ref{zeta-1_Cas}) the supersymmetric zeta values (\ref{Cas_Schur}) give rise to the supersymmetric Casimir energies of the Schur indices \cite{Bobev:2015kza}. 
Similarly, the supersymmetric zeta values (\ref{Cas_Mac}) are expected to produce the supersymmetric Casimir energies of the Macdonald indices. 

From (\ref{Cardy_Schur}) and (\ref{Cas_Schur}) 
the 4d central charges of $\mathcal{N}=2$ SCFTs can be expressed in terms of the supersymmetric zeta functions as
\begin{align}
\label{c4dN=2_zeta}
c_{4d}&=\mathfrak{Z}_{\textrm{Schur}}(-1,0), \\
\label{a4dN=2_zeta}
a_{4d}&=\mathfrak{Z}_{\textrm{Schur}}(-1,0)-\frac{\mathrm{Res}_{s=1}\mathfrak{Z}_{\textrm{Schur}}(s,0)}{48}. 
\end{align}

It is demonstrated \cite{Beem:2013sza} that 
there exists the 2d chiral algebra or vertex operator algebra (VOA) of central charge 
\begin{align}
c_{2d}&=-12c_{4d}. 
\end{align}
associated with any 4d $\mathcal{N}=2$ SCFT. 
The unflavored Schur index is identical to the (normalized) vacuum character of the associated VOA
\begin{align}
\chi_{\textrm{vac}}(q)&=\mathcal{I}_{\textrm{Schur}}(q). 
\end{align}
If the vacuum character is annihilated by a linear modular differential equation (LMDE), 
there exists a vector-valued modular form whose components $\chi_i(\tau)$ obey
\begin{align}
\chi_{i}(\tau)&=\sum_{j}\mathcal{S}_{ij}\chi_j(-1/\tau)
\end{align}
with $\chi_0(\tau)$ $=$ $\chi_{\textrm{vac}}(q=e^{2\pi i\tau})$. 
When the components $\chi_i(\tau)$ are expanded with respect to the nome $q=e^{2\pi i\tau}$, one can find the smallest exponent $h_*$ among the expanded components. 
The exponent $h_*$ is known to encode the leading asymptotic behavior and it is related to the 4d central charges as \cite{Beem:2017ooy}
\begin{align}
a_{4d}&=\frac{1}{48}(24h_*-5c_{2d}). 
\end{align}
From the perspective of the associated VOA, the formulae (\ref{c4dN=2_zeta}) and (\ref{a4dN=2_zeta}) can be rewritten as 
\begin{align}
c_{2d}&=-12 \mathfrak{Z}_{\textrm{Schur}}(-1,0), \\
h_*&=-\frac12  \mathfrak{Z}_{\textrm{Schur}}(-1,0)-\frac{1}{24}\mathrm{Res}_{s=1}\mathfrak{Z}_{\textrm{Schur}}(s,0). 
\end{align}

\subsubsection{4d $\mathcal{N}=2$ hypermultiplet}
The Schur index, the generalized Schur partition function and Macdonald index of the $\mathcal{N}=2$ hypermultiplet are given by
\begin{align}
\label{indSchur_4dN2hyp}
\mathcal{I}_{\textrm{Schur}}^{\textrm{4d $\mathcal{N}=2$ hyp}}(x;q)
&=\frac{1}{(q^{\frac12}x;q)_{\infty}(q^{\frac12}x^{-1};q)_{\infty}}, \\
\label{indSchurP_4dN2hyp}
\mathcal{I}_{\textrm{Schur}_{\alpha}}^{\textrm{4d $\mathcal{N}=2$ hyp}}(x;q)
&=\frac{1}{(q^{\frac12}x;q)_{\infty}^{\alpha}(q^{\frac12}x^{-1};q)_{\infty}^{\alpha}}, \\
\label{indMac_4dN2hyp}
\mathcal{I}_{\textrm{Mac}}^{\textrm{4d $\mathcal{N}=2$ hyp}}(x;\tilde{q};q)
&=\frac{1}{(\tilde{q}^{\frac12}x;q)_{\infty}(\tilde{q}^{\frac12}x^{-1};q)_{\infty}}. 
\end{align}
The single-particle indices are
\begin{align}
\label{sindSchur_4dN2hyp}
i_{\textrm{Schur}}^{\textrm{4d $\mathcal{N}=2$ hyp}}(x;q)
&=\frac{q^{\frac12}(x+x^{-1})}{1-q}, \\
\label{sindSchurP_4dN2hyp}
i_{\textrm{Schur}_{\alpha}}^{\textrm{4d $\mathcal{N}=2$ hyp}}(x;q)
&=\frac{\alpha q^{\frac12}(x+x^{-1})}{1-q}, \\
\label{sindMac_4dN2hyp}
i_{\textrm{Mac}}^{\textrm{4d $\mathcal{N}=2$ hyp}}(x;\tilde{q};q)
&=\frac{\tilde{q}^{\frac12}(x+x^{-1})}{1-q}. 
\end{align}

The supersymmetric zeta functions are
\begin{align}
\label{zetaSchur_4dN2hyp}
\mathfrak{Z}_{\textrm{Schur}}^{\textrm{4d $\mathcal{N}=2$ hyp}}(s,z)
&=2\zeta\left(s,\frac12+z\right), \\
\label{zetaSchurP_4dN2hyp}
\mathfrak{Z}_{\textrm{Schur}_{\alpha}}^{\textrm{4d $\mathcal{N}=2$ hyp}}(s,z)
&=2\alpha \zeta\left(s,\frac12+z\right), \\
\label{zetaMac_4dN2hyp}
\mathfrak{Z}_{\textrm{Mac}}^{\textrm{4d $\mathcal{N}=2$ hyp}}(s,z;\omega_1,\omega_2)
&=2\omega_1^{-s}\zeta\left(s,\frac{\omega_2}{2\omega_1}+\frac{z}{\omega_1}\right). 
\end{align}
The residues at $s=1$ are
\begin{align}
\label{Res1Schur_4dN2hyp}
\mathrm{Res}_{s=1}\mathfrak{Z}_{\textrm{Schur}}^{\textrm{4d $\mathcal{N}=2$ hyp}}(s,z)&=2, \\
\label{Res1SchurP_4dN2hyp}
\mathrm{Res}_{s=1}\mathfrak{Z}_{\textrm{Schur}_{\alpha}}^{\textrm{4d $\mathcal{N}=2$ hyp}}(s,z)&=2\alpha, \\
\label{Res1Mac_4dN2hyp}
\mathrm{Res}_{s=1}\mathfrak{Z}_{\textrm{Mac}}^{\textrm{4d $\mathcal{N}=2$ hyp}}(s,z)&=\frac{2}{\omega_1}. 
\end{align}
For $s=0$ and $z=0$ the supersymmetric zeta values vanish
\begin{align}
\label{0Schur_4dN2hyp}
\mathfrak{Z}_{\textrm{Schur}}^{\textrm{4d $\mathcal{N}=2$ hyp}}(0,0)&=0, \\ 
\label{0SchurP_4dN2hyp}
\mathfrak{Z}_{\textrm{Schur}_{\alpha}}^{\textrm{4d $\mathcal{N}=2$ hyp}}(0,0)&=0, \\ 
\label{0Mac_4dN2hyp}
\mathfrak{Z}_{\textrm{Mac}}^{\textrm{4d $\mathcal{N}=2$ hyp}}(0,0)&=1-\frac{\omega_2}{\omega_1}. 
\end{align}
For $s=-1$ and $z=0$ we have
\begin{align}
\label{-1Schur_4dN2hyp}
\mathfrak{Z}_{\textrm{Schur}}^{\textrm{4d $\mathcal{N}=2$ hyp}}(-1,0)&=\frac{1}{12}, \\ 
\label{-1SchurP_4dN2hyp}
\mathfrak{Z}_{\textrm{Schur}_{\alpha}}^{\textrm{4d $\mathcal{N}=2$ hyp}}(-1,0)&=\frac{\alpha}{12}, \\ 
\label{-1Mac_4dN2hyp}
\mathfrak{Z}_{\textrm{Mac}}^{\textrm{4d $\mathcal{N}=2$ hyp}}(-1,0)&=-\frac{\omega_1}{6}+\frac{\omega_2}{2}-\frac{\omega_2^2}{4\omega_1}.  
\end{align}
Plugging (\ref{Res1Schur_4dN2hyp}) and (\ref{-1Schur_4dN2hyp}) into (\ref{c4dN=2_zeta}) and (\ref{a4dN=2_zeta}), 
we reproduce the 4d central charges (\ref{c_4dN2hyp}) and (\ref{a_4dN2hyp}). 
If  the generalized Schur partition function for rational value $\alpha$ coincides with the Schur index for some SCFT, 
it should have the central charges $a_{4d}=\frac{\alpha}{24}$ and $c_{4d}=\frac{\alpha}{12}$ 
so that they hit the lower bound $a_{4d}/c_{4d}$ $=$ $1/2$. 

The supersymmetric determinants are
\begin{align}
\mathfrak{D}_{\textrm{Schur}}^{\textrm{4d $\mathcal{N}=2$ hyp}}(z)
&=\frac{\Gamma\left(\frac12+z\right)^2}{2\pi}, \\
\mathfrak{D}_{\textrm{Schur}_{\alpha}}^{\textrm{4d $\mathcal{N}=2$ hyp}}(z)
&=\frac{\Gamma\left(\frac12+z\right)^{2\alpha}}{(2\pi)^{\alpha}}, \\
\mathfrak{D}_{\textrm{Mac}}^{\textrm{4d $\mathcal{N}=2$ hyp}}(z;\omega_1,\omega_2)
&=\omega_1^{\frac{2z-\omega_1+\omega_2}{\omega_1}}\frac{\Gamma\left(\frac{\omega_2}{2\omega_1}+\frac{z}{\omega_1}\right)^2}{2\pi}. 
\end{align}
The vacuum exponents are 
\begin{align}
\mathfrak{D}_{\textrm{Schur}}^{\textrm{4d $\mathcal{N}=2$ hyp}}(0)&=\frac12, \\
\mathfrak{D}_{\textrm{Schur}_{\alpha}}^{\textrm{4d $\mathcal{N}=2$ hyp}}(0)&=\frac{1}{2^{\alpha}}, \\
\mathfrak{D}_{\textrm{Mac}}^{\textrm{4d $\mathcal{N}=2$ hyp}}(0;\omega_1,\omega_2)
&=\omega_1^{\frac{\omega_2}{\omega_1}-1}
\frac{\Gamma\left(\frac{\omega_2}{2\omega_1}\right)^2}{2\pi}. 
\end{align}

\subsubsection{4d $\mathcal{N}=2$ vector multiplet}
For the 4d $\mathcal{N}=2$ $U(1)$ vector multiplet we have the Schur and Macdonald indices 
\begin{align}
\label{indSchur_4dN2vec}
\mathcal{I}_{\textrm{Schur}}^{\textrm{4d $\mathcal{N}=2$ vec}}(q)&=(q)_{\infty}^2, \\
\label{indMac_4dN2vec}
\mathcal{I}_{\textrm{Mac}}^{\textrm{4d $\mathcal{N}=2$ vec}}(\tilde{q};q)&=(\tilde{q};q)_{\infty}(q;q)_{\infty}. 
\end{align}
The single-particle indices are 
\begin{align}
\label{sindSchur_4dN2vec}
i_{\textrm{Schur}}^{\textrm{4d $\mathcal{N}=2$ vec}}(q)&=-\frac{2q}{1-q}, \\
\label{sindMac_4dN2vec}
i_{\textrm{Mac}}^{\textrm{4d $\mathcal{N}=2$ vec}}(\tilde{q};q)&=-\frac{\tilde{q}+q}{1-q}. 
\end{align}

The supersymmetric zeta functions are
\begin{align}
\label{zetaSchur_4dN2vec}
\mathfrak{Z}_{\textrm{Schur}}^{\textrm{4d $\mathcal{N}=2$ vec}}(s,z)
&=-2\zeta(s,1+z), \\
\label{zetaMac_4dN2vec}
\mathfrak{Z}_{\textrm{Mac}}^{\textrm{4d $\mathcal{N}=2$ vec}}(s,z;\omega_1,\omega_2)
&=-\omega_1^{-s}\zeta\left(s,\frac{\omega_1+z}{\omega_1}\right)
-\omega_1^{-s}\zeta\left(s,\frac{\omega_2+z}{\omega_1}\right). 
\end{align}
The residues at a simple pole $s=1$ are
\begin{align}
\label{Res1Schur_4dN2vec}
\mathrm{Res}_{s=1}\mathfrak{Z}_{\textrm{Schur}}^{\textrm{4d $\mathcal{N}=2$ vec}}(s,z)&=-2, \\
\label{Res1Mac_4dN2vec}
\mathrm{Res}_{s=1}\mathfrak{Z}_{\textrm{Mac}}^{\textrm{4d $\mathcal{N}=2$ vec}}(s,z;\omega_1,\omega_2)&=-\frac{2}{\omega_1}. 
\end{align} 
The supersymmetric zeta values with $s=0$ and $z=0$ are
\begin{align}
\label{0Schur_4dN2vec}
\mathfrak{Z}_{\textrm{Schur}}^{\textrm{4d $\mathcal{N}=2$ vec}}(0,0)&=1, \\
\label{0Mac_4dN2vec}
\mathfrak{Z}_{\textrm{Mac}}^{\textrm{4d $\mathcal{N}=2$ vec}}(0,0;\omega_1,\omega_2)&=\frac{\omega_2}{\omega_1}. 
\end{align}
The supersymmetric zeta values with $s=-1$ and $z=0$ are 
\begin{align}
\label{-1Schur_4dN2vec}
\mathfrak{Z}_{\textrm{Schur}}^{\textrm{4d $\mathcal{N}=2$ vec}}(-1,0)&=\frac16, \\
\label{-1Mac_4dN2vec}
\mathfrak{Z}_{\textrm{Mac}}^{\textrm{4d $\mathcal{N}=2$ vec}}(-1,0;\omega_1,\omega_2)
&=\frac{\omega_1}{6}-\frac{\omega_2}{2}+\frac{\omega_2^2}{2\omega_1}. 
\end{align}
From (\ref{Res1Schur_4dN2vec}) and (\ref{-1Schur_4dN2vec}) as well as the formulae (\ref{c4dN=2_zeta}) and (\ref{a4dN=2_zeta}), 
the central charges (\ref{c_4dN2vec}) and (\ref{a_4dN2vec}) are reproduced. 

The supersymmetric determinants are evaluated as
\begin{align}
\mathfrak{D}_{\textrm{Schur}}^{\textrm{4d $\mathcal{N}=2$ vec}}(z)
&=\frac{2\pi}{\Gamma(1+z)^2}, \\
\mathfrak{D}_{\textrm{Mac}}^{\textrm{4d $\mathcal{N}=2$ vec}}(z;\omega_1,\omega_2)
&=\frac{2\pi \omega_1^{-\frac{2z+\omega_2}{\omega_1}}}
{\Gamma\left(\frac{z+\omega_1}{\omega_1}\right)\Gamma\left(\frac{z+\omega_2}{\omega_1}\right)}. 
\end{align}
The vacuum exponents are 
\begin{align}
\mathfrak{D}_{\textrm{Schur}}^{\textrm{4d $\mathcal{N}=2$ vec}}(0)&=2\pi, \\
\mathfrak{D}_{\textrm{Mac}}^{\textrm{4d $\mathcal{N}=2$ vec}}(0;\omega_1,\omega_2)
&=\frac{2\pi \omega_1^{-\frac{\omega_2}{\omega_1}}}{\Gamma\left(\frac{\omega_2}{\omega_1}\right)}. 
\end{align}

\subsubsection{4d $\mathcal{N}=4$ vector multiplet}
The Schur and Macdonald indices for the 4d $\mathcal{N}=4$ $U(1)$ vector multiplet are
\begin{align}
\label{indSchur_4dN4vec}
\mathcal{I}_{\textrm{Schur}}^{\textrm{4d $\mathcal{N}=4$ vec}}(t;q)
&=\mathcal{I}_{\textrm{Schur}}^{\textrm{4d $\mathcal{N}=2$ vec}}(q)\times 
\mathcal{I}_{\textrm{Schur}}^{\textrm{4d $\mathcal{N}=2$ hyp}}(x=t^2;q)
\nonumber\\
&=\frac{(q)_{\infty}^2}{(q^{\frac12}t^2;q)_{\infty}(q^{\frac12}t^{-2};q)_{\infty}}, \\
\label{indMac_4dN4vec}
\mathcal{I}_{\textrm{Mac}}^{\textrm{4d $\mathcal{N}=4$ vec}}(t;\tilde{q};q)
&=\mathcal{I}_{\textrm{Mac}}^{\textrm{4d $\mathcal{N}=2$ vec}}(\tilde{q};q)\times 
\mathcal{I}_{\textrm{Mac}}^{\textrm{4d $\mathcal{N}=2$ hyp}}(x=t^2;\tilde{q};q)
\nonumber\\
&=\frac{(\tilde{q};q)_{\infty}(q;q)_{\infty}}
{(\tilde{q}^{\frac12}t^2;q)_{\infty}(\tilde{q}^{\frac12}t^{-2};q)_{\infty}}. 
\end{align}
The single-particle indices read
\begin{align}
\label{sindSchur_4dN4vec}
i_{\textrm{Schur}}^{\textrm{4d $\mathcal{N}=4$ vec}}(t;q)
&=\frac{-2q+q^{\frac12}(t^2+t^{-2})}{1-q}, \\
\label{sindMac_4dN4vec}
i_{\textrm{Mac}}^{\textrm{4d $\mathcal{N}=4$ vec}}(t;\tilde{q};q)
&=\frac{-\tilde{q}-q+\tilde{q}^{\frac12}(t^2+t^{-2})}{1-q}. 
\end{align}

The supersymmetric zeta functions are 
\begin{align}
\label{zetaSchur_4dN4vec}
\mathfrak{Z}_{\textrm{Schur}}^{\textrm{4d $\mathcal{N}=4$ vec}}(s,z)
&=-2\zeta(s,1+z)+2\zeta\left(s,\frac12+z\right), \\
\label{zetaMac_4dN4vec}
\mathfrak{Z}_{\textrm{Mac}}^{\textrm{4d $\mathcal{N}=4$ vec}}(s,z;\omega_1,\omega_2)
&=-\omega_1^{-s}\zeta\left(s,\frac{\omega_1+z}{\omega_1}\right)
-\omega_1^{-s}\zeta\left(s,\frac{\omega_2+z}{\omega_1}\right)
\nonumber\\
&\quad +2\omega_1^{-s}\zeta\left(s,\frac{\omega_2}{2\omega_1}+\frac{z}{\omega_1}\right). 
\end{align}
The residues at $s=1$ vanish
\begin{align}
\label{Res1Schur_4dN4vec}
\mathrm{Res}_{s=1}\mathfrak{Z}_{\textrm{Schur}}^{\textrm{4d $\mathcal{N}=4$ vec}}(s,z)
&=0, \\
\label{Res1Mac_4dN4vec}
\mathrm{Res}_{s=1}\mathfrak{Z}_{\textrm{Mac}}^{\textrm{4d $\mathcal{N}=4$ vec}}(s,z;\omega_1,\omega_2)
&=0. 
\end{align}
For $s=0$ and $z=0$ the supersymmetric zeta values are
\begin{align}
\label{0Schur_4dN4vec}
\mathfrak{Z}_{\textrm{Schur}}^{\textrm{4d $\mathcal{N}=4$ vec}}(0,0)
&=1, \\
\label{0Mac_4dN4vec}
\mathfrak{Z}_{\textrm{Mac}}^{\textrm{4d $\mathcal{N}=4$ vec}}(0,0;\omega_1,\omega_2)
&=1. 
\end{align}
For $s=-1$ and $z=0$ we have
\begin{align}
\label{-1Schur_4dN4vec}
\mathfrak{Z}_{\textrm{Schur}}^{\textrm{4d $\mathcal{N}=4$ vec}}(-1,0)
&=\frac14, \\
\label{-1Mac_4dN4vec}
\mathfrak{Z}_{\textrm{Mac}}^{\textrm{4d $\mathcal{N}=4$ vec}}(-1,0;\omega_1,\omega_2)
&=\frac{\omega_2^2}{4\omega_1}. 
\end{align}
Again we obtain from (\ref{Res1Schur_4dN4vec}) and (\ref{-1Schur_4dN4vec}) the central charges (\ref{ca_4dN4vec}). 

The supersymmetric determinants are 
\begin{align}
\mathfrak{D}_{\textrm{Schur}}^{\textrm{4d $\mathcal{N}=4$ vec}}(z)
&=\frac{\Gamma\left(\frac12+z\right)^2}{\Gamma(1+z)^2}, \\
\mathfrak{D}_{\textrm{Mac}}^{\textrm{4d $\mathcal{N}=4$ vec}}(z;\omega_1,\omega_2)
&=\omega_1^{-1} \frac{\Gamma\left(\frac{\omega_2}{2\omega_1}+\frac{z}{\omega_1}\right)^2}
{\Gamma\left(\frac{z+\omega_1}{\omega_1}\right)\Gamma\left(\frac{z+\omega_2}{\omega_1}\right)}. 
\end{align}
The vacuum exponents are 
\begin{align}
\mathfrak{D}_{\textrm{Schur}}^{\textrm{4d $\mathcal{N}=4$ vec}}(0)
&=\pi, \\
\mathfrak{D}_{\textrm{Mac}}^{\textrm{4d $\mathcal{N}=4$ vec}}(0;\omega_1,\omega_2)
&=\omega_1^{-1} \frac{\Gamma\left(\frac{\omega_2}{2\omega_1}\right)^2}
{\Gamma\left(\frac{\omega_2}{\omega_1}\right)}. 
\end{align}

\subsubsection{4d $\mathcal{N}=2$ pure SYM theories}
Consider the 4d $\mathcal{N}=2$ pure SYM theory of gauge group $G$. 
The Schur indices can be evaluated from the matrix integral of the form 
\begin{align}
\label{indSchur_suN}
\mathcal{I}_{\textrm{Schur}}^{\textrm{4d $\mathcal{N}=2$ $G$}}(q)
&=\frac{1}{|\mathrm{Weyl}(G)|}(q)_{\infty}^{2\mathrm{rank}(G)}
\oint \prod_{\alpha\in \mathrm{root}(G)} \frac{ds}{2\pi is}
(s^{\alpha};q)(qs^{\alpha};q)_{\infty}, 
\end{align}
where $|\mathrm{Weyl(G)}|$ is the order of the Weyl group of $G$. 
They are evaluated as \cite{Okazaki:2024kzo,Lu:2025ecb}
\begin{align}
\label{indSchur_SUN}
\mathcal{I}_{\textrm{Schur}}^{\textrm{4d $\mathcal{N}=2$ $SU(N)$}}(q)
&=\frac{(q^{2N};q^{2N})_{\infty}^{N}}{(q^2;q^2)_{\infty}}, \\
\label{indSchur_SO2N+1}
\mathcal{I}_{\textrm{Schur}}^{\textrm{4d $\mathcal{N}=2$ $SO(2N+1)$}}(q)
&=\frac{(q^{4N-2};q^{4N-2})_{\infty}^{N-1}(q^{8N-4};q^{8N-4})_{\infty}(q^4;q^4)_{\infty}}
{(q^2;q^2)_{\infty}},\\
\label{indSchur_USp2N}
\mathcal{I}_{\textrm{Schur}}^{\textrm{4d $\mathcal{N}=2$ $USp(2N)$}}(q)
&=\frac{(q^{2N+2};q^{2N+2})_{\infty}(q^{4N+4};q^{4N+4})_{\infty}^{N-1}(q^4;q^4)_{\infty}}
{(q^2;q^2)_{\infty}},\\
\label{indSchur_SO2N}
\mathcal{I}_{\textrm{Schur}}^{\textrm{4d $\mathcal{N}=2$ $SO(2N)$}}(q)
&=\frac{(q^{4N-4};q^{4N-4})_{\infty}^{N+1}(q^4;q^4)_{\infty}}
{(q^{2N-2};q^{2N-2})_{\infty}(q^2;q^2)_{\infty}},\\
\label{indSchur_G2}
\mathcal{I}_{\textrm{Schur}}^{\textrm{4d $\mathcal{N}=2$ $G_2$}}(q)
&=\frac{(q^4;q^4)_{\infty}(q^6;q^6)_{\infty}(q^8;q^8)_{\infty}(q^{24};q^{24})_{\infty}}
{(q^2;q^2)_{\infty}(q^{12};q^{12})_{\infty}}, \\
\label{indSchur_F4}
\mathcal{I}_{\textrm{Schur}}^{\textrm{4d $\mathcal{N}=2$ $F_4$}}(q)
&=\frac{(q^4;q^4)_{\infty}(q^6;q^6)_{\infty}(q^{18};q^{18})_{\infty}^2(q^{36};q^{36})_{\infty}^2}
{(q^2;q^2)_{\infty}(q^{12};q^{12})_{\infty}}, \\
\label{indSchur_E6}
\mathcal{I}_{\textrm{Schur}}^{\textrm{4d $\mathcal{N}=2$ $E_6$}}(q)
&=\frac{(q^4;q^4)_{\infty}(q^6;q^6)_{\infty}(q^{24};q^{24})_{\infty}^7}
{(q^2;q^2)_{\infty}(q^8;q^8)_{\infty}(q^{12};q^{12})_{\infty}}, \\
\label{indSchur_E7}
\mathcal{I}_{\textrm{Schur}}^{\textrm{4d $\mathcal{N}=2$ $E_7$}}(q)
&=\frac{(q^4;q^4)_{\infty}(q^6;q^6)_{\infty}(q^{36};q^{36})_{\infty}^8}
{(q^2;q^2)_{\infty}(q^{12};q^{12})_{\infty}(q^{18};q^{18})_{\infty}}, \\
\label{indSchur_E8}
\mathcal{I}_{\textrm{Schur}}^{\textrm{4d $\mathcal{N}=2$ $E_8$}}(q)
&=\frac{(q^4;q^4)_{\infty}(q^6;q^6)_{\infty}(q^{10};q^{10})_{\infty}(q^{60};q^{60})_{\infty}^9}
{(q^2;q^2)_{\infty}(q^{12};q^{12})_{\infty}(q^{20};q^{20})_{\infty}(q^{30};q^{30})_{\infty}}. 
\end{align}
The single-particle indices are
\begin{align}
\label{sindSchur_SUN}
i_{\textrm{Schur}}^{\textrm{4d $\mathcal{N}=2$ $SU(N)$}}(q)
&=-\frac{Nq^{2N}}{1-q^{2N}}+\frac{q^2}{1-q^2}, \\
\label{sindSchur_SO2N+1}
i_{\textrm{Schur}}^{\textrm{4d $\mathcal{N}=2$ $SO(2N+1)$}}(q)
&=-\frac{(N-1)q^{4N-2}}{1-q^{4N-2}}-\frac{q^{8N-4}}{1-q^{8N-4}}-\frac{q^4}{1-q^4}+\frac{q^2}{1-q^2}, \\
\label{sindSchur_USp2N}
i_{\textrm{Schur}}^{\textrm{4d $\mathcal{N}=2$ $USp(2N)$}}(q)
&=-\frac{q^{2N+2}}{1-q^{2N+2}}-\frac{(N-1)q^{4N+4}}{1-q^{4N+4}}-\frac{q^4}{1-q^4}+\frac{q^2}{1-q^2}, \\
\label{sindSchur_SO2N}
i_{\textrm{Schur}}^{\textrm{4d $\mathcal{N}=2$ $SO(2N)$}}(q)
&=-\frac{(N-1)q^{4N-4}}{1-q^{4N-4}}-\frac{q^4}{1-q^4}+\frac{q^{2N-2}}{1-q^{2N-2}}+\frac{q^2}{1-q^2}, \\
\label{sindSchur_G2}
i_{\textrm{Schur}}^{\textrm{4d $\mathcal{N}=2$ $G_2$}}(q)
&=-\frac{q^4}{1-q^4}-\frac{q^6}{1-q^6}-\frac{q^8}{1-q^8}-\frac{q^{24}}{1-q^{24}}
+\frac{q^2}{1-q^2}+\frac{q^{12}}{1-q^{12}}, \\
\label{sindSchur_F4}
i_{\textrm{Schur}}^{\textrm{4d $\mathcal{N}=2$ $F_4$}}(q)
&=-\frac{q^4}{1-q^4}-\frac{q^6}{1-q^6}-\frac{2q^{18}}{1-q^{18}}-\frac{2q^{36}}{1-q^{36}}
+\frac{q^2}{1-q^2}+\frac{q^{12}}{1-q^{12}}, \\
\label{sindSchur_E6}
i_{\textrm{Schur}}^{\textrm{4d $\mathcal{N}=2$ $E_6$}}(q)
&=-\frac{q^4}{1-q^4}-\frac{q^6}{1-q^6}-7\frac{q^{24}}{1-q^{24}}
+\frac{q^2}{1-q^2}+\frac{q^8}{1-q^8}+\frac{q^{12}}{1-q^{12}}, \\
\label{sindSchur_E7}
i_{\textrm{Schur}}^{\textrm{4d $\mathcal{N}=2$ $E_7$}}(q)
&=-\frac{q^4}{1-q^4}-\frac{q^6}{1-q^6}-8\frac{q^{36}}{1-q^{36}}
+\frac{q^2}{1-q^2}+\frac{q^{12}}{1-q^{12}}+\frac{q^{18}}{1-q^{18}}, \\
\label{sindSchur_E8}
i_{\textrm{Schur}}^{\textrm{4d $\mathcal{N}=2$ $E_8$}}(q)
&=-\frac{q^4}{1-q^4}-\frac{q^6}{1-q^6}-\frac{q^{10}}{1-q^{10}}-9\frac{q^{60}}{1-q^{60}}
\nonumber\\
&+\frac{q^2}{1-q^2}+\frac{q^{12}}{1-q^{12}}+\frac{q^{20}}{1-q^{20}}+\frac{q^{30}}{1-q^{30}}. 
\end{align}

We obtain the supersymmetric zeta functions 
\begin{align}
\label{zetaSchur_SUN}
\mathfrak{Z}_{\textrm{Schur}}^{\textrm{4d $\mathcal{N}=2$ $SU(N)$}}(s,z)
&=-2^{-s} N^{1-s}\zeta\left(s,\frac{2N+z}{2N}\right)+2^{-s}\zeta\left(s,\frac{2+z}{2}\right), \\
\label{zetaSchur_SO2N+1}
\mathfrak{Z}_{\textrm{Schur}}^{\textrm{4d $\mathcal{N}=2$ $SO(2N+1)$}}(s,z)
&=-(N-1)(4N-2)^{-s}\zeta\left(s,\frac{4N-2+z}{4N-2}\right)
-4^{-s}\zeta\left(s,\frac{4+z}{4}\right)
\nonumber\\
&-(8N-4)^{-s}\zeta\left(s,\frac{8N-4+z}{8N-4}\right)
+2^{-s}\zeta\left(s,\frac{2+z}{2}\right), \\
\label{zetaSchur_USp2N}
\mathfrak{Z}_{\textrm{Schur}}^{\textrm{4d $\mathcal{N}=2$ $USp(2N)$}}(s,z)
&=
-(2N+2)^{-s}\zeta\left(s,\frac{2N+2+z}{2N+2}\right)
-4^{-s}\zeta\left(s,\frac{4+z}{4}\right)
\nonumber\\
&-(N-1)(4N+4)^{-s}\zeta\left(s,\frac{4N+4+z}{4N+4}\right)
+2^{-s}\zeta\left(s,\frac{2+z}{2}\right), \\
\label{zetaSchur_SO2N}
\mathfrak{Z}_{\textrm{Schur}}^{\textrm{4d $\mathcal{N}=2$ $SO(2N)$}}(s,z)
&=-(N+1)(4N-4)^{-s}\zeta\left(s,\frac{4N-4+z}{4N-4}\right)
-4^{-s}\zeta\left(s,\frac{4+z}{4}\right)
\nonumber\\
&+(2N-2)^{-s}\zeta\left(s,\frac{2N-2+z}{2N-2}\right)
+2^{-s}\zeta\left(s,\frac{2+z}{2}\right), 
\end{align}
\begin{align}
\label{zetaSchur_G2}
\mathfrak{Z}_{\textrm{Schur}}^{\textrm{4d $\mathcal{N}=2$ $G_2$}}(s,z)
&=-4^{-s}\zeta\left(s,\frac{4+z}{4}\right)
-6^{-s}\zeta\left(s,\frac{6+z}{6}\right)
-8^{-s}\zeta\left(s,\frac{8+z}{8}\right)
\nonumber\\
&-24^{-s}\zeta\left(s,\frac{24+z}{24}\right)
+2^{-s}\zeta\left(s,\frac{2+z}{2}\right)
+12^{-s}\zeta\left(s,\frac{12+z}{12}\right), \\
\label{zetaSchur_F4}
\mathfrak{Z}_{\textrm{Schur}}^{\textrm{4d $\mathcal{N}=2$ $F_4$}}(s,z)
&=-2\cdot (36)^{-s}\zeta\left(s,\frac{36+z}{36}\right)
-2\cdot(18)^{-s}\zeta\left(s,\frac{18+z}{18}\right)
\nonumber\\
&
-6^{-s}\zeta\left(s,\frac{6+z}{6}\right)
-4^{-s}\zeta\left(s,\frac{4+z}{4}\right)
\nonumber\\
&+12^{-s}\zeta\left(s,\frac{12+z}{12}\right)
+2^{-s}\zeta\left(s,\frac{2+z}{2}\right), 
\end{align}
\begin{align}
\label{zetaSchur_E6}
\mathfrak{Z}_{\textrm{Schur}}^{\textrm{4d $\mathcal{N}=2$ $E_6$}}(s,z)
&=-7\cdot (24)^{-s}\zeta\left(s,\frac{24+z}{24}\right)
-6^{-s}\zeta\left(s,\frac{6+z}{6}\right)
-4^{-s}\zeta\left(s,\frac{4+z}{4}\right)
\nonumber\\
&+2^{-s}\zeta\left(s,\frac{2+z}{2}\right)
+8^{-s}\zeta\left(s,\frac{8+z}{8}\right)
+12^{-s}\zeta\left(s,\frac{12+z}{12}\right), \\
\label{zetaSchur_E7}
\mathfrak{Z}_{\textrm{Schur}}^{\textrm{4d $\mathcal{N}=2$ $E_7$}}(s,z)
&=-8\cdot (36)^{-s}\zeta\left(s,\frac{36+z}{36}\right)
-6^{-s}\zeta\left(s,\frac{6+z}{6}\right)
-4^{-s}\zeta\left(s,\frac{4+z}{4}\right)
\nonumber\\
&+2^{-s}\zeta\left(s,\frac{2+z}{2}\right)
+12^{-s}\zeta\left(s,\frac{12+z}{12}\right)
+18^{-s}\zeta\left(s,\frac{18+z}{18}\right), \\
\label{zetaSchur_E8}
\mathfrak{Z}_{\textrm{Schur}}^{\textrm{4d $\mathcal{N}=2$ $E_8$}}(s,z)
&=-9\cdot (60)^{-s}\zeta\left(s,\frac{60+z}{60}\right)
-10^{-s}\zeta\left(s,\frac{10+z}{10}\right)
-6^{-s}\zeta\left(s,\frac{6+z}{6}\right)
\nonumber\\
&
-4^{-s}\zeta\left(s,\frac{4+z}{4}\right)
+2^{-s}\zeta\left(s,\frac{2+z}{2}\right)
+12^{-s}\zeta\left(s,\frac{12+z}{12}\right)
\nonumber\\
&
+20^{-s}\zeta\left(s,\frac{20+z}{20}\right)
+30^{-s}\zeta\left(s,\frac{30+z}{30}\right). 
\end{align}
Though each term in the supersymmetric zeta functions has a simple pole at $s=1$, 
the residue vanishes
\begin{align}
\label{Res1_G}
\mathrm{Res}_{s=1}\mathfrak{Z}_{\textrm{Schur}}^{\textrm{4d $\mathcal{N}=2$ $G$}}(s,z)
&=0. 
\end{align}
The supersymmetric zeta value with $s=0$ and $z=0$ is given by
\begin{align}
\label{0Schur_G}
\mathfrak{Z}_{\textrm{Schur}}^{\textrm{4d $\mathcal{N}=2$ $G$}}(0,0)
&=\frac{\mathrm{rank}(G)}{2}. 
\end{align}
The supersymmetric zeta value with $s=-1$ and $z=0$ is
\begin{align}
\label{-1Schur_G}
\mathfrak{Z}_{\textrm{Schur}}^{\textrm{4d $\mathcal{N}=2$ $G$}}(-1,0)
&=\frac{\dim G}{6}. 
\end{align}
According to the formulae (\ref{c4dN=2_zeta}), we get
\begin{align}
\label{caSchur_G}
c_{4d}&=\frac{\dim G}{6}. 
\end{align}
This is the anomaly contribution for the 4d $\mathcal{N}=2$ vector multiplet of gauge group $G$. 
However, the formula (\ref{a4dN=2_zeta}) does not simply give rise to the anomaly contribution $\frac{5}{24}\dim G$ for $a_{4d}$ 
as the theory is not conformal. 

The supersymmetric determinants are 
\begin{align}
\label{detSchur_SUN}
\mathfrak{D}^{\textrm{4d $\mathcal{N}=2$ $SU(N)$}}(z)
&=2^N \pi^{\frac{N-1}{2}} N^{\frac{N-z}{2}}
\frac{\Gamma\left(1+\frac{z}{2}\right)}{z^N \Gamma\left(\frac{z}{2N}\right)^N}, \\
\label{detSchur_SO2N+1}
\mathfrak{D}^{\textrm{4d $\mathcal{N}=2$ $SO(2N+1)$}}(z)
&=\frac{2^{\frac{N+1+z}{2}}(4N-2)^{-\frac{(N-1)(2N-1+z)}{4N-2}}\pi^{\frac{N-1}{2}}\Gamma\left(\frac{2+z}{4}\right)}
{(8N-4)^{\frac{4N-2+z}{8N-4}}\left(\frac{z}{4N-2}\right)^N
\Gamma\left(\frac{z}{8N-4}\right) \Gamma\left(\frac{z}{4N-2}\right)^{N-1}}, \\
\label{detSchur_USp2N}
\mathfrak{D}^{\textrm{4d $\mathcal{N}=2$ $USp(2N)$}}(z)
&=\frac{2^{\frac{N(3N-1)-4+z}{2(N+1)}}\pi^{\frac{N-1}{2}}\Gamma\left(\frac{2+z}{4}\right)}
{(N+1)^{\frac{2N+z}{4}}\left(\frac{z}{N+1}\right)^{N-1}\Gamma\left(1+\frac{z}{2N+2}\right)\Gamma\left(\frac{z}{4+4N}\right)^{N-1}}, \\
\label{detSchur_SO2N}
\mathfrak{D}^{\textrm{4d $\mathcal{N}=2$ $SO(2N)$}}(z)
&=\frac{\pi^{\frac{N-1}{2}}\Gamma\left(\frac{2+z}{4}\right)\Gamma\left(1+\frac{z}{2(N-1)}\right)}
{2^{\frac{N^2+N-2+z}{2(N-1)}} (N-1)^{\frac{2N+z}{4}} \Gamma\left(1+\frac{z}{4(N-1)}\right)^{N+1}}, \\
\label{detSchur_G2}
\mathfrak{D}^{\textrm{4d $\mathcal{N}=2$ $G_2$}}(z)
&=\frac{\pi\Gamma\left(\frac{2+z}{4}\right)}
{2^{2+\frac{z}{6}}3^{\frac12+\frac{z}{8}}
\Gamma\left(1+\frac{z}{24}\right)\Gamma\left(1+\frac{z}{8}\right)\Gamma\left(\frac{6+z}{12}\right)}, \\
\label{detSchur_F4}
\mathfrak{D}^{\textrm{4d $\mathcal{N}=2$ $F_4$}}(z)
&=\frac{2^{5+\frac{z}{9}} 3^{4-\frac{5z}{12}}\pi^2 \Gamma\left(\frac{2+z}{4}\right)}
{z^4 \Gamma\left(\frac{z}{36}\right)^2 \Gamma\left(\frac{z}{18}\right)^2 \Gamma\left(\frac{6+z}{12}\right)}, \\
\label{detSchur_E6}
\mathfrak{D}^{\textrm{4d $\mathcal{N}=2$ $E_6$}}(z)
&=\frac{\pi^3 \Gamma\left(1+\frac{z}{8}\right)\Gamma\left(\frac{2+z}{4}\right)}
{2^{6+\frac{z}{6}} 3^{\frac72+\frac{3z}{8}} \Gamma\left(1+\frac{z}{24}\right)^7 \Gamma\left(\frac{6+z}{12}\right)}, \\
\label{detSchur_E7}
\mathfrak{D}^{\textrm{4d $\mathcal{N}=2$ $E_7$}}(z)
&=\frac{\pi^3 \Gamma\left(\frac{2+z}{4}\right)\Gamma\left(\frac{18+z}{36}\right)}
{16\cdot 3^{7+\frac{5z}{12}}\Gamma\left(1+\frac{z}{36}\right)^7 \Gamma\left(\frac{6+z}{12}\right)}, \\
\label{detSchur_E8}
\mathfrak{D}^{\textrm{4d $\mathcal{N}=2$ $E_8$}}(z)
&=\frac{\pi^4 \Gamma\left(\frac{2+z}{4}\right) \Gamma\left(\frac{30+z}{60}\right)}
{16\cdot 3^{4+\frac{z}{5}}5^{4+\frac{z}{6}} \Gamma\left(1+\frac{z}{60}\right)^8 \Gamma\left(\frac{6+z}{12}\right) \Gamma\left(\frac{10+z}{20}\right)}. 
\end{align}
The vacuum exponents are evaluated as
\begin{align}
\label{vacSchur_SUN}
\mathfrak{D}^{\textrm{4d $\mathcal{N}=2$ $SU(N)$}}(0)
&=\frac{\pi^{\frac{N-1}{2}}}{N^{\frac{N}{2}}}, \\
\label{vacSchur_SO2N+1}
\mathfrak{D}^{\textrm{4d $\mathcal{N}=2$ $SO(2N+1)$}}(0)
&=\frac{\pi^{\frac{N}{2}}}{2(2N-1)^{\frac{N}{2}}}, \\
\label{vacSchur_USp2N}
\mathfrak{D}^{\textrm{4d $\mathcal{N}=2$ $USp(2N)$}}(0)
&=\frac{\pi^{\frac{N}{2}}}{2^{\frac{N}{2}} (N+1)^{\frac{N}{2}}}, \\
\label{vacSchur_SO2N}
\mathfrak{D}^{\textrm{4d $\mathcal{N}=2$ $SO(2N)$}}(0)
&=\frac{\pi^{\frac{N}{2}}}{2^{\frac{N+2}{2}}(N-1)^{\frac{N}{2}}}, \\
\mathfrak{D}^{\textrm{4d $\mathcal{N}=2$ $G_2$}}(0)
&=\frac{\pi}{4\sqrt{3}}, \\
\mathfrak{D}^{\textrm{4d $\mathcal{N}=2$ $F_4$}}(0)
&=\frac{\pi^2}{162}, \\
\mathfrak{D}^{\textrm{4d $\mathcal{N}=2$ $E_6$}}(0)
&=\frac{\pi^3}{1728\sqrt{3}}, \\
\mathfrak{D}^{\textrm{4d $\mathcal{N}=2$ $E_7$}}(0)
&=\frac{\pi^{\frac72}}{34992}, \\
\mathfrak{D}^{\textrm{4d $\mathcal{N}=2$ $E_8$}}(0)
&=\frac{\pi^4}{810000}. 
\end{align}

\subsubsection{$J^{h^{\vee}}[k]$ AD theories}
\label{sec_AD}
Argyres-Douglas (AD) theories \cite{Argyres:1995jj,Argyres:1995xn} were originally found as the IR fixed points 
at special singular points on the Coulomb branches of 4d $\mathcal{N}=2$ gauge theories. 
They typically have no Lagrangian description constructed by the $\mathcal{N}=2$ supermultiplets 
and contain local operators with fractional scaling dimensions. 
The AD theories can be engineered from 6d $\mathcal{N}=(2,0)$ theory of type $J$ $=$ $\{A_n, D_n, E_n\}$ on a Riemann surface $\Sigma$ \cite{Xie:2012hs}. 
The Coulomb branch of the theory is described by a Hitchin system on $\Sigma$, which involves the Higgs field $\Phi$ \cite{Gaiotto:2009hg}. 
The irregular singularity is associated with an irregular solution to Hitchin's equation with the singular form 
\begin{align}
\Phi&=\frac{\mathfrak{t}}{z^{2+k/b}}+\cdots, 
\end{align}
where $\mathfrak{t}$ is a regular semi-simple element of the Lie algebra $J$. 
The possible values of $b$ are listed in \cite{Wang:2015mra}. 
The resulting AD theories are labeled by the irregular singularities $J^{b}[k]$. 

It is conjectured  \cite{Song:2017oew} 
that the single-particle Schur index of the $J^{h^{\vee}}[k]$ AD theory is given by\footnote{Also see \cite{Cordova:2015nma,Song:2015wta} for earlier conjectures with $J=A_{N-1}$. }
\begin{align}
\label{sind_J^{h}[k]}
i_{\textrm{Schur}}^{J^{h^{\vee}}[k]}(q)
&=\sum_{i=1}^{\mathrm{rank}(J)}
\frac{q^{d_i}}{1-q}
-\frac{q^{h^{\vee}+k}}{1-q^{h^{\vee}+k}}
\sum_{i=1}^{\mathrm{rank}(J)}
\sum_{j=0}^{2d_i-2}
q^{1-d_i+j}, 
\end{align}
where $\mathrm{rank}(J)$ is the rank of $J$, $h^{\vee}$ is the dual Coxeter number of $J$, 
and $d_i$ are degrees of the Casimirs of $J$. 

The supersymmetric zeta function is evaluated as
\begin{align}
\label{zeta_J^{h}[k]}
\mathfrak{Z}_{\textrm{Schur}}^{J^{h^{\vee}}[k]}(s,z)
&=\sum_{i=1}^{\mathrm{rank}(J)}\zeta(s,d_i+z)
-\sum_{i=1}^{\mathrm{rank}(J)}\sum_{j=0}^{2d_i-2}(h^{\vee}+k)^{-s}
\zeta\left(s,\frac{h^{\vee}+k+1-d_i+j+z}{h^{\vee}+k}\right). 
\end{align}
It has a simple pole at $s=1$ with residue
\begin{align}
\label{Res1_J^{h}[k]}
\mathrm{Res}_{s=1}\mathfrak{Z}_{\textrm{Schur}}^{J^{h^{\vee}}[k]}(s,z)
&=\mathrm{rank}(J)-\frac{\dim J}{h^{\vee}+k}
\nonumber\\
&=\frac{(k-1)\mathrm{rank}(J)}{k+h^{\vee}}. 
\end{align}
Here we have used the relation (\ref{d_Casimir_relation1}) as well as the relation
\begin{align}
\label{dim_rank_dh_ADE}
\dim J&=\mathrm{rank}(J) (h^{\vee}+1)
\end{align}
for the simply-laced Lie algebra $J$. 
With $s=0$ and $z=0$ the supersymmetric zeta value vanishes
\begin{align}
\label{0_J^{h}[k]}
\mathfrak{Z}_{\textrm{Schur}}^{J^{h^{\vee}}[k]}(0,0)&=0. 
\end{align}
The supersymmetric zeta value for $s=-1$ and $z=0$ is evaluated as
\begin{align}
\label{-1_J^{h}[k]}
\mathfrak{Z}_{\textrm{Schur}}^{J^{h^{\vee}}[k]}(-1,0)
&=\sum_{i=1}^{\mathrm{rank}(J)}
\frac{(h^{\vee}+k-2d_i+1) \left[2d_i(h^{\vee}+k-d_i+1)-h^{\vee}-k\right]}{12(h^{\vee}+k)}
\nonumber\\
&=\frac{k^2 \dim J}{12(h^{\vee}+k)}-\frac{1}{12}\mathrm{rank}(J), 
\end{align}
where we have used the relations
\begin{align}
\frac13 h^{\vee} \dim J&=\sum_{i=1}^{\mathrm{rank}(J)}d_i(d_i-1), \\
\frac43 h^{\vee} \dim J+\mathrm{rank}(J)&=\sum_{i=1}^{\mathrm{rank}(J)}(2d_i-1)^2, \\
\frac12 (h^{\vee})^2 \dim J&=\sum_{i=1}^{\mathrm{rank}}d_i(2d_i-1)(d_i-1), 
\end{align}
as well as (\ref{d_Casimir_relation1}). 
Substituting (\ref{Res1_J^{h}[k]}) and (\ref{-1_J^{h}[k]}) into the formulae (\ref{c4dN=2_zeta}) and (\ref{a4dN=2_zeta}), 
we get 
\begin{align}
\label{c_J^{h}[k]}
c_{4d}&=\frac{k^2 \dim J}{12(h^{\vee}+k)}-\frac{1}{12}\mathrm{rank}(J), \\
\label{a_J^{h}[k]}
a_{4d}&=\frac{(4k^2+1)\dim J}{48(h^{\vee}+k)}-\frac{5}{48}\mathrm{rank}(J). 
\end{align}
The explicit values are summarized as follows: 
\begin{align}
\label{c_J^{h}[k]_ref}
\begin{array}{c|c|c|c|c|c}
J^{h^{\vee}}[k]&\mathrm{rank}(J)&h^{\vee}&d_i&c_{4d}&a_{4d}\\ \hline
\scalebox{0.7}{$A_{N-1}^{N}[k]$}&\scalebox{0.7}{$N-1$}&\scalebox{0.7}{$N$}&\scalebox{0.7}{$i+1$}&\frac{(N-1)(k-1)(N+k+kN)}{12(N+k)}&\frac{(N-1)(k-1)(4kN+4N+4k-1)}{48(N+k)} \\
\scalebox{0.7}{$D_N^{2N-2}[k]$}&\scalebox{0.7}{$N$}&\scalebox{0.7}{$2N-2$}&\scalebox{0.7}{$2,4,\cdots,2(N-1),N$}&\frac{N(k-1)(2kN+2N-k-2)}{12(2N+k-2)}&\frac{N(k-1)(8kN+8N-4k-9)}{48(2N+k-2)} \\
\scalebox{0.7}{$E_{6}^{12}[k]$}&\scalebox{0.7}{$6$}&\scalebox{0.7}{$12$}&\scalebox{0.7}{$2,5,6,8,9,12$}&\frac{(k-1)(12+13k)}{2(12+k)}&\frac{(k-1)(47+52k)}{8(12+k)}\\
\scalebox{0.7}{$E_{7}^{18}[k]$}&\scalebox{0.7}{$7$}&\scalebox{0.7}{$18$}&\scalebox{0.7}{$2,6,8,10,12,14,18$}&\frac{7(k-1)(18+19k)}{12(18+k)}&\frac{7(k-1)(71+76k)}{48(18+k)}\\
\scalebox{0.7}{$E_{8}^{30}[k]$}&\scalebox{0.7}{$8$}&\scalebox{0.7}{$30$}&\scalebox{0.7}{$2,8,12,14,18,20,24,30$}&\frac{2(k-1)(30+31k)}{3(30+k)}&\frac{(k-1)(119+124k)}{6(30+k)}\\
\end{array}. 
\end{align}
We see that all these values of $c_{4d}$ precisely agree with the central charges computed in \cite{Xie:2016evu}. 

The supersymmetric determinant is 
\begin{align}
&
\mathfrak{D}_{\textrm{Schur}}^{J^{h^{\vee}}[k]}(z)
\nonumber\\
&=(2\pi)^{\frac{\dim J-\mathrm{rank}(J)}{2}}
(h^{\vee}+k)^{-\frac{(h^{\vee}+k+2z)\dim J}{2(h^{\vee}+k)}}
\frac{\prod_{i=1}^{\mathrm{rank}(J)} \Gamma(d_i+z)}
{\prod_{i=1}^{\mathrm{rank}(J)}\prod_{j=0}^{2d_i-2}\Gamma\left(\frac{h^{\vee}+k+1-d_i+j+z}{h^{\vee}+k}\right)}.
\end{align}
The vacuum exponent is
\begin{align}
&
\mathfrak{D}_{\textrm{Schur}}^{J^{h^{\vee}}[k]}(0)
\nonumber\\
&=(2\pi)^{\frac{\dim J-\mathrm{rank}(J)}{2}}
(h^{\vee}+k)^{-\frac{\dim J}{2}}
\frac{\prod_{i=1}^{\mathrm{rank}(J)} \Gamma(d_i)}
{\prod_{i=1}^{\mathrm{rank}(J)}\prod_{j=0}^{2d_i-2}\Gamma\left(\frac{h^{\vee}+k+1-d_i+j}{h^{\vee}+k}\right)}. 
\end{align}

\subsubsection{The generalized $(G_1,G_2)$ AD theories}
The generalized $(G_1,G_2)$ Argyres-Douglas theories are constructed in \cite{Cecotti:2010fi} 
by geometric engineering Type IIB superstring theory on the singular Calabi-Yau hypersurfaces of $\mathbb{C}^4$ 
defined as the zero locus of the polynomial associated with a pair of the ADE Lie algebras $G_1$ and $G_2$. 
They are symmetric under an exchange of $G_1$ and $G_2$. 

For $\mathrm{gcd}(h_1^{\vee},h_2^{\vee})$ $=$ $1$, 
where $h_i^{\vee}$ are the dual Coxeter numbers of $G_i$, 
we have the single-particle index of the form \cite{Xie:2019zlb}
\begin{align}
i_{\textrm{Schur}}^{(G_1,G_2)}(q)
&=\frac{
\left(\sum_{i=1}^{\mathrm{rank}(G_1)} q^{d_{i}^{(1)}-1} \right)
\left(\sum_{j=1}^{\mathrm{rank}(G_2)} q^{d_{j}^{(2)}-1} \right)
}{(1-q^{h_1^{\vee}+h_2^{\vee}})}, 
\end{align}
where $d_i^{(i)}$ are degrees of the Casimirs of $G_i$. 
The condition $\mathrm{gcd}(h_1^{\vee},h_2^{\vee})$ $=$ $1$ is satisfied 
when $G_1=J$ and $G_2=A_{k-1}$ with even $k$, which is equivalent to the AD theory $J^{h^{\vee}}[k]$. 
So we can reproduce the results in the previous section. 

The supersymmetric zeta function is given by
\begin{align}
\mathfrak{Z}_{\textrm{Schur}}^{(G_1,G_2)}(s,z)
&=\sum_{i=1}^{\mathrm{rank}(G_1)}
\sum_{j=1}^{\mathrm{rank}(G_2)}
(h_1^{\vee}+h_2^{\vee})^{-s}
\zeta\left(s,\frac{d_i^{(1)}+d_j^{(2)}-2+z}{h_1^{\vee}+h_2^{\vee}}\right). 
\end{align}
The residue at a simple pole $s=1$ is evaluated as
\begin{align}
\mathrm{Res}_{s=1}\mathfrak{Z}_{\textrm{Schur}}^{(G_1,G_2)}(s,z)
&=\frac{\mathrm{rank}(G_1)\mathrm{rank}(G_2)}{h_1^{\vee}+h_2^{\vee}}. 
\end{align}
The Zeta-index is evaluated as
\begin{align}
&
\mathfrak{Z}_{\textrm{Schur}}^{(G_1,G_2)}(0,0)
\nonumber\\
&=\frac{
\mathrm{rank}(G_1)\mathrm{rank}(G_2)(h_1^{\vee}+h_2^{\vee}+2)
-\mathrm{rank}(G_1)\dim G_2-\mathrm{rank}(G_2)\dim G_1
}{2(h_1^{\vee}+h_2^{\vee})}
\nonumber\\
&=0, 
\end{align}
where we have used the relations (\ref{d_Casimir_relation1}) and (\ref{dim_rank_dh_ADE}). 
The special supersymmetric zeta value with $s=-1$ and $z=0$ is evaluated as
\begin{align}
&
\mathfrak{Z}_{\textrm{Schur}}^{(G_1,G_2)}(-1,0)
\nonumber\\
&=\frac{1}{12(h_1^{\vee}+h_2^{\vee})}
\Biggl[
(h_1^{\vee}+3h_2^{\vee}+6)\dim G_1 \mathrm{rank}(G_2)
+
(h_2^{\vee}+3h_1^{\vee}+6)\dim G_2 \mathrm{rank}(G_1)
\nonumber\\
&-3\dim G_1 \dim G_2-(h_1^{\vee}+h_2^{\vee}+3)^2 \mathrm{rank}(G_1)\mathrm{rank}(G_2)
\Biggr]
\nonumber\\
&=\frac{(h_1^{\vee}+h_2^{\vee}+h_1^{\vee}h_2^{\vee})\mathrm{rank}(G_1)\mathrm{rank}(G_2)}{12(h_1^{\vee}+h_2^{\vee})}. 
\end{align}
Again we have used the relations (\ref{d_Casimir_relation1}) and (\ref{dim_rank_dh_ADE}). 
According to the formulae (\ref{c4dN=2_zeta}) and (\ref{a4dN=2_zeta}) we obtain 
\begin{align}
c_{4d}&=\frac{(h_1^{\vee}+h_2^{\vee}+h_1^{\vee}h_2^{\vee})\mathrm{rank}(G_1)\mathrm{rank}(G_2)}{12(h_1^{\vee}+h_2^{\vee})}, \\
a_{4d}&=\frac{(4h_1^{\vee}+4h_2^{\vee}+4h_1^{\vee}h_2^{\vee}-1)\mathrm{rank}(G_1)\mathrm{rank}(G_2)}{48(h_1^{\vee}+h_2^{\vee})}. 
\end{align}
For $G_1=J$ and $G_2=A_{k-1}$ they agree with the expressions (\ref{c_J^{h}[k]}) and (\ref{a_J^{h}[k]}) for the 4d central charges as expected. 

\subsubsection{$(J^{h^{\vee}}[k],F)$ AD matters}
\label{sec_ADmatter}
The Argyres Douglas (AD) matters can be engineered by taking 6d $\mathcal{N}=(2,0)$ theory on a sphere with an irregular puncture and a regular puncture. 
They have non-Abelian flavor symmetries and contain the Coulomb branch operators with fractional scaling dimensions. 
They are labeled by $(J^{b}[k],Y)$ \cite{Wang:2015mra} 
where $J^{b}[k]$ stands for the irregular singularity 
and $Y$ the regular singularity \cite{Gaiotto:2009we,Chacaltana:2012zy}. 
When $b$ is the dual Coxeter number $h^{\vee}$ of the Lie algebra $J$ and $Y$ is full, 
the associated VOA is conjectured to be the affine Kac-Moody algebra $\widehat{F}_{k_F}$ 
with level $k_F$ $=$ $h^{\vee}(-1+\frac{1}{h^{\vee}+k})$ \cite{Xie:2016evu}. 

The Schur index for the AD matters $(J^{h^{\vee}}[k],F)$ is given by the vacuum character of the affine Kac-Moody algebra \cite{Song:2017oew,MR3612699}. 
The single-particle index reads
\begin{align}
\label{sind_J^{h}[k],F}
i_{\textrm{Schur}}^{(J^{h^{\vee}}[k],F)}(q)
&=\dim F \frac{q-q^{h^{\vee}+k}}{(1-q)(1-q^{h^{\vee}+k})}
\nonumber\\
&=\dim F\left(\frac{1}{1-q}-\frac{1}{1-q^{h^{\vee}+k}}\right). 
\end{align}

The supersymmetric zeta function is evaluated as
\begin{align}
\label{zeta_J^{h}[k],F}
\mathfrak{Z}_{\textrm{Schur}}^{(J^{h^{\vee}}[k],F)}(s,z)
&=\dim F 
\left[
\zeta(s,z)-(h^{\vee}+k)^{-s}\zeta\left(s,\frac{z}{h^{\vee}+k}\right)
\right]. 
\end{align}
It has a simple pole at $s=1$ with residue 
\begin{align}
\label{Res1_J^{h}[k],F}
\mathrm{Res}_{s=1}\mathfrak{Z}_{\textrm{Schur}}^{(J^{h^{\vee}}[k],F)}(s,z)
&=\dim F \frac{h^{\vee}+k-1}{h^{\vee}+k}. 
\end{align}
The supersymmetric zeta value vanishes for $s=0$ and $z=0$
\begin{align}
\label{0_J^{h}[k],F}
\mathfrak{Z}_{\textrm{Schur}}^{J^{h^{\vee}}[k],F}(0,0)&=0. 
\end{align}
On the other hand, the supersymmetric zeta value at $s=-1$ is non-trivial. 
We find 
\begin{align}
\label{-1_J^{h}[k],F}
\mathfrak{Z}_{\textrm{Schur}}^{(J^{h^{\vee}}[k],F)}(-1,0)
&=\frac{\dim F (h^{\vee}+k-1)}{12}. 
\end{align}
According to the formulae (\ref{c4dN=2_zeta}) and (\ref{a4dN=2_zeta}), we get 
\begin{align}
\label{c_J^{h}[k],F}
c_{4d}&=\frac{\dim F (h^{\vee}+k-1)}{12}, \\
\label{a_J^{h}[k],F}
a_{4d}&=\frac{\dim F(h^{\vee}+k-1)(4k+4h^{\vee}-1)}{48(h^{\vee}+k)}. 
\end{align}
The actual values (\ref{c_J^{h}[k],F}) and (\ref{a_J^{h}[k],F}) are listed in the following: 
\begin{align}
\begin{array}{c|c|c|c|c|c}
(J^{h^{\vee}}[k],F)&\dim F&h^{\vee}&c_{4d}&a_{4d}&k_F\\ \hline
\scalebox{0.7}{$(A_{N-1}^{N}[k],F)$}&\scalebox{0.7}{$N^2-1$}&\scalebox{0.7}{$N$}&\frac{(N^2-1)(N+k-1)}{12}&\frac{(N^2-1)(N+k-1)(4N+4k-1)}{48(N+k)}&\frac{N(N+k-1)}{N+k} \\
\scalebox{0.7}{$(D_N^{2N-2}[k],F)$}&\scalebox{0.7}{$N(2N-1)$}&\scalebox{0.7}{$2N-2$}&\frac{N(2N-1)(2N+k-3)}{12}&\frac{N(2N-1)(2N+k-3)(8N+4k-9)}{48(2N+k-2)}&\frac{2(N-1)(2N+k-3)}{2N+k-2} \\
\scalebox{0.7}{$(E_{6}^{12}[k],F)$}&\scalebox{0.7}{$78$}&\scalebox{0.7}{$12$}&\frac{13(k+11)}{2}&\frac{13(k+1)(4k+47)}{8(k+12)}&-\frac{12(k+11)}{k+12}\\
\scalebox{0.7}{$(E_{7}^{18}[k],F)$}&\scalebox{0.7}{$133$}&\scalebox{0.7}{$18$}&\frac{133(k+17)}{12}&\frac{133(k+17)(4k+71)}{48(k+18)}&-\frac{18(k+17)}{k+18}\\
\scalebox{0.7}{$(E_{8}^{30}[k],F)$}&\scalebox{0.7}{$248$}&\scalebox{0.7}{$30$}&\frac{62(k+29)}{3}&\frac{31(k+29)(4k+119)}{6(k+30)}&-\frac{30(k+29)}{k+30}\\
\end{array}
\end{align}
We see that they precisely agree with the central charges $c_{4d}$ computed in \cite{Xie:2016evu}. 

The supersymmetric determinant is 
\begin{align}
\mathfrak{D}_{\textrm{Schur}}^{J^{h^{\vee}}[k],F}(z)
&=(h^{\vee}+k)^{\dim F(\frac12-\frac{z}{h^{\vee}+k})}
\frac{\Gamma(z)^{\dim F}}{\Gamma\left(\frac{z}{h^{\vee}+k}\right)^{\dim F}}
\end{align}
The vacuum exponent is evaluated as
\begin{align}
\mathfrak{D}_{\textrm{Schur}}^{J^{h^{\vee}}[k],F}(0)
&=(h^{\vee}+k)^{-\frac12\dim F}. 
\end{align}
where we have used the fact that
\begin{align}
\lim_{z\rightarrow 0}\frac{\Gamma(z)}{\Gamma(z/a)}&=\frac{1}{a}. 
\end{align}

\subsubsection{$(A_{N-1}^{N}[k],[N-1,1])$ AD matter}
\label{sec_ADmatterS}
Next consider the $(A_{N-1}^{N}[k],[N-1,1])$ Argyres Douglas (AD) matter 
which arises from 6d $\mathcal{N}=(2,0)$ theory of type $A_{N-1}$ wrapped on a sphere with 
an irregular puncture of type $A_{N-1}^{N}[k]$ and a simple regular puncture associated with the Young diagram $[N-1,1]$. 

The single-particle Schur index is given by \cite{Song:2017oew}
\begin{align}
\label{sind_A_{N-1}^{N}[k][N-1,1]}
i_{\textrm{Schur}}^{A_{N-1}^{N}[k][N-1,1]}(q)
&=\sum_{i=1}^{N-1}\frac{q^i}{1-q}+\frac{2q^{\frac{N}{2}}}{1-q}-\frac{q^{N+k}}{1-q^{N+k}}
\left(
\sum_{i=1}^{N-1}\sum_{j=1}^{N-1}q^{j-i}
+2\sum_{i=1}^{N-1}q^{\frac{N-2i}{2}}
\right). 
\end{align}

The supersymmetric zeta function is evaluated as
\begin{align}
\label{zeta_A_{N-1}^{N}[k][N-1,1]}
\mathfrak{Z}_{\textrm{Schur}}^{A_{N-1}^{N}[k][N-1,1]}(s,z)
&=\sum_{i=1}^{N-1}\zeta(s,i+z)+2\zeta\left(s,\frac{N}{2}+z\right)
\nonumber\\
&-\sum_{i=1}^{N-1}\sum_{j=1}^{N-1}(k+N)^{-s}\zeta\left(s,\frac{j-i+k+N+z}{k+N}\right)
\nonumber\\
&-2\sum_{i=1}^{N-1}(k+N)^{-s}
\zeta\left(s,\frac{\frac{N-2i}{2}+k+N+z}{k+N}\right). 
\end{align}
It has a simple pole at $s=1$ with residue 
\begin{align}
\label{Res1_A_{N-1}^{N}[k][N-1,1]}
\mathrm{Res}_{s=1}\mathfrak{Z}_{\textrm{Schur}}^{A_{N-1}^{N}[k][N-1,1]}(s,z)
&=\frac{(N+1)(k+1)}{N+k}. 
\end{align}
When $s=0$ and $z=0$, the supersymmetric zeta function (\ref{zeta_A_{N-1}^{N}[k][N-1,1]}) becomes zero
\begin{align}
\label{0_A_{N-1}^{N}[k][N-1,1]}
\mathfrak{Z}_{\textrm{Schur}}^{A_{N-1}^{N}[k][N-1,1]}(0,0)&=0. 
\end{align}
On the other hand, the supersymmetric zeta value for $s=-1$ and $z=0$ is 
\begin{align}
\label{-1_A_{N-1}^{N}[k],[N-1,1]}
\mathfrak{Z}_{\textrm{Schur}}^{A_{N-1}^{N}[k][N-1,1]}(-1,0)
&=\frac{(N^2(k+2)-N-k)(k+1)}{12(N+k)}. 
\end{align}
From (\ref{c4dN=2_zeta}) and (\ref{a4dN=2_zeta}), we find 
\begin{align}
\label{c_A_{N-1}^{N}[k],[N-1,1]}
c_{4d}&=\frac{(N^2(k+2)-N-k)(k+1)}{12(N+k)}, \\
\label{a_A_{N-1}^{N}[k],[N-1,1]}
a_{4d}&=\frac{(k+1)(4k(N^2-1)+8N^2-5N-1)}{48(N+k)}. 
\end{align}
When $k=1$, the $(A_{N-1}^{N}[k],[N-1,1])$ AD matter is identified with the $(A_1,A_{2N-1})$ AD theory. 
As expected, the expressions (\ref{c_A_{N-1}^{N}[k],[N-1,1]}) and (\ref{a_A_{N-1}^{N}[k],[N-1,1]}) 
agree with the central charges for the $(A_1,A_{2N-1})$ AD theory \cite{Xie:2012hs,Xie:2013jc}.\footnote{The 4d central charges for the $(A_{k-1},A_{kn-1})$ AD theory are \cite{Xie:2012hs,Xie:2013jc}
\begin{align}
c_{4d}&=\frac{(k-1)(k^2n^2+kn^2-2n-2)}{12(n+1)} \\
a_{4d}&=\frac{(k-1)(2k^2n^2+2kn^2-5n-5)}{24{n+1}}. 
\end{align}
} 

The supersymmetric determinant reads
\begin{align}
\mathfrak{D}_{\textrm{Schur}}^{A_{N-1}^{N}[k][N-1,1]}(z)
&=(2\pi)^{\frac{(N+1)(N-2)}{2}}
(k+N)^{-\frac{(N^2-1)(N+k+2z)}{2(N+k)}}
\nonumber\\
&\times 
\frac{\Gamma\left(\frac{N}{2}+z\right)^2
\prod_{i=1}^{N-1}\Gamma(i+z)}
{\prod_{i=1}^{N-1}\Gamma\left(\frac{\frac{N-2i}{2}+N+k+z}{N+k}\right)^2
\prod_{i,j=1}^{N-1} \Gamma\left(\frac{j-i+N+k+z}{N+k}\right)}. 
\end{align}
We have the vacuum exponent
\begin{align}
\mathfrak{D}_{\textrm{Schur}}^{A_{N-1}^{N}[k][N-1,1]}(0)
&=(2\pi)^{\frac{(N+1)(N-2)}{2}}
(k+N)^{-\frac{N^2-1}{2}}
\nonumber\\
&\times 
\frac{\Gamma\left(\frac{N}{2}\right)^2
\prod_{i=1}^{N-1}\Gamma(i)}
{\prod_{i=1}^{N-1}\Gamma\left(\frac{\frac{N-2i}{2}+N+k}{N+k}\right)^2
\prod_{i,j=1}^{N-1} \Gamma\left(\frac{j-i+N+k}{N+k}\right)}. 
\end{align}

\subsubsection{rank $2$ AD theory $\textrm{AD}(C_2)$}
In \cite{Kaidi:2021tgr} Kaidi and Martone constructed the rank $2$ Argyres Douglas theory $\textrm{AD}(C_2)$ 
that arises as a non-trivial IR fixed point on the Coulomb branch of the mass deformed Lagrangian theory 
with gauge group $G_2$ and $4$ fundamental hypermultiplets. 

The single-particle Schur index is given by \cite{Kaidi:2021tgr}
\begin{align}
\label{sind_ADC2}
i_{\textrm{Schur}}^{\textrm{AD}(C_2)}(q)
&=\frac{1}{1-q}\sum_{i=0}^{\infty}
(10q^{3i+1}+4q^{3i+\frac32}-4q^{3i+\frac52}-10q^{3i+3})
\nonumber\\
&=\frac{10q+4q^{\frac32}+10q^2}{1-q^3}. 
\end{align}

We find the supersymmetric zeta function 
\begin{align}
\label{zeta_ADC2}
\mathfrak{Z}_{\textrm{Schur}}^{\textrm{AD}(C_2)}(s,z)
&=10\cdot 3^{-s}\zeta\left(s,\frac{1+z}{3}\right)
+4\cdot 3^{-s}\zeta\left(s,\frac{\frac32+z}{3}\right)
+10\cdot 3^{-s}\zeta\left(s,\frac{2+z}{3}\right). 
\end{align}
The residue at a simple pole $s=1$ is
\begin{align}
\label{Res1_ADC2}
\mathrm{Res}_{s=1}\mathfrak{Z}_{\textrm{Schur}}^{\textrm{AD}(C_2)}(s,z)
&=8. 
\end{align}
The Zeta-index vanishes
\begin{align}
\label{0_ADC2}
\mathfrak{Z}_{\textrm{Schur}}^{\textrm{AD}(C_2)}(0,0)&=0. 
\end{align}
The special supersymmetric zeta value with $s=-1$ and $z=0$ is 
\begin{align}
\label{-1_ADC2}
\mathfrak{Z}_{\textrm{Schur}}^{\textrm{AD}(C_2)}(-1,0)&=\frac{13}{6}. 
\end{align}
Plugging (\ref{Res1_ADC2}) and (\ref{-1_ADC2}) into the formulae (\ref{c4dN=2_zeta}) and (\ref{a4dN=2_zeta}), 
we get 
\begin{align}
c_{4d}&=\frac{13}{6}, \\
a_{4d}&=2, 
\end{align}
which agree with the central charges of the AD theory $\textrm{AD}(C_2)$ \cite{Kaidi:2021tgr}. 

The supersymmetric determinant is
\begin{align}
\mathfrak{D}_{\textrm{Schur}}^{\textrm{AD}(C_2)}(z)
&=\frac{3^{5-2z}\Gamma(z)^{10}\Gamma\left(\frac{3+2z}{6}\right)^4}
{4\pi^2 \Gamma\left(\frac{z}{3}\right)^{10}}. 
\end{align}
The vacuum exponent is
\begin{align}
\mathfrak{D}_{\textrm{Schur}}^{\textrm{AD}(C_2)}(0)
&=\frac{1}{972}. 
\end{align}

\section{6d supersymmetric field theories}
\label{sec_6d}

\subsection{Zeta functions for supersymmetric indices}

The supersymmetric index for the 6d $\mathcal{N}=(1,0)$ supersymmetric field theories is defined as follows \cite{Bhattacharya:2008zy,Imamura:2012efi,Kim:2012qf}:
\begin{align}
\label{6d_ind_DEF}
\mathcal{I}^{\textrm{6d $(1,0)$}}(x;p;q;t)
&=\Tr (-1)^F p^{j_{12}+R}q^{j_{34}+R}t^{j_{56}+R}\prod_{\alpha}x_{\alpha}^{f_{\alpha}}, 
\end{align}
where $j_{12}$, $j_{34}$ and $j_{56}$ are the Cartan generators of the rotation group, 
$R$ is the Cartan generator of the $SU(2)_{R}$ R-symmetry, and
$f_{\alpha}$ are the Cartan generators of other global symmetries.  
The index can be computed as an integral over a product of topological string amplitudes \cite{Lockhart:2012vp}. 

The $8$-form anomaly polynomial takes the form \cite{Ohmori:2014kda}:
\begin{align}
\label{6d_anomaly}
\mathcal{A}_8&=
\frac{1}{4!}
\left[
\alpha c_2(R)^2+\beta c_2(R)p_1(T)+\gamma p_1(T)^2+\delta p_2(T)
\right]. 
\end{align}
Here $c_2(R)$ is the second Chern class of the $SU(2)_R$ R-symmetry bundle 
and $p_{1,2}(T)$ are the Pontryagin classes of the tangent bundle. 

We define the multiple supersymmetric zeta functions $\mathfrak{Z}^{\textrm{6d $(1,0)$}}(s,z;\omega_1,\omega_2,\omega_3)$ 
for the 6d $\mathcal{N}=(1,0)$ supersymmetric field theories 
by taking the Mellin transform (\ref{Mellin_transf1}) 
of the plethystic logarithm of the 6d supersymmetric index (\ref{6d_ind_DEF}) with the fugacities $p$, $q$, and $t$ being replaced with $p=q^{\omega_1}$, $q=q^{\omega_2}$, and $t=q^{\omega_3}$.
We also define by (\ref{sDet_DEF}) the associated supersymmetric determinants $\mathfrak{D}^{\textrm{6d $(1,0)$}}(z;\omega_1,\omega_2,\omega_3)$. These supersymmetric zeta functions $\mathfrak{Z}^{\textrm{6d $(1,0)$}}(s,z;\omega_1,\omega_2,\omega_3)$ may have simple poles 
at $s=3$, $2$, and $1$.

It then follows from (\ref{Cardy_lim0}) that 
\begin{align}
\label{6d_asymp1}
&
\log \mathcal{I}^{\textrm{6d $(1,0)$}}(x;q^{\omega_1};q^{\omega_2};q^{\omega_3})
\nonumber\\
&\sim \frac{\pi^4 \mathrm{Res}_{s=3}\mathfrak{Z}^{\textrm{6d $(1,0)$}}(s,0;\omega_1,\omega_2,\omega_3)}
{45\beta^3}
+\frac{\zeta(3) \mathrm{Res}_{s=2}\mathfrak{Z}^{\textrm{6d $(1,0)$}}(s,0;\omega_1,\omega_2,\omega_3)}
{\beta^2}
\nonumber\\
&+\frac{\pi^2 \mathrm{Res}_{s=1}\mathfrak{Z}^{\textrm{6d $(1,0)$}}(s,0;\omega_1,\omega_2,\omega_3)}
{6\beta}
-\mathfrak{Z}^{\textrm{6d $(1,0)$}}(0,0;\omega_1,\omega_2,\omega_3)\log\beta
\nonumber\\
&+\log \mathfrak{D}^{\textrm{6d $(1,0)$}}(0;\omega_1,\omega_2,\omega_3). 
\end{align}
Also, when we expand the indices (\ref{6d_ind_DEF}) as
\begin{align}
\mathcal{I}^{\textrm{6d $(1,0)$}}(1;q;q;q)
&=\sum_{n}d^{\textrm{6d $(1,0)$}}(n) q^n, 
\end{align}
we obtain the asymptotic degeneracy (\ref{Asymptotic_Dege2B})
\begin{align}
\label{6d_asymp2}
&
d^{\textrm{6d $(1,0)$}}(n)
\nonumber\\
&\sim 
\frac{1}{2\sqrt{2}}\left(\frac{ R_3 }{15n^5}\right)^{\frac18}
\left(\frac{15n}{\pi^4 R_3}\right)^{\frac{\mathfrak{Z}^{\textrm{6d $(1,0)$}}(0,0;1,1,1)}{4}}
\nonumber\\
&\times 
\mathfrak{D}^{\textrm{6d $(1,0)$}}(0;1,1,1)
\exp
\Biggl[
\frac{4\pi}{3}\left(\frac{R_3}{15}\right)^{\frac14} n^{\frac34}
+\frac{R_2 \zeta(3)}{\pi^2}\left(\frac{15}{R_3}\right)^{\frac12}n^{\frac12}
\nonumber\\
&+\frac{\pi^6R_1R_3-45R_2^2\zeta(3)^2}{2\pi^5}\left(\frac{15}{R_3}\right)^{\frac54}n^{\frac14}
-\frac{5(\pi^6 R_1R_2R_3\zeta(3)-60R_2^3\zeta(3)^3)}{4\pi^8}\left(\frac{15}{R_3}\right)^{2}
\Biggr], 
\end{align}
where we have introduced the abbreviations
\begin{align}
R_i&:=\mathrm{Res}_{s=i}\mathfrak{Z}^{\textrm{6d $(1,0)$}}(s,0;1,1,1). 
\end{align}

Di Pietro and Komargodski \cite{DiPietro:2014bca} proposed that the leading and subleading terms in the Cardy-like limit (\ref{6d_asymp1}) can be universally fixed by the anomaly coefficients in (\ref{6d_anomaly}). 
The proof was presented in \cite{Chang:2019uag} for the superconformal field theories with pure Higgs branches, 
including rank-$N$ E-string theories \cite{Seiberg:1996vs,Ganor:1996mu} and $(G,G)$-type minimal conformal matter theories \cite{DelZotto:2014hpa}. 
Assuming that the conjecture is true, the residues of the supersymmetric zeta functions are given by 
the anomaly coefficients 
\begin{align}
\label{Cardy_6dSCI1}
\mathrm{Res}_{s=3}\mathfrak{Z}^{\textrm{6d $(1,0)$}}(s,0;\omega_1,\omega_2,\omega_3)&=\frac{40\gamma+10\delta}{\omega_1\omega_2\omega_3}, \\
\label{Cardy_6dSCI2}
\mathrm{Res}_{s=2}\mathfrak{Z}^{\textrm{6d $(1,0)$}}(s,0;\omega_1,\omega_2,\omega_3)&=0, \\
\label{Cardy_6dSCI3}
\mathrm{Res}_{s=1}\mathfrak{Z}^{\textrm{6d $(1,0)$}}(s,0;\omega_1,\omega_2,\omega_3)&=
\frac{3(\sum_{i=1}^3\omega_i)^2\beta-16(\sum_{i=1}^3\omega_i^2)\gamma+2(\sum_{i=1}^3\omega_i^2)\delta}
{6\omega_1\omega_2\omega_3}. 
\end{align}

In addition, we propose that 
the supersymmetric zeta values with $s=-1$ and $z=0$ can be determined by the anomaly coefficients in (\ref{6d_anomaly}) 
\begin{align}
\label{Cas_6dSCI}
&
\mathfrak{Z}^{\textrm{6d $(1,0)$}}(-1,0;\omega_1,\omega_2,\omega_3)
\nonumber\\
&=-\frac{(\omega_1+\omega_2+\omega_3)^4}{192\omega_1\omega_2\omega_3}\alpha
+\frac{(\omega_1+\omega_2+\omega_3)^2(\omega_1^2+\omega_2^2+\omega_3^2)}{48\omega_1\omega_2\omega_3}\beta
\nonumber\\
&-\frac{(\omega_1^2+\omega_2^2+\omega_3^2)^2}{12\omega_1\omega_2\omega_3}\gamma
-\frac{\omega_1^2\omega_2^2+\omega_1^2\omega_3^2+\omega_2^2\omega_3^2}{12\omega_1\omega_2\omega_3}\delta. 
\end{align}
We explicitly check the proposed formula (\ref{Cas_6dSCI}) for the hypermultiplet, tensor multiplet, and vector multiplet. Comparing these values with the supersymmetric Casimir energies given by the equivariant integral proposed in \cite{Bobev:2015kza} 
(also see \cite{Yankielowicz:2017xkf}), We find that they are precisely related by the formula (\ref{zeta-1_Cas}), providing another justification for the proposed relation between the supersymmetric zeta values and the supersymmetric Casimir energies.

\subsubsection{6d $\mathcal{N}=(1,0)$ hypermultiplet}

The supersymmetric index of the 6d $\mathcal{N}=(1,0)$ free hypermultiplet takes the form \cite{Bhattacharya:2008zy,Imamura:2012efi}
\begin{align}
\label{ind_6d10hyp}
\mathcal{I}^{\textrm{6d $(1,0)$ hyp}}(x;p;q;t)
&=\frac{1}{\Gamma((pqt)^{\frac12}x;p,q,t)}, 
\end{align}
where 
\begin{align}
\Gamma(z;p,q,t)&=\prod_{n=0}^{\infty}\prod_{m=0}^{\infty}\prod_{l=0}^{\infty}
(1-zp^n q^m t^l)(1-z^{-1}p^{n+1}q^{m+1}t^{l+1})
\end{align}
is the double elliptic gamma function \cite{Spiridonov:2012de}. 
We have the single-particle index
\begin{align}
\label{sind_6d10hyp}
i^{\textrm{6d $(1,0)$ hyp}}(x;p;q;t)
&=\frac{(pqt)^{\frac12}(x+x^{-1})}{(1-p)(1-q)(1-t)}. 
\end{align}
Setting $p=t=$ $q$ and $x=1$, it becomes 
\begin{align}
\label{sind_6d10hyp_lim}
i^{\textrm{6d $(1,0)$ hyp}}(1;q;q;q)
&=\frac{2q^{\frac32}}{(1-q)^3}
\nonumber\\
&=\sum_{n=1}^{\infty}(n^2-n)q^{n-\frac12}. 
\end{align}

The supersymmetric zeta function is given by
\begin{align}
\mathfrak{Z}^{\textrm{6d $(1,0)$ hyp}}(s,z;\omega_1,\omega_2,\omega_3)
&=2\zeta_3\left(s,\frac12(\omega_1+\omega_2+\omega_3)+z;\omega_1,\omega_2,\omega_3\right), 
\end{align}
where
\begin{align}
\zeta_3(s,z;\omega_1,\omega_2,\omega_3)
&=\sum_{n_1=0}^{\infty}\sum_{n_2=0}^{\infty}\sum_{n_3=0}^{\infty}
\frac{1}{(\omega_1n_1+\omega_2n_2+\omega_3n_3+z)^s}
\end{align}
is the Barnes triple zeta function. 
When $\omega_1=\omega_2=\omega_3$ $=$ $1$, it reduces to 
\begin{align}
\label{zeta_6d10_hyp_lim}
&
\mathfrak{Z}^{\textrm{6d $(1,0)$ hyp}}(s,z;1,1,1)
\nonumber\\
&=\zeta\left(s-2,\frac12+z\right)-2z\zeta\left(s-1,\frac12+z\right)+\left(z^2-\frac14\right)\zeta\left(s,\frac12+z\right). 
\end{align}
The residues at simple poles at $s=3$, $2$ and $1$ are given by
\begin{align}
\label{Res3_6d10hyp}
\mathrm{Res}_{s=3}\mathfrak{Z}^{\textrm{6d $(1,0)$ hyp}}(s,z;\omega_1,\omega_2,\omega_3)&=\frac{1}{\omega_1\omega_2\omega_3}, 
\\
\label{Res2_6d10hyp}
\mathrm{Res}_{s=2}\mathfrak{Z}^{\textrm{6d $(1,0)$ hyp}}(s,z;\omega_1,\omega_2,\omega_3)&=-\frac{2z}{\omega_1\omega_2\omega_3}, 
\\
\label{Res1_6d10hyp}
\mathrm{Res}_{s=1}\mathfrak{Z}^{\textrm{6d $(1,0)$ hyp}}(s,z;\omega_1,\omega_2,\omega_3)&=\frac{12z^2-\omega_1^2-\omega_2^2-\omega_3^2}{12\omega_1\omega_2\omega_3}. 
\end{align}
For $s=0$ and $z=0$ the supersymmetric zeta value is zero
\begin{align}
\label{0_6d10hyp}
\mathfrak{Z}^{\textrm{6d $(1,0)$ hyp}}(0,0;\omega_1,\omega_2,\omega_3)&=0. 
\end{align}
For $s=-1$ and $z=0$ we obtain 
\begin{align}
\label{-1_6d10hyp}
&
\mathfrak{Z}^{\textrm{6d $(1,0)$ hyp}}(-1,0;\omega_1,\omega_2,\omega_3)
\nonumber\\
&=-\frac{7(\omega_1^4+\omega_2^4+\omega_3^4)+10(\omega_1^2\omega_2^2+\omega_1^2\omega_3^2+\omega_2^2\omega_3^2)}
{2880\omega_1\omega_2\omega_3}. 
\end{align}
The anomaly coefficients for the free hypermultiplet are \cite{Ohmori:2014kda}
\begin{align}
\left(\alpha,\beta,\gamma,\delta\right)&=\left(0,0,\frac{7}{240},-\frac{1}{60}\right). 
\end{align}
Plugging these values into the formulae (\ref{Cardy_6dSCI1}), (\ref{Cardy_6dSCI3}) and (\ref{Cas_6dSCI}), 
we reproduce the residues (\ref{Res3_6d10hyp}), (\ref{Res1_6d10hyp}) with $z=0$ and the supersymmetric zeta value (\ref{-1_6d10hyp}). 

The supersymmetric determinant is 
\begin{align}
\mathfrak{D}^{\textrm{6d $(1,0)$ hyp}}(z;\omega_1,\omega_2,\omega_3)
&=\Gamma_3\left(\frac{1}{2}(\omega_1+\omega_2+\omega_3)+z;\omega_1,\omega_2,\omega_3\right)^2. 
\end{align}
The vacuum exponent with $\omega_1=\omega_2=\omega_3$ $=$ $1$ is 
\begin{align}
\mathfrak{D}^{\textrm{6d $(1,0)$ hyp}}(0;1,1,1)
&=2^{\frac18} e^{\frac{3\zeta(3)}{16\pi^2}}=1.1157\ldots
\end{align}

The degeneracy $d^{\textrm{6d $(1,0)$ hyp}}(n)$ obeys the exponential growth. 
From the formula (\ref{6d_asymp2}) we obtain the asymptotic growth
\begin{align}
\label{growth_6d10hyp}
d^{\textrm{6d $(1,0)$ hyp}}(n)
&\sim \frac{1}{2^{\frac{11}{8}} \cdot 15^{\frac{1}{8}} n^{\frac{5}{8}}}
\exp \left[
\frac{4\pi}{3\cdot 15^{\frac14}}n^{\frac34}-\frac{5^{\frac14}\pi}{8\cdot 3^{\frac34}}n^{\frac14}+\frac{3\zeta(3)}{16\pi^2}
\right]. 
\end{align}
The exact numbers $d_{\textrm{exact}}^{\textrm{6d $(1,0)$ hyp}}(n)$ and the values $d_{\textrm{asymp}}^{\textrm{6d $(1,0)$ hyp}}(n)$ 
evaluated from (\ref{growth_6d10hyp}) are shown as follows: 
\begin{align}
\begin{array}{c|c|c}
n&d_{\textrm{exact}}^{\textrm{6d $(1,0)$ hyp}}(n)&d_{\textrm{asymp}}^{\textrm{6d $(1,0)$ hyp}}(n)\\ \hline 
10&2964&655.87\\
100&5.8501\times 10^{26}&1.19298\times 10^{27}\\
1000&1.05043\times 10^{161}&2.11485\times 10^{161}\\
\end{array}. 
\end{align}

\subsubsection{6d $\mathcal{N}=(1,0)$ tensor multiplet}

The single-particle index for the 6d $\mathcal{N}=(1,0)$ tensor multiplet is given by \cite{Bhattacharya:2008zy}
\begin{align}
\label{ind_6d10tensor}
i^{\textrm{6d $(1,0)$ tensor}}(p;q;t)
&=\frac{pqt-pq-pt-qt}{(1-p)(1-q)(1-t)}. 
\end{align}
When $p=t$ $=$ $q$, it reduces to 
\begin{align}
\label{sind_6d10tensor}
i^{\textrm{6d $(1,0)$ tensor}}(q;q;q)
&=\frac{-3q^2+q^3}{(1-q)^3}
\nonumber\\
&=-\sum_{n=1}^{\infty}(n^2-1)q^n. 
\end{align}

The supersymmetric zeta function is 
\begin{align}
\label{zeta_6d10tensor}
&
\mathfrak{Z}^{\textrm{6d $(1,0)$ tensor}}(s,z;\omega_1,\omega_2,\omega_3)
\nonumber\\
&=\zeta_3(s,\omega_1+\omega_2+\omega_3+z;\omega_1,\omega_2,\omega_3)
-\zeta_3(s,\omega_1+\omega_2+z;\omega_1,\omega_2,\omega_3)
\nonumber\\
&
-\zeta_3(s,\omega_1+\omega_3+z;\omega_1,\omega_2,\omega_3)
-\zeta_3(s,\omega_2+\omega_3+z;\omega_1,\omega_2,\omega_3). 
\end{align}
When $\omega_1=\omega_2=\omega_3$ $=$ $1$, it becomes 
\begin{align}
\label{zeta_6d10tensor_lim}
&
\mathfrak{Z}^{\textrm{6d $(1,0)$ tensor}}(s,z)
\nonumber\\
&=-\zeta(s-2,1+z)+2z\zeta(s-1,1+z)+(1-z^2)\zeta(s,1+z). 
\end{align}
The supersymmetric zeta function (\ref{zeta_6d10tensor}) has simple poles at $s=3$, $2$ and $1$ with residues 
\begin{align}
\label{Res3_6d10tensor}
\mathrm{Res}_{s=3}\mathfrak{Z}^{\textrm{6d $(1,0)$ tensor}}(s,z;\omega_1,\omega_2,\omega_3)
&=-\frac{1}{\omega_1\omega_2\omega_3}, \\
\label{Res2_6d10tensor}
\mathrm{Res}_{s=2}\mathfrak{Z}^{\textrm{6d $(1,0)$ tensor}}(s,z;\omega_1,\omega_2,\omega_3)
&=\frac{2z}{\omega_1\omega_2\omega_3}, \\
\label{Res1_6d10tensor}
\mathrm{Res}_{s=1}\mathfrak{Z}^{\textrm{6d $(1,0)$ tensor}}(s,z;\omega_1,\omega_2,\omega_3)
&=-\frac{6z^2+\omega_1^2+\omega_2^2+\omega_3^2-3(\omega_1\omega_2+\omega_1\omega_3+\omega_2\omega_3)}
{6\omega_1\omega_2\omega_3}. 
\end{align}
The Zeta-index is given by
\begin{align}
\label{0_6d10tensor}
\mathfrak{Z}^{\textrm{6d $(1,0)$ tensor}}(0,0;\omega_1,\omega_2,\omega_3)
&=-\frac12. 
\end{align}
For $s=-1$ and $z=0$ we have 
\begin{align}
\label{-1_6d10tensor}
&
\mathfrak{Z}^{\textrm{6d $(1,0)$ tensor}}(-1,0;\omega_1,\omega_2,\omega_3)
\nonumber\\
&=-\frac{1}{360\omega_1\omega_2\omega_3}
\Biggl[
\omega_1^4+\omega_2^4+\omega_3^4
-5(\omega_1^2\omega_2^2+\omega_1^2\omega_3^2+\omega_2^2\omega_3^2)
+15\omega_1\omega_2\omega_3(\omega_1+\omega_2+\omega_3)
\Biggr]. 
\end{align}
The anomaly coefficients for the 6d $\mathcal{N}=(1,0)$ tensor multiplet is given by \cite{Ohmori:2014kda}
\begin{align}
(\alpha,\beta,\gamma,\delta)&=\left(1,\frac12,\frac{23}{240},-\frac{29}{60}\right). 
\end{align}
Substituting these values into the formulae (\ref{Cardy_6dSCI1}), (\ref{Cardy_6dSCI3}) and (\ref{Cas_6dSCI}), 
we find the residues (\ref{Res3_6d10tensor}), (\ref{Res1_6d10tensor}) for $z=0$ and the supersymmetric zeta value (\ref{-1_6d10tensor}). 

The supersymmetric determinant is
\begin{align}
&
\mathfrak{D}^{\textrm{6d $(1,0)$ tensor}}(z;\omega_1,\omega_2,\omega_3)
\nonumber\\
&=\frac{
\Gamma_3\left(\omega_1+\omega_2+\omega_3+z;\omega_1,\omega_2,\omega_3\right)
}
{
\Gamma_3\left(\omega_1+\omega_2+z;\omega_1,\omega_2,\omega_3\right)
\Gamma_3\left(\omega_1+\omega_3+z;\omega_1,\omega_2,\omega_3\right)
\Gamma_3\left(\omega_2+\omega_3+z;\omega_1,\omega_2,\omega_3\right)
}. 
\end{align}
The vacuum exponent for $\omega_1=\omega_2=\omega_3$ $=$ $1$ is 
\begin{align}
\mathfrak{D}^{\textrm{6d $(1,0)$ tensor}}(0;1,1,1)
&=\frac{e^{\frac{\zeta(3)}{4\pi^2}}}{\sqrt{2\pi}}
=0.411276\ldots
\end{align}

\subsubsection{6d $\mathcal{N}=(1,0)$ vector multiplet}
The 6d $\mathcal{N}=(1,0)$ free vector multiplet is not a conformal field theory 
while the free vector field is scale invariant \cite{El-Showk:2011xbs}. 

The single-particle index for the 6d $\mathcal{N}=(1,0)$ vector multiplet reads \cite{Imamura:2012efi,Spiridonov:2012de}
\begin{align}
\label{sind_6d10vec}
i^{\textrm{6d $(1,0)$ vec}}(p;q;t)
&=\frac{pq+pt+qt-p-q-t-2pqt}{(1-p)(1-q)(1-t)}
\end{align}
For $p=q=t$ it reduces to 
\begin{align}
\label{sind_6d10vec_lim}
i^{\textrm{6d $(1,0)$ vec}}(q)
&=\frac{3q^2-3q-2q^3}{(1-q)^3}
\nonumber\\
&=-\frac32\sum_{n=1}^{\infty}(n^2-n)q^{n-1}
+\frac32\sum_{n=1}^{\infty}(n^2-n)q^{n}
-\sum_{n=1}^{\infty}(n^2-n)q^{n+1}. 
\end{align}

The supersymmetric zeta function is
\begin{align}
\label{zeta_6d10vec}
&\mathfrak{Z}^{\textrm{6d $(1,0)$ vec}}(s,z;\omega_1,\omega_2,\omega_3)
\nonumber\\
&=-\sum_{i=1}^3\zeta_3(s,\omega_i+z;\omega_1,\omega_2,\omega_3)
+\sum_{i<j}^3 \zeta_{3}(s,\omega_i+\omega_j+z;\omega_1,\omega_2,\omega_3)
\nonumber\\
&\quad -2\zeta_3(s,\omega_1+\omega_2+\omega_3+z;\omega_1,\omega_2,\omega_3). 
\end{align}
For $\omega_1=\omega_2=\omega_3$ $=$ $1$, it becomes 
\begin{align}
\label{zeta_6d10vec_lim}
&
\mathfrak{Z}^{\textrm{6d $(1,0)$ vec}}(s,z)
\nonumber\\
&=-\frac32\left[
\zeta(s-2,z)-(2z-1)\zeta(s-1,z)+(z^2-z)\zeta(s,z)
\right]
\nonumber\\
&+\frac32\left[
\zeta(s-2,z+1)-(2z+1)\zeta(s-1,z+1)+(z^2+z)\zeta(s,z+1)
\right]
\nonumber\\
&-\left[
\zeta(s-2,z+2)-(2z+3)\zeta(s-1,z+2)+(z^2+3z+2)\zeta(s,z+2)
\right]. 
\end{align}
It has simple poles at $s=3$, $2$ and $1$ with residues 
\begin{align}
\label{Res3_6d10vec}
\mathrm{Res}_{s=3}\mathfrak{Z}^{\textrm{6d $(1,0)$ vec}}(s,z;\omega_1,\omega_2,\omega_3)
&=-\frac{1}{\omega_1\omega_2\omega_3}, \\
\label{Res2_6d10vec}
\mathrm{Res}_{s=2}\mathfrak{Z}^{\textrm{6d $(1,0)$ vec}}(s,z;\omega_1,\omega_2,\omega_3)
&=\frac{2z}{\omega_1\omega_2\omega_3}, \\
\label{Res1_6d10vec}
\mathrm{Res}_{s=1}\mathfrak{Z}^{\textrm{6d $(1,0)$ vec}}(s,z;\omega_1,\omega_2,\omega_3)
&=-\frac{6z^2+\omega_1^2+\omega_2^2+\omega_3^2+3(\omega_1\omega_2+\omega_1\omega_3+\omega_2\omega_3)}
{6\omega_1\omega_2\omega_3}. 
\end{align}
The Zeta-index is 
\begin{align}
\label{0_6d10vec}
\mathfrak{Z}^{\textrm{6d $(1,0)$ vec}}(0,0;\omega_1,\omega_2,\omega_3)&=1. 
\end{align}
For $s=-1$ and $z=0$ the supersymmetric zeta value is 
\begin{align}
\label{-1_6d10vec}
&
\mathfrak{Z}^{\textrm{6d $(1,0)$ vec}}(-1,0;\omega_1,\omega_2,\omega_3)
\nonumber\\
&=-\frac{\omega_1^4+\omega_2^4+\omega_3^4
-5(\omega_1^2\omega_2^2+\omega_1^2\omega_3^2+\omega_2^2\omega_3^2)
-15\omega_1\omega_2\omega_3(\omega_1+\omega_2+\omega_3)}
{360\omega_1\omega_2\omega_3}. 
\end{align}

The anomaly coefficients for the 6d $\mathcal{N}=(1,0)$ vector multiplet are \cite{Ohmori:2014kda}
\begin{align}
(\alpha,\beta,\gamma,\delta)&=\left(-1,-\frac12,-\frac{7}{240},\frac{1}{60}\right). 
\end{align}
We recover the residues (\ref{Res3_6d10vec}), (\ref{Res1_6d10vec}) for $z=0$ and the supersymmetric zeta value (\ref{-1_6d10vec}) 
by inserting these values into the formulae (\ref{Cardy_6dSCI1}), (\ref{Cardy_6dSCI3}) and (\ref{Cas_6dSCI}). 

The supersymmetric determinant is
\begin{align}
&
\mathfrak{D}^{\textrm{6d $(1,0)$ vec}}(z;\omega_1,\omega_2,\omega_3)
\nonumber\\
&=\frac{\prod_{i<j}^3 \Gamma_3\left(\omega_i+\omega_j+z;\omega_1,\omega_2,\omega_3\right)}
{\Gamma_3\left(\omega_1+\omega_2+\omega_3+z;\omega_1,\omega_2,\omega_3\right)^2
\prod_{i=1}^3 \Gamma_3(\omega_i+z;\omega_1,\omega_2,\omega_3)}. 
\end{align}
When $\omega_1=\omega_2=\omega_3$ $=$ $1$, the vacuum exponent is
\begin{align}
\mathfrak{D}^{\textrm{6d $(1,0)$ vec}}(0;1,1,1)&=2\pi e^{\frac{\zeta(3)}{4\pi^2}}=6.47744\ldots
\end{align}

\subsubsection{6d $\mathcal{N}=(2,0)$ tensor multiplet}

The 6d $\mathcal{N}=(2,0)$ free tensor multiplet consists of the 6d $\mathcal{N}=(1,0)$ free tensor multiplet and hypermultiplet. 

The single-particle index is given by \cite{Bhattacharya:2008zy,Kim:2012qf}
\begin{align}
\label{sind_6d20tensor}
i^{\textrm{6d $(2,0)$ tensor}}(x;p,q,t)
&=i^{\textrm{6d $(1,0)$ tensor}}(p;q;t)+i^{\textrm{6d $(1,0)$ hyp}}(x;p;q;t)
\nonumber\\
&=\frac{(pqt)^{\frac12}(x+x^{-1})-pq-pt-qt+pqt}{(1-p)(1-q)(1-t)}. 
\end{align}
With $p=t=q$ and $x=1$, it becomes 
\begin{align}
\label{sind_6d20tensor_lim}
i^{\textrm{6d $(2,0)$ tensor}}(1;q;q;q)
&=i^{\textrm{6d $(1,0)$ tensor}}(q;q;q)+i^{\textrm{6d $(1,0)$ hyp}}(1;q;q;q)
\nonumber\\
&=\frac{2q^{\frac32}-3q^2+q^3}{(1-q)^3}. 
\end{align}

The supersymmetric zeta function is 
\begin{align}
\label{zeta_6d20tensor}
&
\mathfrak{Z}^{\textrm{6d $(2,0)$ tensor}}(s,z;\omega_1,\omega_2,\omega_3)
\nonumber\\
&=\zeta_3(s,\omega_1+\omega_2+\omega_3+z;\omega_1,\omega_2,\omega_3)
-\zeta_3(s,\omega_1+\omega_2+z;\omega_1,\omega_2,\omega_3)
\nonumber\\
&-\zeta_3(s,\omega_1+\omega_3+z;\omega_1,\omega_2,\omega_3)
-\zeta_3(s,\omega_2+\omega_3+z;\omega_1,\omega_2,\omega_3)
\nonumber\\
&+2\zeta_3\left(s,\frac12(\omega_1+\omega_2+\omega_3)+z;\omega_1,\omega_2,\omega_3\right). 
\end{align}
When $\omega_1=\omega_2=\omega_3$ $=$ $1$, it becomes 
\begin{align}
\label{zeta_6d20tensor_lim}
&
\mathfrak{Z}^{\textrm{6d $(2,0)$ tensor}}(s,z;1,1,1)
\nonumber\\
&=-\zeta(s-2,1+z)+2z\zeta(s-1,1+z)+(1-z^2)\zeta(s,1+z)
\nonumber\\
&+\zeta\left(s-2,\frac12+z\right)-2z\zeta\left(s-1,\frac12+z\right)+\left(z^2-\frac14\right)\zeta\left(s,\frac12+z\right). 
\end{align}
The supersymmetric zeta function (\ref{zeta_6d20tensor}) has simple poles at $s=3$, $2$ and $1$ with residues 
\begin{align}
\label{Res3_6d20tensor}
\mathrm{Res}_{s=3}\mathfrak{Z}^{\textrm{6d $(2,0)$ tensor}}(s,z;\omega_1,\omega_2,\omega_3)
&=0, \\
\label{Res2_6d20tensor}
\mathrm{Res}_{s=2}\mathfrak{Z}^{\textrm{6d $(2,0)$ tensor}}(s,z;\omega_1,\omega_2,\omega_3)
&=0, \\
\label{Res1_6d20tensor}
\mathrm{Res}_{s=1}\mathfrak{Z}^{\textrm{6d $(2,0)$ tensor}}(s,z;\omega_1,\omega_2,\omega_3)
&=-\frac{\omega_1^2+\omega_2^2+\omega_3^2-2(\omega_1\omega_2+\omega_1\omega_3+\omega_2\omega_3)}
{4\omega_1\omega_2\omega_3}. 
\end{align}
We have the Zeta-index 
\begin{align}
\label{0_6d20tensor}
\mathfrak{Z}^{\textrm{6d $(2,0)$ tensor}}(0,0;\omega_1,\omega_2,\omega_3)&=-\frac12.
\end{align}
The supersymmetric zeta value for $s=-1$ and $z=0$ is 
\begin{align}
\label{-1_6d20tensor}
&
\mathfrak{Z}^{\textrm{6d $(2,0)$ tensor}}(-1,0;\omega_1,\omega_2,\omega_3)
\nonumber\\
&=-\frac{\omega_1^4+\omega_2^4+\omega_3^4-
2(\omega_1^2\omega_2^2+\omega_1^2\omega_3^2+\omega_2^2\omega_3^2)
+8\omega_1\omega_2\omega_3(\omega_1+\omega_2+\omega_3)
}
{192\omega_1\omega_2\omega_3}. 
\end{align}
Substituting the supersymmetric zeta value (\ref{-1_6d20tensor}) into (\ref{zeta-1_Cas}), 
we reproduce the supersymmetric Casimir energy in \cite{Kim:2012qf,Bobev:2015kza}.
The anomaly coefficients of the 6d $\mathcal{N}=(2,0)$ free tensor multiplet are \cite{Witten:1996hc}
\begin{align}
(\alpha,\beta,\gamma,\delta)&=\left(1,\frac12,\frac18,-\frac12\right). 
\end{align} 
Again the expressions (\ref{Res1_6d20tensor}), (\ref{Res3_6d20tensor}) and (\ref{-1_6d20tensor}) 
can be found by plugging the anomaly coefficients into the formulae (\ref{Cardy_6dSCI1}), (\ref{Cardy_6dSCI3}) and (\ref{Cas_6dSCI}). 

The supersymmetric determinant is
\begin{align}
&
\mathfrak{D}^{\textrm{6d $(2,0)$ tensor}}(z;\omega_1,\omega_2,\omega_3)
\nonumber\\
&=\mathfrak{D}^{\textrm{6d $(1,0)$ tensor}}(z;\omega_1,\omega_2,\omega_3)
\mathfrak{D}^{\textrm{6d $(1,0)$ hyp}}(z;\omega_1,\omega_2,\omega_3)
\nonumber\\
&=\frac{
\Gamma_3\left(\omega_1+\omega_2+\omega_3;\omega_1,\omega_2,\omega_3\right)
\Gamma_3\left(\frac{1}{2}(\omega_1+\omega_2+\omega_3);\omega_1,\omega_2,\omega_3\right)^2
}
{
\Gamma_3\left(\omega_1+\omega_2;\omega_1,\omega_2,\omega_3\right)
\Gamma_3\left(\omega_1+\omega_3;\omega_1,\omega_2,\omega_3\right)
\Gamma_3\left(\omega_2+\omega_3;\omega_1,\omega_2,\omega_3\right)
}. 
\end{align}
The vacuum exponent with $\omega_1=\omega_2=\omega_3$ $=$ $1$ is
\begin{align}
\mathfrak{D}^{\textrm{6d $(2,0)$ tensor}}(0;1,1,1)&=\frac{e^{\frac{7\zeta(3)}{16\pi^2}}}{2^{\frac38}\sqrt{\pi}}
=0.45886\ldots
\end{align}

\subsection{Zeta functions for unrefined indices}

\subsubsection{6d $\mathcal{N}=(2,0)$ theories}
It is conjectured \cite{Beem:2014kka} that 6d $\mathcal{N}=(2,0)$ theories of type $\mathfrak{g}$ $=$ $\{A_n, D_n, E_n\}$ admit 
the protected chiral algebra that is isomorphic to the W-algebra $\mathcal{W}(\mathfrak{g})$ of central charge 
\begin{align}
c&=\mathrm{rank}(\mathfrak{g})+4\dim \mathfrak{g} \cdot h^{\vee}. 
\end{align}

The relevant index is called the unrefined index \cite{Kim:2012qf}. 
In our conventions it is obtained from the supersymmetric index of the 6d $\mathcal{N}=(2,0)$ theory of type $\mathfrak{g}$ 
by setting $x=q^{\frac12}$ and $p=t=q$. 
The unrefined index is identified with the vacuum character of the $\mathcal{W}(\mathfrak{g})$. 
The associated single-particle index takes the form \cite{Kim:2012qf,Beem:2014kka}
\begin{align}
i_{\textrm{unref}}^{\textrm{6d $(2,0)$ $\mathfrak{g}$}}(q)
&=\frac{\sum_{i=1}^{\mathrm{rank}(\mathfrak{g})} q^{d_i}}{1-q}, 
\end{align}
where $d_i$ are the degrees of the Casimir of $\mathfrak{g}$. 
This expression can be explicitly checked for type $\mathfrak{g}$ $=$ $\{A_n, D_n\}$ by analyzing the supersymmetric partition functions \cite{Kim:2012qf}, 
while it is conjectural for type $E_n$. 

The supersymmetric zeta function is given by
\begin{align}
\mathfrak{Z}_{\mathrm{unref}}^{\textrm{6d $(2,0)$ $\mathfrak{g}$}}(s,z)
&=\sum_{i=1}^{\mathrm{rank}(\mathfrak{g})}\zeta(s,d_i+z). 
\end{align}
It has a simple pole at $s=1$ with residue 
\begin{align}
\mathrm{Res}_{s=1}\mathfrak{Z}_{\mathrm{unref}}^{\textrm{6d $(2,0)$ $\mathfrak{g}$}}(s,z)
&=\mathrm{rank}(\mathfrak{g}). 
\end{align}
The supersymmetric zeta value for $s=0$ and $z=0$ is given by
\begin{align}
\mathfrak{Z}_{\mathrm{unref}}^{\textrm{6d $(2,0)$ $\mathfrak{g}$}}(0,0)
&=-\frac{\dim \mathfrak{g}}{2}. 
\end{align}
With $s=-1$ and $z=0$ the supersymmetric zeta value is 
\begin{align}
\mathfrak{Z}_{\mathrm{unref}}^{\textrm{6d $(2,0)$ $\mathfrak{g}$}}(-1,0)
&=-\frac{1}{12}
\left[ \mathrm{rank}(\mathfrak{g})+2\dim \mathfrak{g}\cdot h^{\vee} \right]. 
\end{align}

The supersymmetric determinant is 
\begin{align}
\mathfrak{D}_{\textrm{unref}}^{\textrm{6d $(2,0)$ $\mathfrak{g}$}}(z)
&=\frac{1}{(2\pi)^{\frac{\mathrm{rank}(\mathfrak{g})}{2}}}
\prod_{i=1}^{\mathrm{rank}(\mathfrak{g})}
\Gamma(d_i+z). 
\end{align}
The vacuum exponent is
\begin{align}
\mathfrak{D}_{\textrm{unref}}^{\textrm{6d $(2,0)$ $\mathfrak{g}$}}(0)
&=\frac{1}{(2\pi)^{\frac{\mathrm{rank}(\mathfrak{g})}{2}}} 
\prod_{i=1}^{\mathrm{rank}(\mathfrak{g})}
\Gamma(d_i). 
\end{align}
where the factor $\prod_i \Gamma(d_i)$ is the classical Lie superfactorial associated with $\mathfrak{g}$. 

As the degeneracy $d_{\textrm{unref}}^{\textrm{6d $(2,0)$ $\mathfrak{g}$}}(n)$ grows exponentially, 
we find from (\ref{Asymptotic_Dege}) the asymptotic growth
\begin{align}
\label{growth_6d20W}
d_{\textrm{unref}}^{\textrm{6d $(2,0)$ $\mathfrak{g}$}}(n)
&\sim \frac12 
\left(
\frac{\mathrm{rank}(\mathfrak{g})}{6n^3}
\right)^{\frac14}
\left(
\frac{\pi^2 \mathrm{rank}(\mathfrak{g})}{6n}
\right)^{\frac{\dim \mathfrak{g}}{4}}
\frac{1}{(2\pi)^{\frac{\mathrm{rank}(\mathfrak{g})}{2}}} 
\prod_{i=1}^{\mathrm{rank}(\mathfrak{g})}
\Gamma(d_i)
\nonumber\\
&\times 
\exp\left[
2\pi\sqrt{\frac{\mathrm{rank}(\mathfrak{g})}{6}}
n^{\frac12}
\right]. 
\end{align}
For example, for $\mathfrak{g}$ $=$ $A_{4}$ 
the exact values $d_{\textrm{exact}}^{\textrm{6d $(2,0)$ $A_4$}}(n)$ 
and the values $d_{\textrm{asymp}}^{\textrm{6d $(2,0)$ $A_4$}}(n)$ estimated from (\ref{growth_6d20W}) are shown as follows: 
\begin{align}
\begin{array}{c|c|c}
n&d_{\textrm{exact}}^{\textrm{6d $(2,0)$ $A_4$}}(n)&d_{\textrm{asymp}}^{\textrm{6d $(2,0)$ $A_4$}}(n) \\ \hline 
10&72&528251\\
100&4.35542\times 10^{12}&1.61214\times 10^{14}\\
1000&1.20684\times 10^{55}&4.29856\times 10^{55}\\
\end{array}. 
\end{align}

\subsection*{Acknowledgements}
YN is in part supported by JSPS KAKENHI Grant Number 21K03581. TO was supported by the Startup Funding no.\ 4007012317 of Southeast University.  

\appendix

\section{Notations}
\label{app_convention}
We define 
\begin{align}
\label{qpoch_def}
(a;q)_{0}&:=1,\qquad
(a;q)_{n}:=\prod_{k=0}^{n-1}(1-aq^{k}),\qquad 
(q)_{n}:=\prod_{k=1}^{n}(1-q^{k}),
\nonumber \\
(a;q)_{\infty}&:=\prod_{k=0}^{\infty}(1-aq^{k}),\qquad 
(q)_{\infty}:=\prod_{k=1}^{\infty} (1-q^k), 
\end{align}
with $a, q \in \mathbb{C}$ and $|q|<1$. 
We also introduce the following notations: 
\begin{align}
(x^{\pm};q)_{n}&:=(x;q)_{n}(x^{-1};q)_{n}, \\
(x_1,\cdots,x_k;q)_{n}&:=(x_1;q)_{n}\cdots (x_k;q)_{n}. 
\end{align}

\section{Residue formulae}
\label{app_res}
The Barnes multiple zeta function $\zeta_N(s,z;\{\omega_i\})$ has simple poles at $s=1$, $2$, $\cdots$, $N$. 
The residues at these poles are given by (\ref{res_multipleBzeta}). 
Explicitly, the residues at the poles $s=N$, $N-1$, $N-2$ and $N-3$ are 
\begin{align}
\mathrm{Res}_{s=N}\ \zeta_{N}(s,z;\{\omega_i\})
&=\frac{1}{\Gamma(N)}\prod_{i=1}^{N}\frac{1}{\omega_i}, \\
\mathrm{Res}_{s=N-1}\ \zeta_{N}(s,z;\{\omega_i\})
&=\frac{1}{\Gamma(N-1)}\frac{\sum_{i=1}^{N}\omega_i-2z}{2\prod_{i=1}^{N}\omega_i}, \\
\mathrm{Res}_{s=N-2}\ \zeta_{N}(s,z;\{\omega_i\})
&=\frac{1}{\Gamma(N-2)}\frac{\sum_{i=1}^{N}\omega_i^2+3\sum_{i<j}\omega_i\omega_j-6\sum_{i=1}^N\omega_i z+6z^2}
{12\prod_{i=1}^{N}\omega_i}, \\
\mathrm{Res}_{s=N-3}\ \zeta_{N}(s,z;\{\omega_i\})
&=\frac{1}{\Gamma(N-3)}\frac{1}
{24\prod_{i=1}^{N}\omega_i}
\nonumber\\
&\times 
\left[
(\sum_{i=1}^{N}\omega_i)(\sum_{i<j}\omega_i\omega_j)
-2\sum_{i=1}^{N}\omega_i^2 z-6\sum_{i<j}\omega_i\omega_j z
+6\sum_{i=1}^N\omega_i z^2-4z^3
\right]. 
\end{align}
When $N=1$ and $\omega_1=1$, the Barnes multiple zeta function becomes the Hurwitz zeta function
\begin{align}
\zeta(s,z)&=\sum_{n=0}^{\infty}\frac{1}{(n+z)^s}. 
\end{align}
Furthermore, it reduces to the Riemann zeta function when $z=1$. 
The Hurwitz zeta function and the Riemann zeta function have a simple pole at $s=1$ with residue $1$. 

\section{Special zeta values}
\label{app_values}

\subsection{Riemann zeta function}
The zeta values for any positive even integer $2k$ are given by
\begin{align}
\zeta(2k)&=-\frac{B_{2k}}{2(2k)!}(2\pi i)^{2k},  
\end{align}
where $B_{2k}$ are the Bernoulli numbers. 
For negative integers $-k<0$ we have
\begin{align}
\zeta(-k)&=-\frac{B_{k+1}}{k+1}. 
\end{align}
The special values of the $s$-derivative of the Riemann zeta function $\zeta(s)$ at $s=-1$ and $s=-2$ are given by
\begin{align}
\label{RzetaD_-1}
\zeta'(-1)&=\frac{1}{12}-\log A=-0.165421\ldots, \\
\zeta'(-2)&=-\frac{\zeta(3)}{4\pi^2}=-0.0304485\ldots, 
\end{align}
where $A$ is the Glaisher-Kinkelin constant. 

\subsection{Hurwitz zeta function}
The special values of the Hurwitz zeta function are
\begin{align}
\label{values_Hurzeta}
\zeta(1-m,a)&=-\frac{B_m(a)}{m}, 
\end{align}
where $B_j(a)$ are the Bernoulli polynomials. 

\begin{figure}
\begin{center}
\includegraphics[width=10cm]{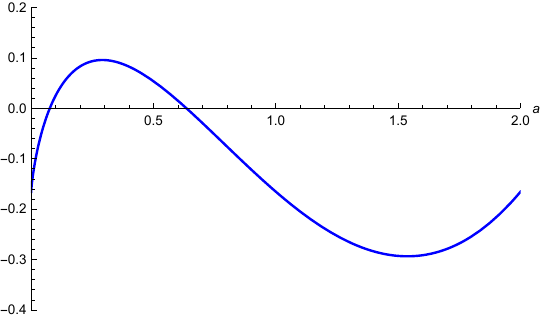}
\caption{The special value $\zeta'(-1,a)$. }
\label{fig_HurwitzD}
\end{center}
\end{figure}

For example, at $s=0$, $-1$ and $-2$ they are 
\begin{align}
\zeta(0,a)&=\frac12-a, \\
\zeta(-1,a)&=-\frac12\left(a^2-a+\frac16\right), \\
\zeta(-2,a)&=-\frac13\left(a^3-\frac32a^2+\frac{1}{2}a\right). 
\end{align}
The special value of the $s$-derivative of the Hurwitz zeta function at $s=0$ is
\begin{align}
\zeta'(0,a)&=\log \Gamma(a)-\frac12 \log(2\pi). 
\end{align}
The explicit formulae for the other special values of the $s$-derivative of the Hurwitz zeta function 
are obtainable from the functional equation for the Hurwitz zeta function (see e.g. \cite{MR1809963,MR1659109}). 
For example, we have
\begin{align}
\zeta'(-1,\frac16)&=\frac{\log(12)}{144}-\frac{\pi}{12\sqrt{3}}+\frac{\psi^{(1)}(\frac13)}{8\sqrt{3}\pi}+\frac16\left(\frac{1}{12}-\log A\right)
=0.0704526\ldots, \\
\zeta'(-1,\frac14)&=\frac{G}{4\pi}-\frac{1}{8}\left(\frac{1}{12}-\log A\right)
=0.0935679\ldots, \\
\zeta'(-1,\frac13)&=-\frac{\log(3)}{72}-\frac{\pi}{18\sqrt{3}}
+\frac{\psi^{(1)}(\frac13)}{12\sqrt{3}\pi}-\frac13\left(\frac{1}{12}-\log A\right)
=0.0937262\ldots, \\
\zeta'(-1,\frac12)&=-\frac{\log(2)}{24}-\frac12\left(\frac{1}{12}-\log A\right)
=0.0538294\ldots, \\
\zeta'(-1,\frac23)&=-\frac{\log(3)}{72}+\frac{\pi}{18\sqrt{3}}-\frac{\psi^{(1)}(\frac13)}{12\sqrt{3}\pi}-\frac13\left(\frac{1}{12}-\log A\right)
=-0.0139624\ldots, 
\end{align}
where $G=0.915965\ldots$ is Catalan's constant. 
We present numerical plots of the special values $\zeta'(-1,a)$ in Figure \ref{fig_HurwitzD}. 
The values at $a=0$ and $a=2$ are equal to (\ref{RzetaD_-1})
\begin{align}
\zeta'(-1,0)=\zeta'(-1,2)=\frac{1}{12}-\log(A)=-0.165421\ldots 
\end{align}

\subsection{Barnes multiple zeta function}
For the Barnes multiple zeta function $\zeta_N(s,z;\{\omega_i\})$ with $z>0$ let 
\begin{align}
z&=z_1\omega_1+\cdots+z_N\omega_N
\end{align}
with $z_i$ $\in$ $\mathbb{R}$. 
Then the special values of the Barnes multiple zeta function are given by
\begin{align}
\label{values_multipleBzeta}
&
\zeta_N(1-m,z;\{\omega_i\})
\nonumber\\
&=(-1)^N (m-1)! \sum_{
\begin{smallmatrix}
m_1+\cdots+m_N=m+N-1\\
m_1\ge0,\cdots, m_N\ge0\\
\end{smallmatrix}
}\left(
\prod_{i=1}^N \frac{B_{m_i}(z_i)}{m_i!}
\right)
\omega_1^{m_1-1}\cdots 
\omega_N^{m_N-1}
\end{align}
for $m\in \mathbb{N}$. 
Here $B_j(a)$ are the Bernoulli polynomials. 
When $N=1$ and $\omega_1=1$, it reduces to (\ref{values_Hurzeta}). 

In particular, the special values of the Barnes double zeta function for non-positive integers 
can be obtained using the result in \cite{MR1607965} (also see \cite{MR2078341}). 
We have 
\begin{align}
\label{values_doubleBzeta}
\zeta_2(-m,z;a,b)
&=a^m
\left[
-\frac{B_{m+1}(\frac{z}{a})}{m+1}
+(-1)^m m! \sum_{j=0}^{m+2}
\frac{B_j B_{m+2-j}(1-\frac{z}{a})}{j!(m+2-j)!}\left(\frac{b}{a}\right)^{j-1}
\right]
\end{align}
for $m\ge 0$, 
where $B_j$ are the Bernoulli numbers.

\bibliographystyle{utphys}
\bibliography{ref}

\end{document}